\documentclass[12pt]{article}
\usepackage{theorem}
\usepackage{amsmath,amssymb}
\usepackage{graphicx,makeidx}
\usepackage{caption,subcaption}
\usepackage[mathscr]{eucal}

\usepackage{titlesec}
\titleformat{\subsection}[runin]
 {\normalfont\normalsize\bfseries}{\thesubsection}{1em}{\!\!}
\titleformat{\subsubsection}[runin]
 {\normalfont\normalsize\bfseries}{\thesubsubsection}{1em}{\!\!}
{\theorembodyfont{\upshape} }
\newcommand{\boma}[1]{{\mbox{\boldmath $#1$} }}

\begin{document}
\def\ov{\overline}
\def\raL{\ra}
\def\Cd{C_d}
\def\E0{E_0}
\def\Dik{\Dir^{(\mm)}}
\def\vi{z}
\def\zi{v}
\def\vt{v}
\def\En{\mathfrak{E}}
\def\TAM{A_{(>)}}
\def\TAm{A_{(<)}}
\def\InB{\mathfrak{B}^\s}
\def\eu{\varepsilon^{\s}}
\def\eren{\varepsilon^{ren}}
\def\TT{\mathscr{T}}
\def\TTs{\TT^{\s}}
\def\an{\lambda}
\def\ET{{}_T\mathcal{E}}
\def\Ep{{}_p\mathcal{E}}
\def\b0{\textbf{0}}
\def\Ta{\mathsf{T}}
\def\Bp{{B_{\sp,n}}}
\def\Bs{{B_{\si,n}}}
\def\sp{{\bar{\si}}}
\def\Sp{{\bar{S}}}
\def\ba{{\bf a}}
\def\bap{\mathfrak{a}}
\def\sip{\mathfrak{p}}
\def\bj{\sip(j)}
\def\Op{\mathfrak{O}}
\def\fo{\mathfrak{F}}
\def\HE{H}
\def\uu{v}
\def\AM{A}
\def\Am{a}
\def\DD{D}
\def\blp{\bar{\bf l}}
\def\bkl{b_{\bk\bl}}
\def\bklp{\bar{b}_{\bk\blp}}
\def\bxp{\bar{\bx}}
\def\bkp{\bar{\bk}}
\def\CB{C}
\def\aa{\alpha}
\def\bxk{\bx_{\star}}
\def\xk{x_\star}
\def\Co{\lozenge}
\def\NCo{\blacksquare}
\def\Toro{{\bf T}}
\def\atanh{\mbox{arcth}\,}
\def\Iun{\mathfrak{J}}
\def\GaL{\mathfrak{g}}
\def\Mk{M_{\mm,k}}
\def\ri{\mathfrak{r}}
\def\oms{\omega_{*}}
\def\Fock{\mathfrak{F}}
\def\Ee{\mathfrak{E}}
\def\ee{\epsilon}
\def\Ome{\Om_\ee}
\def\bxe{\bx_\ee}
\def\FI{\boma{S}}
\def\dFI{\mathfrak{S}}
\def\ff{\mathcal{F}}
\def\hh{\mathcal{H}}
\def\Mel{\mathfrak{M}}
\def\Tr{\mbox{Tr}\,}
\def\xb{\bar{x}}
\def\rb{\bar{\rho}}
\def\rr{\bar{r}}
\def\uno{\mbox{\textbf{1}}}
\def\tez{\al}
\def\Res{\mbox{Res}}
\def\L2m0{L^2_0}
\def\Tm{\tilde{T}}
\def\II{\mathfrak{I}}
\def\Ns{\mathfrak{N}}
\def\Tti{\tilde{T}}
\def\DF{\mathscr{D}}
\def\F{{\mathcal F}}
\def\mm{\kappa}
\def\oo{\varpi}
\def\0{{\bf 0}}
\def\UU{{\mathcal U}}
\def\Hank{\mathfrak{H}}
\def\t{\mathfrak{t}}
\def\Dir{D}
\def\l{\left}
\def\r{\right}
\def\ha{\widehat{a}}
\def\ak{\ha_k}
\def\ah{\ha_h}
\def\had{\ha^{\dagger}}
\def\akd{\had_k}
\def\ahd{\had_h}
\def\GD{\mathfrak{G}}
\def\Fk{F_k}
\def\Fh{F_h}
\def\Fkc{\overline{\Fk}}
\def\Fhc{\overline{\Fh}}
\def\Fkd{\mathfrak{F}_{k_1}}
\def\fk{f_k}
\def\fh{f_h}
\def\fkc{\overline{\fk}}
\def\fhc{\overline{\fh}}
\def\bl{{\bf l}}
\def\bn{{\bf n}}
\def\bk{{\bf k}}
\def\bh{{\bf h}}
\def\bx{{\bf x}}
\def\by{{\bf y}}
\def\bz{{\bf z}}
\def\bq{{\bf q}}
\def\bp{{\bf p}}
\def\Nab{\square}
\def\Fi{\widehat{\phi}}
\def\s{u}
\def\Fis{\Fi^{\s}}
\def\Fiseps{\Fi^{\eps \s}}
\def\Ti{\widehat{T}}
\def\Tis{\Ti^{\s}}
\def\Tiseps{\Ti^{\eps \s}}
\def\Aa{\widehat{A}}
\def\Bb{\widehat{B}}
\def\al{\alpha}
\def\be{\beta}
\def\de{\delta}
\def\eps{\varepsilon}
\def\ga{\gamma}
\def\lam{\lambda}
\def\om{\omega}
\def\si{\sigma}
\def\te{\theta}
\def\Ga{\Gamma}
\def\Om{\Omega}
\def\Si{\Sigma}
\def\dd{\displaystyle}
\def\la{\langle}
\def\ra{\rangle}
\def\leqs{\leqslant}
\def\geqs{\geqslant}
\def\sc{\cdot}
\def\restriction{\upharpoonright}
\def\parn{\par\noindent}
\def\complessi{{\bf C}}
\def\reali{{\bf R}}
\def\razionali{{\bf Q}}
\def\interi{{\bf Z}}
\def\naturali{{\bf N}}
\def\AA{{\mathcal A}}
\def\BB{{\mathcal B}}
\def\FF{{\mathcal F}}
\def\EE{{\mathcal E}}
\def\GG{{\mathcal G} }
\def\HH{{\mathcal H}}
\def\JJ{{\mathcal J}}
\def\KK{{\mathcal K}}
\def\LL{{\mathcal L}}
\def\MM{{\mathcal M}}
\def\OO{{\mathcal O}}
\def\PP{{\mathcal P}}
\def\QQ{{\mathcal Q}}
\def\RR{{\mathcal R}}
\def\SS{{\mathcal S}}
\def\cir{{\scriptscriptstyle \circ}}
\def\parn{\par \noindent}
\def\salto{\vskip 0.2truecm \noindent}
\def\beq{\begin{equation}}
\def\feq{\end{equation}}
\def\barray{\begin{array}}
\def\farray{\end{array}}
\newcommand{\rref}[1]{(\ref{#1})}
%%%%%%%%%%%%%%%%%%%%%%%%%%%%%%%%%
\def\vain{\rightarrow}
\def\spazio{\vskip 0.5truecm \noindent}
\def\fine{\hfill $\square$ \vskip 0.2cm \noindent}
\def\ffine{\hfill $\lozenge$ \vskip 0.2cm \noindent}

\setcounter{secnumdepth}{5}

%%%%%%%%% THIS NUMBERS EQUATIONS BY SECTIONS %%%%%%%%%%%%%
\makeatletter \@addtoreset{equation}{section}
\renewcommand{\theequation}{\thesection.\arabic{equation}}
%\thesection instead of \arabic{section} for correct equation numbering
% in appendices
\makeatother
%%%%%%%%%%%%%%%%%%%%%%%%%%%INTESTAZIONE%%%%%%%%%%%%%%%%%%%%%%%%%%%%%%%
\begin{titlepage}
{~}
\vspace{-2cm}
\begin{center}
{\Large \textbf{Local zeta regularization}}
\vskip 0.2cm
{~}
\hskip -0.4cm
{\Large \textbf{and the scalar Casimir effect II.}}
\vskip 0.2cm
{~}
\hskip -0.4cm
{\Large\textbf{Some explicitly solvable cases}}
\end{center}
\vspace{0.5truecm}
\begin{center}
{\large
Davide Fermi$\,{}^a$, Livio Pizzocchero$\,{}^b$({\footnote{Corresponding author}})} \\
\vspace{0.5truecm}
${}^a$ Dipartimento di Matematica, Universit\`a di Milano\\
Via C. Saldini 50, I-20133 Milano, Italy\\
e--mail: davide.fermi@unimi.it \\
\vspace{0.2truecm}
${}^b$ Dipartimento di Matematica, Universit\`a di Milano\\
Via C. Saldini 50, I-20133 Milano, Italy\\
and Istituto Nazionale di Fisica Nucleare, Sezione di Milano, Italy \\
e--mail: livio.pizzocchero@unimi.it
\end{center}
\begin{abstract}
In Part I of this series of papers we have described a general formalism
to compute the vacuum effects of a scalar field via local (or global) zeta
regularization. In the present Part II we exemplify the general formalism
in a number of cases which can be solved explicitly by analytical means.
More in detail we deal with configurations involving parallel or perpendicular
planes and we also discuss the case of a three-dimensional wedge.
\end{abstract}
\vspace{0.2cm} \noindent
\hfill \parn
\par \vspace{0.3truecm} \noindent \textbf{AMS Subject classifications:} 81T55, 83C47.
\par \vspace{0.3truecm} \noindent \textbf{PACS}: 03.70.+k, 11.10.Gh, 41.20.Cv~.
\end{titlepage}
%%%%%%%%%%%%%%%%%%%%%%%%%%%%% INDICE %%%%%%%%%%%%%%%%%%%%%%%%%%%%%
\tableofcontents
%%%%%%%%%%%%%%%%%%%%%%%%%% FINE INDICE %%%%%%%%%%%%%%%%%%%%%%%%%%%
\vfill \eject \noindent
\section{Introduction}
This is the second part of our series of papers about zeta regularization
and vacuum effects for a scalar field.
In Part I \cite{PI} we have considered a neutral, scalar quantum field
on $(d+1)$-dimensional Minkowski spacetime, assuming the field to be
confined within a spatial domain $\Om$ and to fulfill suitable boundary
conditions; we have also indicated the possibility to replace $\Om$
with a Riemannian manifold, or either an open subset of it (and the
Minkowski environment with a curved ultrastatic spacetime). In the same
work we have been presenting general methods for the zeta regularization
of the vacuum expectation value (VEV) of the stress-energy tensor, of the
total energy and of the boundary forces, emphasizing the fact that we were
producing a set of general rules to be applied almost mechanically in
specific configurations.
The illustration of such mechanical rules was started in Part I with some
simple examples in one spatial dimension, and it is continued in the present
Part II with more engaging configurations. Here we consider a number of
cases, in which the necessary computations can be carried over by purely
analytical means; this makes a difference with respect to the subsequent
Parts III \cite{PIII} and IV \cite{PIV}, where application of the general
rules will require a mixture of analytical and numerical methods. \parn
Most of the completely solvable cases considered here have been
previously treated in the literature, typically in the case of
spatial dimension three, often with special choices of the boundary
conditions and, in most cases, considering only the conformal part
of the stress-energy VEV; for each one of the cases already treated
a specific approach has been employed, possibly different from zeta
regularization. \parn
The present work attains more generality as for the spatial dimension,
the boundary conditions and the presence of a non-conformal part
in the stress-energy VEV; the second, perhaps more significant,
contribution of this paper is the unified viewpoint mentioned in
the previous comments, that is, the presentation of all cases as
applications of the same apparatus. \salto
To be more specific, the configurations that we analyse are the
following. \vspace{0.1cm}\\
i) First of all we consider a massless field in odd spatial
dimension confined between two parallel (hyper-)planes, for several kinds of
boundary conditions; see Section \ref{PPSubsec}. We have already considered
this model in spatial dimension $d = 3$ for Dirichlet boundary conditions
in \cite{ptp}; therein, the analytic continuation required by zeta
regularization was performed using ad hoc, known results on the special
functions related to this specific configuration (namely, the Riemann
zeta and the polylogarithm). As a matter of fact, the literature on the
configuration with two parallel planes is immense, both regarding local
and global aspects; here we only cite a few references. In his seminal
paper \cite{Casimir}, using an exponential cut-off regularization along
with Abel-Plana resummation, Casimir was the first to compute the total
energy and the boundary forces for the case of two parallel planes;
concerning local aspects, the foremost derivation of the full stress-energy
tensor VEV was given by Brown and Maclay \cite{Bro}, using a point-splitting technique
({\footnote{Actually, both Casimir and Brown-Maclay considered the case
of electromagnetic field; yet the methods employed by these authors can
be trivially adapted to the case of a scalar field.}}).
Computation of both global and local quantities for this model was later
reproposed by several authors, using various regularization techniques:
see, e.g., the monographies by Milton \cite{Mil}, Elizalde et al.
\cite{AAP0,10AppZ}, Bordag et al. \cite{Bord} (see, as well, the works
cited therein) and the papers by Zimerman et al. \cite{Zim1} and by
Esposito et al. \cite{Esp}. We will give further bibliographical
references in Section \ref{PPSubsec}. In the present paper we resort
to the formalism of integral kernels developed in Part I in order to
derive automatically the analytic continuations required by the use
of zeta regularization. \vspace{0.1cm}\\
ii) Next, we consider a massive field between an arbitrary
number of perpendicular (hyper-)planes, fulfilling either Dirichlet or
Neumann boundary conditions on each one of them; see Section \ref{SecPerp}.
To the best of our knowledge, this type of configuration was only
considered by Actor \cite{ActBox2} and by Actor and Bender \cite{Actor};
both these papers deal with a scalar field fulfilling Dirichlet boundary
conditions on at most three perpendicular planes (along with other similar
models in three spatial dimensions, all involving boundaries consisting
of flat, perpendicular parts). More precisely, in \cite{ActBox2} the
author fixes $d = 3$ and evaluates the renormalized effective Lagrangian,
along with the VEV $\la 0 |\Fi^2(x)|0\ra_{ren}$ (plus their
analogues for non-zero temperature); in \cite{Actor} $d$ is arbitrary
and it is computed the VEV of each component of the stress-energy tensor.
In both works cited above, a zeta type approach based on the use of heat
kernels is employed; however, contrary to our regularization scheme, this
approach also involves the subtraction of terms (corresponding essentially
to Minkowski spacetime contributions) which diverge for any value of the
regulator parameter. Besides, let us stress once more that our method is
more systematic and the construction of the required analytic continuations
descends automatically from a general framework. \vspace{0.1cm}\\
iii) Finally, we consider a massless field confined within
a wedge of arbitrary width in spatial dimension $d = 3$, for several types
of boundary conditions; see Section \ref{secwedge}. We also consider a
variation of this configuration, which corresponds essentially to identify
the sides of the wedge; this is the so-called case of the ``cosmic string''
(see subsection \ref{string}). Some of these cases have already been
treated by Dowker et al. \cite{DowKen,Dow}, Deutsch and Candelas
\cite{Deu} (also discussing the electromagnetic case) and, more recently,
by Saharian et al. \cite{SahWed00,SahWed} and by Fulling et al. \cite{FulWed}
(see also the citations in these works and \cite{BrevWed1,BrevWed2,NestWed});
nearly all of these authors use the point splitting approach, or some
variant of it. More in detail, in \cite{DowKen} and \cite{Deu} attention
is restricted to the conformal part of the stress-energy VEV for either
Dirichlet or Neumann boundary conditions, while in \cite{SahWed00,SahWed}
also the non-conformal part is considered, but in the Dirichlet case only;
in \cite{FulWed}, instead, the authors only show the graphs of the energy
density and of the pressure components (for which no explicit expression is
given), derived via a point-splitting approach for several configurations
and various choices of the parameters describing the theory.
Our approach via zeta regularization considers several types of
boundary conditions, both in the conformal and in the non-conformal case;
in the subcases already analysed in the literature cited above, we obtain
the same results. \salto
Before describing the applications (i-iii), in the forthcoming
Section \ref{Gloss} we summarize the general scheme of Part I
(more precisely, the parts of it required by the applications
analysed in the present paper). This summary has been written just
for the comfort of the reader who, in absence of it, would be forced
to skip continuously to Part I to recover the basic identities on
integral kernels and analytic continuation employed here.
\salto
Finally, we point out that some of the computations presented in this paper
have been performed using $\tt{Mathematica}$ in the symbolic mode.
\vspace{-0.2cm}
\section{A summary of results from Part I} \label{Gloss} \vspace{-0.1cm}
\subsection{General setting.}
Throughout the paper we use natural units, so that
\beq c = 1 ~, \qquad \hbar = 1 ~. \feq
Our approach works in $(d+1)$-dimensional Minkowski spacetime, which is
identified with $\reali^{d+1}$ using a set of inertial coordinates
\beq x = (x^\mu)_{\mu=0,1,...,d} \equiv (x^0,\bx) \equiv (t,\bx) ~; \feq
the Minkowski metric is $(\eta_{\mu \nu}) = \mbox{diag} (-1,1,...\,,1)$\,.
We fix a spatial domain $\Om \subset \reali^d$ and a background static
potential $V\!:\!\Om\!\to\!\reali$. We consider a quantized neutral,
scalar field $\Fi : \reali \times \Om \to \LL_{s a}(\Fock)$ ($\Fock$
is the Fock space and $\LL_{s a}(\Fock)$ are the selfadjoint operators on it);
suitable boundary conditions are prescribed on $\partial \Om$. The field
equation reads
\beq 0 = (-\partial_{tt}+\Delta-V(\bx)) \Fi(\bx,t) \label{daquan} \feq
($\Delta := \sum_{i=1}^d \partial_{ii}$ is the $d$-dimensional Laplacian).
We put
\beq \AA := - \Delta + V ~, \label{defaa} \feq
keeping into account the boundary conditions on $\partial\Om$, and consider
the Hilbert space $L^2(\Om)$ with inner product $\la f|g\raL :=
\int_{\Om}d\bx\,\overline{f}(\bx)g(\bx)$. We assume $\AA$ to be selfadjoint
in $L^2(\Om)$ and strictly positive (i.e., with spectrum $\si(\AA) \subset
[\eps^2,+\infty)$ for some $\eps >0$); the latter assumption is sometimes
relaxed requiring only that $\AA$ is non-negative ($\si(\AA) \subset [0,+\infty)$). \parn
We often refer to a complete orthonormal set $(\Fk)_{k \in \KK}$ of
(proper or improper) eigenfunctions of $\AA$ with eigenvalues
$(\om^2_k)_{k \in \KK}$ ($\om_k\!\geqs\!\eps$ for all $k\!\in\!\KK$; with
the relaxed condition of non-negativity, we only have $\om_k \geqs 0$). Thus
\begin{equation}\begin{split}
& \Fk : \Om \to \complessi; \qquad \AA\Fk= \om^2_k \Fk ~; \\
& \hspace{-0.7cm} \la \Fk | \Fh \raL = \de(k, h) \quad
\mbox{for all $k,h \in \KK$} ~. \label{eigenf}
\end{split}\end{equation}
The labels $k\!\in\!\KK$ can include both discrete and continuous
parameters; $\int_\KK dk$ indicates summation over all labels and
$\de(h,k)$ is the Dirac delta function on $\KK$. \parn
We expand the field $\Fi$ in terms of destruction and creation
operators corresponding to the above eigenfunctions, and assume the
canonical commutation relations; $|0 \ra \in \Fock$ is the vacuum
state and, as already indicated, VEV stands for ``vacuum expectation value''. \parn
The quantized stress-energy tensor reads ($\xi\!\in\! \reali$ is a parameter)
\beq \Ti_{\mu \nu} := \l(1 - 2\xi \r) \partial_\mu \Fi \circ \partial_\nu \Fi
- \l({1\over 2} - 2\xi \r)\eta_{\mu\nu}(\partial^\lam\Fi \, \partial_\lam \Fi + V\Fi^2)
- 2 \xi \, \Fi \circ \partial_{\mu \nu} \Fi \,; \label{tiquan} \feq
in the above we put $\Aa \circ \Bb := (1/2) (\Aa \Bb + \Bb \Aa)$
for all $\Aa, \Bb \in \LL_{s a}(\Fock)$, and all the bilinear terms
in the field are evaluated on the diagonal (e.g., $\partial_\mu \Fi
\circ \partial_\nu \Fi$ indicates the map $x \mapsto \partial_\mu \Fi(x)
\circ \partial_\nu \Fi(x)$). The VEV $\la 0|\Ti_{\mu\nu}|0\ra$ is
typically divergent.
\vspace{-0.4cm}
\subsection{Zeta regularization.}
The \textsl{zeta-regularized field operator} is
\beq \Fis := (\mm^{-2} \AA)^{-\s/4} \Fi ~, \label{Fis} \feq
where $\AA$ is the operator \rref{defaa}, $\s \in \complessi$
and $\mm > 0$ is a ``mass scale'' parameter; note that
$\Fis|_{\s = 0} = \Fi$, at least formally. The \textsl{zeta
regularized stress-energy tensor} is
\beq \Tis_{\mu \nu} := (1\!-\!2\xi)\partial_\mu \Fis\!\circ\partial_\nu \Fis\!
- \!\l({1\over 2}\!-\!2\xi\!\r)\!\eta_{\mu\nu}\!
\l(\!\partial^\lam\Fis\partial_\lam \Fis\!+\!V(\Fis)^2\r)
- 2 \xi \, \Fis\!\circ \partial_{\mu \nu} \Fis \,. \label{tiquans} \feq
The VEV $\la 0|\Tis_{\mu\nu}|0\ra$ is well defined for $\Re\s$ large enough
(see the forthcoming subsection \ref{DirT}); moreover, in the region of definition
it is an analytic function of $\s$. The same can be said of many related observables
(including global objects, such as the total energy VEV). \parn
For any one of these observables, let us denote with $\F(\s)$ its
zeta-regularized version and assume this to be analytic for $\s$
in a suitable domain $\UU_0 \subset \complessi$.
The zeta approach to renormalization can be formulated in two versions. \parn
i) \textsl{Restricted version}. Assume the map $\UU_0\!\to\!\complessi$,
$\s\!\mapsto\!\F(\s)$ to admit an analytic continuation (indicated with the
same notation) to an open subset $\UU\!\subset\!\complessi$ with $\UU\!\ni\!0$\,;
then we define the renormalized observables as
\beq \F_{ren} := \F(0) ~. \label{ren} \feq
ii) \textsl{Extended version}. Assume that there exists an open subset
$\UU\!\subset\!\complessi$ with $\UU_0\!\subset\!\UU$, such that
$0\!\in\!\UU$ and the map $\s\!\in\!\UU_0 \mapsto \F(\s)$ has an
analytic continuation to $\UU\!\setminus\!\{0\}$ (still denoted with $\F$).
Starting from the Laurent expansion $\F(\s) = \sum_{k = -\infty}^{+\infty}
\F_k \s^k$, we introduce the \textsl{regular part} $(RP\,\F)(\s) :=
\sum_{k=0}^{+\infty} \F_k \s^k$ and define
\beq \F_{ren} := (RP\,\F)(0)~. \label{renest} \feq
Of course, if $\F$ is regular at $\s = 0$ the defnitions \rref{ren}
\rref{renest} coincide. \parn
In the case of the stress-energy VEV, the prescriptions (i) and (ii)
read, respectively,
\beq \la 0 | \Ti_{\mu \nu}(x) | 0 \ra_{ren} :=
\la 0 | \Tis_{\mu \nu}(x) | 0 \ra \Big|_{\s=0} ~, \label{pri} \feq
\beq \la 0 | \Ti_{\mu \nu}(x) | 0 \ra_{ren} :=
RP \Big|_{\s=0} \la 0 | \Tis_{\mu \nu}(x) | 0 \ra ~. \label{prii} \feq
\vspace{-0.8cm}
\subsection{Conformal and non-conformal parts of the stress-energy VEV.} \label{ConfSubsec}
These are indicated by the superscripts $(\Co)$ and $(\NCo)$, respectively;
they are defined by
\beq \la 0|\Ti_{\mu\nu}|0\ra_{ren} = \la 0|\Ti^{(\Co)}_{\mu\nu}|0\ra_{ren}
+ (\xi\!-\!\xi_d)\,\la 0|\Ti^{(\NCo)}_{\mu\nu}|0\ra_{ren} ~, \label{TRinCo}\feq
where we are considering for the parameter $\xi$ the critical value
\beq \xi_d := {d\!-\!1 \over 4d} ~. \label{xic} \feq
\vspace{-0.9cm}
\subsection{Integral kernels.} If $\BB$ is a linear operator in $L^2(\Om)$,
its integral kernel is the (generalized) function $(\bx,\by) \in \Om\!
\times\!\Om \mapsto \BB(\bx,\by) := \la \de_{\bx}|\BB\,\de_{\by}\raL$
($\de_\bx$ is the Dirac delta at $\bx$). The trace of $\BB$, assuming
it exists, fulfills $\Tr \BB = \int_\Om d\bx\,\BB(\bx,\bx)$\,. \parn
In the following subsections $\AA$ is a strictly positive selfadjoint
operator in $L^2(\Om)$, with a complete orthonormal set of eigenfunctions
as in Eq. \rref{eigenf}; in some situations (explicitly indicated) we
only require $\AA$ to be non-negative. In typical applications, $\AA$
is the operator \rref{defaa}.
\vspace{-0.4cm}
\subsection{The Dirichlet kernel and its relations with the stress-energy VEV.}\label{DirT}
For (suitable) $s \in \complessi$, the $s$-th Dirichlet kernel of
$\AA$ is \beq \Dir_s(\bx, \by) := \AA^{-s}(\bx,\by) = \int_\KK {dk
\over \om_k^{2s}}\; \Fk(\bx) \Fkc(\by) ~. \label{eqkerdi} \feq If
$\AA = -\Delta + V$ (with $V$ a smooth potential) is strictly
positive, the map $\Dir_s(~,~):\Om \times \Om \to \complessi$,
$(\bx,\by) \mapsto \Dir_s(\bx,\by)$ is continuous along with all
its partial derivatives up to order $j \in \naturali$, for all $s
\in \complessi$ with $\Re s > d/2+j/2$\,; in particular,
$\Dir_s$ and its derivatives up to order $j$ are continuous
on the diagonal $\by=\bx$, a fact of special interest for
our purposes.
(Under stronger
assumptions on $\AA$, one can give results of absolute
and uniform convergence of the eigenfunction expansion in Eq.
\rref{eqkerdi}, for suitable values of $s$; the same can be
said for all the corresponding derivatives).
\parn Recalling Eq. \rref{tiquans}, the regularized stress-energy
VEV can be expressed as follows: \beq {~}\hspace{-0.5cm} \la 0 |
\Tis_{0 0}(\bx) | 0 \ra \! = \!\mm^\s
\!\!\l[\!\l(\!\frac{1}{4}\!+\!\xi\!\r)\!\!\Dir_{{\s - 1\over
2}}(\bx,\by) \!+\!\l(\!\frac{1}{4}\!-\!\xi\!\r)\!\!
(\partial^{x^\ell}\!\partial_{y^\ell}\!+\!V(\bx)) \Dir_{{\s + 1
\over 2}}(\bx,\by)\r]_{\by = \bx}\!, \label{Tidir00} \feq \beq \la
0 | \Tis_{0 j}(\bx) | 0 \ra = \la 0 | \Tis_{j 0}(\bx) | 0 \ra = 0
~, \label{Tidiri0} \feq
\begin{equation}\begin{split}
& {~}\hspace{3.8cm} \la 0 | \Tis_{i j}(\bx) | 0 \ra =
\la 0 | \Tis_{j i}(\bx) | 0 \ra = \\
& = \mm^\s \!\l[\!\Big({1\over 4} - \xi\Big) \de_{i j}
\Big(\!\Dir_{{\s - 1 \over 2}}(\bx,\by) -
(\partial^{\,x^\ell}\!\partial_{y^\ell}\!+\!V(\bx))
\Dir_{{\s + 1 \over 2}}(\bx,\by) \Big) \, + \r. \\
& \hspace{5cm} \l. + \l(\!\Big({1\over 2} - \xi\Big)\partial_{x^i y^j}
- \xi\,\partial_{x^i x^j}\!\r)\!
\Dir_{{\s + 1 \over 2}}(\bx,\by) \r]_{\by = \bx}
\label{Tidirij}
\end{split}\end{equation}
($\la 0|\Tis_{\mu \nu}(\bx)|0\ra$ is short for $\la 0|\Tis_{\mu\nu}(t,\bx)|0\ra$;
indeed, the VEV does not depend on $t$). Of course, the map $\Om \to \complessi$,
$\bx \mapsto \la 0|\Tis_{\mu \nu}(\bx)|0\ra$ possesses the same regularity as the
functions $\bx \in \Om \mapsto \Dir_{\s\pm 1 \over 2}(\bx,\bx),
\partial_{zw}\Dir_{\s+1 \over 2}(\bx,\bx)$ ($z,w$ any two spatial variables);
so, due to the previously mentioned results, $\bx \mapsto \la 0|\Tis_{\mu \nu}(\bx)|0\ra$
is continuous for $\Re \s > d+1$. \parn
The renormalized stress-energy VEV is $\la 0|\Ti_{\mu \nu}(\bx)|0\ra_{ren} :=
RP|_{\s = 0}\la 0|\Tis_{\mu \nu}(\bx)|0\ra$; introducing the functions
\beq \Dik_{\pm{1 \over 2}}(\bx,\by) := RP\Big|_{\s= 0}
\l(\mm^\s \Dir_{\s \pm 1 \over 2}(\bx,\by)\r) \,, \label{Dik} \feq
\beq \partial_{z w} \Dik_{{1 \over 2}}(\bx,\by) := RP\Big|_{\s= 0}
\l(\mm^\s \partial_{z w} \Dir_{\s + 1 \over 2}(\bx,\by)\r) \label{Dikzw} \feq
(with $z, w$ any two spatial variables), this can be expressed as follows:
\beq {~}\hspace{-0.3cm} \la 0 | \Ti_{0 0}(\bx)|0\ra_{ren}\! =
\!\l[\!\l(\!\frac{1}{4}\!+\!\xi\!\r)\!\Dik_{\!-{1\over 2}}(\bx,\by)
\!+\!\l(\!\frac{1}{4}\!-\!\xi\!\r)\!
(\partial^{x^\ell}\!\partial_{y^\ell}\!+\!V(\bx))
\Dik_{\!+{1 \over 2}}(\bx,\by)\r]_{\by = \bx} \!,\! \label{Tidir00R} \feq
\beq \la 0 | \Ti_{0 j}(\bx) | 0 \ra_{ren} =
\la 0 | \Ti_{j 0}(\bx) | 0 \ra_{ren} = 0 ~, \label{Tidiri0R} \feq
\begin{equation}\begin{split}
& \hspace{4.1cm} \la 0 | \Ti_{i j}(\bx) | 0 \ra_{ren} =
\la 0 | \Ti_{j i}(\bx) | 0 \ra_{ren} = \\
& = \!\l[\!\Big({1\over 4} - \xi\Big) \de_{i j}
\Big(\!\Dik_{\!-{1 \over 2}}(\bx,\by) -
(\partial^{\,x^\ell}\!\partial_{y^\ell}\!+\!V(\bx))
\Dik_{\!+{1 \over 2}}(\bx,\by) \Big)\, + \r. \\
& \hspace{5.4cm} \l. + \l(\!\Big({1\over 2} - \xi\Big)\partial_{x^i y^j}
- \xi\,\partial_{x^i x^j}\!\r)\!
\Dik_{\!+{1 \over 2}}(\bx,\by) \r]_{\by = \bx} . \label{TidirijR}
\end{split}\end{equation}
If $\Dir_{\s \pm 1 \over 2}(\bx,\by)$ and $\partial_{z w}
\Dir_{{\s + 1 \over 2}}(\bx,\by)$ have analytic continuations
regular at $\s=0$, indicated with $\Dir_{\pm {1 \over 2}}(\bx,\by)$
and $\partial_{z w} \Dir_{{1 \over 2}}(\bx,\by)$, for any $\mm > 0$
one has
\begin{equation}\begin{split}
& \hspace{0.34cm} \Dik_{\pm{1 \over 2}}(\bx,\by) =
\Dir_{\pm {1 \over 2}}(\bx,\by) ~, \\
& \partial_{z w} \Dik_{{1 \over 2}}(\bx,\by) =
\partial_{z w} \Dir_{{1 \over 2}}(\bx,\by) ~. \label{DikAC}
\end{split}\end{equation}
In the sequel we will consider the total energy VEV and express it
in terms of the trace $\Tr\AA^{-s}$, fulfilling
\beq \Tr \AA^{-s} = \int_\Om d\bx \; \Dir_s(\bx, \bx)~. \label{TrAs} \feq
\vspace{-0.8cm}
\subsection{The heat and cylinder kernels.} For $\t \in [0,+\infty)$,
these are given by \beq K(\t\,;\bx,\by) := e^{-\t\AA}(\bx,\by) =
\int_\KK dk\;e^{- \t\,\om_k^2}\,\Fk(\bx)\Fkc(\by) ~;
\label{eqheat} \feq \beq T(\t\,;\bx,\by) :=
e^{-\t\sqrt{\AA}}(\bx,\by) = \int_\KK
dk\;e^{-\t\,\om_k}\,\Fk(\bx)\Fkc(\by) ~. \label{eqcyl} \feq
Sometimes we also consider the modified cylinder kernel \beq
\Tm(\t\,;\bx,\by) := (\sqrt{\AA}^{\;-1}e^{- \t
\sqrt{\AA}})(\bx,\by) = \int_\KK
{dk\over\om_k}\;e^{-\t\,\om_k}\,\Fk(\bx)\Fkc(\by) ~; \label{SKer}
\feq it turns out that $T(\t\,;\bx,\by) =
-\,\partial_\t\Tm(\t\,;\bx,\by)$\,. \salto If $\AA = -\Delta + V$
($V$ smooth) is strictly positive, the map $K(\t\,;~,~) :\Om
\times \Om \to \complessi$, $(\bx,\by) \mapsto K(\t\,;\bx,\by)$ is
continuous along with all its partial derivatives of any order,
for all $\t > 0$ (the same holds for $T$ and $\Tm$)\,; with
stronger assumptions on $\AA$, the eigenfunction expansions in
Eq.s \rref{eqheat} \rref{eqcyl} \rref{SKer} converge absolutely
and uniformly, along with all their derivatives. \parn If $\AA$ is
not strictly positive, but non-negative, Eq.s \rref{eqheat}
\rref{eqcyl} continue to make sense;
moreover, if $0$ is in the continuous spectrum of $\AA$,
its spectral measure vanishes and the operator
$\sqrt{\AA}^{-1} e^{- \t \sqrt{\AA}}$ can still be defined;
in this case, typically, the modified cylinder kernel
and its eigenfunction expansion in \rref{SKer} still
make sense.
\salto The \textsl{cylinder trace}, if it exists, is \beq
T(\t) := \Tr e^{-\t\sqrt{\AA}} = \int_\Om d\bx\;T(\t\,;\bx,\bx) ~.
\label{TTr} \feq
%% \vspace{-0.8cm}
\subsection{The Dirichlet kernel as Mellin transform of the heat or cylinder kernel.}
For suitable values of $s\!\in\!\complessi$ (see Part I), there hold
\beq \Dir_s(\bx,\by) = {1 \over \Ga(s)} \int_0^{+\infty}
\!d\t \; \t^{s-1}\,K(\t\,;\bx,\by) ~; \label{DirHeat} \feq
\beq \Dir_s(\bx,\by) = {1 \over \Ga(2s)}\int_0^{+\infty}
\!d\t \; \t^{2s-1}\,T(\t\,;\bx,\by) ~. \label{DirCyl} \feq
Similar results hold for $\Tr \AA^{-s}$; for example, using the
cylinder trace $T(\t)$ of Eq. \rref{TTr}, we obtain
\beq \Tr \AA^{-s} = {1 \over \Ga(2s)}\int_0^{+\infty}
\!d\t \; \t^{2s-1}\,T(\t) ~. \label{DirCylTr} \feq
\vspace{-0.8cm}
\subsection{Analytic continuation of $\boma{\Dir_s}$ via complex integration.}
Let us assume \beq T(\t\,;\bx,\by) = {1 \over
\t^q}\;J(\t\,;\bx,\by) ~, \label{espk} \feq for some $q \in
\interi$, where the map $J:[0,+\infty)\times \Om \times \Om \to
\reali$ admits an extension $J : \UU([0,+\infty)) \times \Om
\times \Om \to \complessi$ ($\UU([0,+\infty)) \subset \complessi$
is an open neigbourhood of $[0,+\infty)$); besides, for fixed
$\bx,\by \in \Om$, assume the function $\t\in\UU([0,+\infty))
\mapsto J(\t\,;\bx,\by)$ to be analytic and exponentially
vanishing for $\Re \t \to + \infty$. Then, one can infer from Eq.
\rref{DirCyl} that \beq \Dir_s(\bx,\by) = {e^{-2i\pi
s}\,\Ga(1\!-\!2s) \over 2 \pi i} \int_\Hank
d\t\;\t^{2s-1}\,T(\t\,;\bx,\by) \label{DirHankCyl} \feq where
$\Hank$ denotes the \textsl{Hankel contour}, that is a simple path
in the complex plane that starts in the upper half-plane near
$+\infty$, encircles the origin counterclockwise and returns to
$+\infty$ in the lower half-plane (see Part I for an illustration
of $\Hank$ and for a precise definition of the complex powers of
$\t$). \parn Eq. \rref{DirHankCyl} gives the analytic continuation
of $\Dir_s$ to a meromorphic function of $s$ on the whole complex
plane, possibly with simple poles for $s \in
\{q/2,(q\!-\!1)/2,(q\!-\!2)/2,\,...\} \setminus
\{0,-1/2,-1,-3/2,...\}$. For half-integer values of $s$, the
integral in Eq. \rref{DirHankCyl} can be computed via the residue
theorem; in this way for example, for $s=-n/2$ and $n \in
\{0,1,2,...\}$, we obtain \beq \Dir_{-{n \over 2}}(\bx,\by) =
(-1)^n \,\Ga(n\!+\!1)\, \Res\Big(\t^{-(n +1)}\,T(\t\,;\bx,\by)\,;
0\Big) ~. \label{DirHankCylRes} \feq Using the modified cylinder
kernel $\Tm$, we deduce a variant of Eq. \rref{DirHankCyl}: \beq
\Dir_s(\bx,\by) = -\,{e^{-2 i \pi s}\,\Ga(2\!-\!2s) \over 2 \pi i}
\int_\Hank d\t\;\t^{2s-2}\,\Tm(\t\,;\bx,\by) ~. \label{DirHankS}
\feq Again, for half-integer $s$ the above analytic continuation
can be computed explicitly by the residue theorem; more precisely,
for $n \in \{-1,0,1,2,...\}$, Eq. \rref{DirHankS} gives \beq
\Dir_{-{n \over 2}}(\bx,\by) = (-1)^{n+1} \,\Ga(n\!+\!2)\,
\Res\Big(\t^{-(n +2)}\,\Tm(\t\,;\bx,\by)\,; 0\Big) ~.
\label{DirHankCylResTm} \feq Similar results hold for the spatial
derivatives of $\Dir_s$; moreover an analogous discussion can
be made for the trace $\Tr\AA^{-s}$ starting from Eq.
\rref{DirCylTr} and using the cylinder trace $T(\t)$. Assuming the
latter to admit a meromorphic extension to a neighborhood of
$[0,+\infty)$, with only a pole at $\t = 0$ and vanishing
exponentially for $\Re\t \to +\infty$, we have \beq \Tr\AA^{-s} =
{e^{-2i\pi s}\,\Ga(1\!-\!2s) \over 2 \pi i} \int_\Hank
d\t\;\t^{2s-1}\,T(\t) ~. \label{TrAHankCyl} \feq For $s = -n/2$
and $n\!\in\!\{0,1,2,...\}$, the above relation and the residue
theorem give \beq \Tr \AA^{n/2} = (-1)^n \,\Ga(n\!+\!1)\,
\Res\Big(\t^{-(n+1)}\,T(\t)\,; 0\Big) ~. \label{DirHankCylTr} \feq
\vspace{-0.8cm}
\subsection{The case of product domains. Factorization of the heat kernel.} \label{prodDom}
Let $\AA := -\Delta + V$ and consider the case where
\beq \Om = \Om_1 \times \Om_2 \ni \bx = (\bx_1, \bx_2)\,,
\by = (\by_1,\by_2) ~, \label{omfact} \feq
\beq V(\bx) = V_1(\bx_1) + V_2(\bx_2) \feq
($\Om_a\!\subset\!\reali^{d_a}$ is an open subset, for $a \in \{1,2\}$;
$d_1\!+\!d_2=d$); assume the boundary conditions on $\partial \Om$
to arise from suitable boundary conditions prescribed separately
on $\partial \Om_1$ and $\partial \Om_2$ so that,
for $a=1,2$, the operators
\beq \AA_a := - \Delta_a + V(\bx_a) \feq
(with $\Delta_a$ the Laplacian on $\Om_a$) are selfadjoint and strictly
positive (or at least, non-negative) in $L^2(\Om_a)$. Then, the Hilbert space $L^2(\Om)$ and the
operator $\AA$ can be represented as
\beq L^2(\Om) = L^2(\Om_1) \otimes L^2(\Om_2) ~, \qquad
\AA = \AA_1 \otimes \uno + \uno \otimes \AA_2 ~. \label{prodOmAA}\feq
This implies, amongst else, that the heat kernels $K(\t\,;\bx,\by) :=
e^{\t \AA}(\bx, \by)$, $K_a(\t\,;\bx_a,\by_a)$ $:= e^{\t \AA_a}(\bx_a,\by_a)$
($a = 1,2$) are related by
\beq K(\t\,;\bx,\by) = K_1(\t\,;\bx_1,\by_1)\,K_2(\t\,;\bx_1,\bx_2) ~. \label{prodK} \feq
%% \vspace{-0.8cm}
\subsection{The subcase of a slab: reduction to a lower-dimensional problem.} \label{slabSubsec}
By definition, we have a \textsl{slab} if
\beq \Om = \Om_1 \times \reali^{d_2} \ni \bx = (\bx_1, \bx_2)\,,
\by = (\by_1,\by_2)~, \qquad V(\bx) = V(\bx_1) \feq
with $\Om_1\!\subset\!\reali^{d_1}$ an open subset ($d_1 +d_2 = d$),
and if the boundary conditions prescribed for the field only refer
to $\partial \Om_1 \times \reali^{d_2}$.
We write $\Dir_s(\bx, \by)$ for the Dirichlet kernel of $\AA:=-\Delta+V(\bx_1)$
at $\bx =(\bx_1,\bx_2)$, $\by = (\by_1,\by_2)\!\in\!\Om$\,; $\Dir^{(1)}_s(\bx_1,\by_1)$
is the Dirichlet kernel of $\AA_1 := - \Delta_1\!+\!V(\bx_1)$. There hold
\beq \Dir_{\s \pm 1 \over 2}(\bx, \by) \Big|_{\by = \bx} =
{\Ga({\s- d_2 \pm 1 \over 2}) \over (4\pi)^{d_2/2}\,\Ga({\s \pm 1 \over 2})}\;
\Dir_{\s - d_2 \pm 1 \over 2}^{(1)} (\bx_1,\by_1)\Big|_{\by_1 = \bx_1} ~;
\label{TmnDirRid1} \feq
\begin{equation}\begin{split}
& \l.\partial_{x_a^i y_b^j}\Dir_{\s+1\over 2}(\bx,\by)\r|_{\by = \bx}\!\! =
\l.\partial_{x_a^i x_b^j}\Dir_{\s+1\over 2}(\bx,\by)\r|_{\by = \bx}\!\! =
\l.\partial_{y_a^i y_b^j}\Dir_{\s+1\over 2}(\bx,\by)\r|_{\by = \bx}\!\! = 0 \\
& \; \mbox{for $(a,b)=(1,2)$ or $(a,b)=(2,1)$ and
$i \in \{1,...d_a\}$, $j \in \{1,...,d_b\}$}; \label{TmnDirRid2a}
\end{split}\end{equation}
\begin{equation}\begin{split}
& \l.\partial_{z_1^i w_1^j}\Dir_{\s+1 \over 2}(\bx, \by)\r|_{\by = \bx} =
{\Ga({\s- d_2 + 1 \over 2}) \over (4\pi)^{d_2/2}\,\Ga({\s+1 \over 2})} \;
\partial_{z_1^i w_1^j}\Dir_{\!\s - d_2 + 1 \over 2}^{(1)} (\bx_1,\by_1)
\Big|_{\by_1 = \bx_1} \\
& \hspace{2.5cm} \mbox{for $z,w \in \{x,y\}$ and $i,j \in \{1,...,d_1\}$} ~;
\label{TmnDirRid2}
\end{split}\end{equation}
\begin{equation}\begin{split}
& \l.\partial_{x_2^i y_2^j}\Dir_{\s+1\over 2}(\bx,\by)\r|_{\by = \bx}\! =
- \l.\partial_{x_2^i x_2^j}\Dir_{\s+1\over 2}(\bx,\by)\r|_{\by = \bx}\! =
- \l.\partial_{y_2^i y_2^j}\Dir_{\s+1\over 2}(\bx,\by)\r|_{\by = \bx}\! = \\
& \hspace{0.5cm} = \de_{ij}\;{\Ga({\s - d_2 -1 \over 2}) \over
(4\pi)^{d_2/2}\,2\,\Ga({\s+1 \over 2})}\,\Dir_{\s-d_2 -1 \over 2}^{(1)}
(\bx_1,\by_1)\Big|_{\by_1 = \bx_1} \quad \mbox{for $i,j \in \{1,...,d_2\}$} ~.
\label{TmnDirRid3}
\end{split}\end{equation}
The above relations, along with Eq.s (\ref{Tidir00}-\ref{Tidirij}), imply
\begin{equation}\begin{split}
& \hspace{1.9cm} \la 0|\Tis_{ij}(\bx)|0\ra = 0 \qquad
\mbox{for $i,j\in \{d_1\!+\!1,...,d\}$, $i \neq j$} ~; \\
& \la 0|\Tis_{ij}(\bx)|0\ra = \la 0|\Tis_{ji}(\bx)|0\ra = 0 \qquad
\mbox{for $i\!\in\!\{1,...,d_1\}$, $j\!\in\!\{d_1\!+\!1,...,d\}$} ~.
\end{split}\end{equation}
Clearly, analogous relations hold for the renormalized VEV
$\la 0|\Ti_{ij}(\bx)|0\ra_{ren}$\,.
\vspace{-0.4cm}
\subsection{Reduced energy for a slab configuration.} \label{totenslab}
Consider the slab configuration of subsection \ref{slabSubsec}; the
\textsl{reduced regularized energy} (i.e., the total energy per unit
volume in the ``free'' dimensions) is
\beq \EE_1^\s := \int_{\Om_1} d\bx_1\; \la 0|\Tis_{00}|0\ra = E_1^\s + B_1^\s ~. \feq
The second equality is proved after defining the \textsl{regularized
reduced bulk} and \textsl{boundary energies}, which are
\begin{equation}\begin{split}
& E_1^\s := {\mm^\s\,\Ga({\s-d_2-1 \over 2})\over 2\,(4\pi)^{d_2/2}\,\Ga({\s-1\over 2})}
\int_{\Om_1}d\bx_1\;\Dir^{(1)}_{{\s -d_2- 1\over 2}}(\bx_1,\bx_1) = \\
& \hspace{1.7cm} = {\mm^\s\,\Ga({\s-d_2-1 \over 2})\over 2\,(4\pi)^{d_2/2}\,
\Ga({\s-1\over 2})}\; \Tr\,\AA_1^{{d_2 + 1 -\s \over 2}} ~, \label{Ered1}
\end{split}\end{equation}
\beq B_1^\s := {\mm^\s\,\Ga({\s-d_2+1 \over 2})\over (4\pi)^{d_2/2}\,\Ga({\s+1\over 2})\!}
\l(\!{1 \over 4}-\xi\!\r)\! \int_{\partial \Om_1}\!\!d a(\bx_1)
\l.{\partial \over \partial n_{\by_1}}\,
\Dir^{(1)}_{{\s - d_2 + 1\over 2}}(\bx_1,\by_1)\r|_{\by_1 = \bx_1} ~.
\label{Bred1} \feq
One has $B_1^\s = 0$ for $\Om_1$ bounded and either Dirichlet or Neummann
boundary conditions on $\partial\Om_1$. Assuming the functions \rref{Ered1}
\rref{Bred1} to be finite and analytic for suitable $\s \in \complessi$,
we define the renormalized, reduced bulk energy by the generalized
(or restricted) zeta approach:
\beq E_1^{ren} := RP \Big|_{\s=0} E_1^\s \qquad \l(\mbox{or } \;
E_1^{ren} := E_1^\s \Big|_{\s=0}\,\r)\,. \feq
\vspace{-0.9cm}
\subsection{Pressure on the boundary.} \label{pressuretmunu}
This is the force per unit area produced by the quantized field
inside $\Om$ at a point $\bx \in \partial\Om$. We first consider,
for $\Re \s$ large, the \textsl{regularized pressure} $\bp^\s(\bx)$
with components
\beq p^\s_i(\bx) := \la 0|\Tis_{i j}(\bx)|0\ra \, n^j(\bx) ~; \label{press1} \feq
here and in the remainder of this paper, $\bn(\bx) \equiv (n^i(\bx))$
is the unit outer normal at $\bx \in \partial\Om$. For Dirichlet
boundary conditions the above definition implies
\beq p^{\s}_i(\bx) = \mm^\s \l[\l(\!-{1 \over 4}\,\de_{ij}\,
\partial^{x^\ell}\!\partial_{y^\ell} + {1 \over 2}\,\partial_{x^i y^j}\!\r)\!
\Dir_{\s+1 \over 2}(\bx,\by)\r]_{\by=\bx} n^j(\bx) ~. \label{pEeT} \feq
We can define the \textsl{renormalized pressure} by analytic continuation as
\beq p^{ren}_i(\bx) := RP \Big|_{\s=0}\; p^\s_i(\bx) \label{preren} \feq
(first compute the regularized pressure at the boundary, and then
analytically continue at $\s=0$); alternatively, we could put
\beq p^{ren}_i(\bx) := \l(\lim_{\bx'\in\Om, \bx'\to\bx}
\la 0|\Ti_{ij}(\bx')|0\ra_{ren} \r) n^j(\bx) \label{alt} \feq
(first renormalize the stress-energy VEV at inner points of $\Om$,
and then move to the boundary). \parn
Prescriptions \rref{preren} \rref{alt} do not always agree (for a
counterexample, see Section \ref{secwedge}). In Part I we conjectured
that the two approaches agree when both of them give a finite result
(this is true in all the examples of the present Part II).
\vspace{-0.4cm}
\subsection{The Hilbert space when $\boma{0}$ is an isolated point of $\boma{\si(\AA)}$; the case of Neumann and periodic boundary conditions.} \label{HiNPBC}
Assume $\AA = -\Delta\!+\!V$, acting on $L^2(\Om)$, to have its spectrum
contained in $[0,+\infty)$, with $0$ an isolated point (hence a proper
eigenvalue); in this case, we replace the basic Hilbert space $L^2(\Om)$ with
\beq \L2m0(\Om) := (\ker \AA)^{\perp} \;\quad(\,\subset L^2(\Om)\,) ~. \label{L2m0} \feq
The restriction of $\AA$ to $L^2_0(\Om)$ is selfadjoint and strictly
positive; we take $L^2_0(\Om)$ as the basic space even for the field
quantization. \parn
For example, if $\AA = \!-\Delta$\,, $\Om$ is bounded and either Neumann
or periodic boundary conditions are prescribed on $\partial\Om$, one finds
\beq \L2m0(\Om) = \l\{f \in L^2(\Om) ~\Big|~
\int_\Om d\bx\, f(\bx) = 0 \r\} \,. \label{mean0} \feq
For slab configurations where $\Om = \Om_1 \times \reali^{d_2}$ and
Neumann or periodic boundary conditions are prescribed on $\partial\Om_1
\times \reali^{d_2}$, we set ($\AA_1$ is the reduced operator
in $L^2(\Om_1)$)
\beq \L2m0(\Om_1) := (\ker \AA_1)^{\perp} = \l\{f \in L^2(\Om_1) ~\Big|~
\int_{\Om_1} d\bx_1\, f(\bx_1) = 0 \r\} \,; \label{mean0slab} \feq
the basic Hilber space for the full theory on $\Om$ is
$\L2m0(\Om_1)\otimes L^2(\reali^{d_2})$\,.
%% \vspace{-0.4cm}
\subsection{The case where $\boma{0}$ is in the continuous spectrum of $\boma{\AA}$.} \label{AAeps}
Assume the fundamental operator $\AA = -\Delta+V$ to be non-negative
($\si(\AA) \subset [0,+\infty)$), with $0$ in the continuous spectrum of $\AA$.
The approach we consider in this case is to represent $\AA$ as
\beq \AA := ``{\lim_{\eps \to 0^+}}" \AA_{\eps} \feq
where, for $\eps \in (0, \eps_0)$, $\AA_{\eps}$ is a selfadjoint operator
in $L^2(\Om)$ with $\si(\AA_{\eps}) \subset [\eps^2, +\infty)$\,. \parn
We define a \textsl{deformed, smeared field operator} $\Fiseps :=
(\mm^{-2}\AA_\eps)^{-\s/4} \Fi$ and a \textsl{deformed, regularized
stress-energy tensor operator} $\Tiseps_{\mu\nu}$, whose VEV is
$$ \la 0 | \Tiseps_{\mu \nu}(x) | 0 \ra =  $$
$$ = \! \l. \l({1 \over 2}\!-\!\xi\!\r)\!(\partial_{x^\mu y^\nu}\!
+ \partial_{x^\nu y^\mu}\!)\! -\!\l({1\over 2}\!-\!2\xi\!\r)\!
\eta_{\mu\nu}\!\l(\partial^{x^\lam}\!\partial_{y^\lam}\!+\!V\r)\!
- \xi (\partial_{x^ \mu x^\nu}\!+ \partial_{y^\mu y^\nu}\!) \r|_{y=x}\!\!\cdot $$
\beq \cdot ~ \la 0|\Fiseps(x) \Fiseps(y) | 0 \ra ~; \label{tispropep} \feq
similarly, for $\bx \in \partial\Om$, we consider the \textsl{deformed,
regularized pressure}
\beq p^{\eps\s}_i(\bx) := \la 0 | \Tiseps_{i j}(\bx) | 0 \ra\,n^{j}(\bx)~.
\label{altmreg} \feq
In the end, we put
\beq \la 0|\Ti_{\mu\nu}(x)|0\ra_{ren} := \lim_{\eps \to 0^+} RP \Big|_{\s = 0}\,
\la 0|\Tiseps_{\mu\nu}(x)|0\ra ~; \label{TmnepRen} \feq
\beq p^{ren}_i(\bx) := \lim_{\eps \to 0^+}RP \Big|_{\s=0}\;p^{\eps\s}_i(\bx) ~.
\label{altm} \feq
We could alternatively put
\beq p^{ren}_i(\bx) := \l(\lim_{\bx'\in\Om, \bx'\to\bx}
\la 0|\Ti_{ij}(\bx')|0\ra_{ren} \r) n^j(\bx) \label{altt} \feq
($\la 0|\Ti_{ij}(\bx')|0\ra_{ren}$ is defined via Eq. \rref{TmnepRen}).
For the VEV \rref{tispropep} we have an expression analogous to
(\ref{Tidir00}-\ref{Tidirij}) in terms of the \textsl{deformed
Dirichlet kernel}
\beq \Dir^{\eps}_s(\bx, \by) := \AA^s_{\eps}(\bx,\by)
= \la \de_\bx | \AA^s_{\eps}\, \de_\by \raL ~, \feq
with $s = (u \pm 1)/2$\,. For the renormalized stress-energy VEV \rref{TmnepRen}
we have an expression of the form (\ref{Tidir00R}-\ref{TidirijR}), where
\beq {~}\hspace{-0.4cm} \Dik_{\!\pm {1 \over 2}}(\bx,\by) := \lim_{\eps \to 0^+}\!
RP \Big|_{\s=0}\Big(\mm^{\s} \Dir^{\eps}_{{\s \pm 1 \over 2}}(\bx,\by)\Big)\!
= \lim_{\eps \to 0^+}\! RP \Big|_{s= \pm {1 \over 2}}\Big(\mm^{2 s \mp 1}
\Dir^{\eps}_{s}(\bx,\by)\Big) \,,\! \label{DirRen} \feq
and analogous definitions hold for the spatial derivatives in the cited equations. \parn
In Part I we showed that two useful choices of $\AA_\eps$
are the following ($K^\eps,T^\eps$ and $K,T$ denote the heat and
cylinder kernels associated to $\AA_\eps$ and $\AA$, respectively):
\beq \AA_\eps := \AA +\eps^2 \quad \Rightarrow \quad
K^{\eps}(\t\,;\bx,\by) = e^{-\eps^2 \t}\; K(\t\,;\bx,\by) ~; \label{Kep}\feq
\beq \AA_\eps := (\sqrt{\AA} +\eps)^2 \quad \Rightarrow \quad
T^{\eps}(\t\,;\bx,\by) = e^{-\eps \t}\; T(\t\,;\bx,\by) ~. \label{Tep}\feq
Let $n\!\in\!\{-1,0,1,2,...\}$. If the modified cylinder kernel $\Tm(\t\,;\bx,\by)$
of $\AA$ admits a meromorphic extension in $\t$ to a
neighborhood of $[0,+\infty)$ fulfilling
\beq |\Tm(\t\,;\bx,\by)| \leqs C\,|\t|^{-a - n + 1} \qquad
\mbox{for $\Re\t \to +\infty$ and some $C,a\!>\!0$} ~, \label{bout} \feq
then, assuming $\AA_\eps$ to be as in Eq. \rref{Tep}, we have
\begin{equation}\begin{split}
& \Dik_{-{n \over 2}}(\bx,\by) := \lim_{\eps \to 0^+}\!RP \Big|_{s = -{n \over 2}}
\Big(\mm^{2s+n}\Dir^{\eps}_s(\bx,\by)\Big)\! = \\
& \hspace{0.3cm} (-1)^{n+1} \,\Ga(n\!+\!2)\,
\Res\Big(\t^{-(n +2)}\,\Tm(\t\,;\bx,\by)\,; 0\Big)\,. \label{Res1}
\end{split}\end{equation}
With analogous hypoteses for $\partial_{z w}\Tm(\t\,;\bx,\by)$
($z,w$ any pair of spatial variables), similar relations
for the spatial derivatives can be deduced.
\vspace{-0.4cm}
\subsection{Some variations involving the spatial domain.} \label{curvSubsec}
Configurations involving periodic boundary conditions are formulated
rigorously describing $\Om$ in terms of tori; e.g., the domain
$\Om = (0,a)^d$ with periodic boundary conditions is viewed as the
torus $\Toro^d_a := \reali^d/(a \interi)^d \simeq (\reali/a \interi)^d$
({\footnote{\label{L20Tor}The considerations of subsection \ref{HiNPBC} for
the periodic case are easily rephrased in terms of the torus $\Toro^d_a$
%% see footnote \ref{L20Tor}
(see footnote 21 in Part I).}}). \parn
Sometimes (see Section \ref{secwedge}) we employ on $\Om$ some
set of curvilinear coordinates $(q^i)_{i = 1,...,d} \equiv \bq$,
inducing a set of spacetime coordinates $q \equiv (q^\mu) \equiv (t,\bq)$;
the spatial and space-time line elements are, respectively
\begin{equation}\begin{split}
& \hspace{1.2cm} d\ell^2 = a_{i j}(\bq)dq^i dq^j ~; \qquad
ds^2 = - dt^2\!+\!d\ell^2 = g_{\mu \nu}(q)\, dq^\mu dq^\nu ~, \\
& g_{00} := -1 ~, ~\quad g_{i 0} = g_{0 i} := 0~, ~\quad
g_{i j}(q) := a_{i j}(\bq) \qquad \mbox{for $i,j \in \{1,...,d\}$}~. \label{dsq}
\end{split}\end{equation}
The analogue of Eq. \rref{tiquans} in the coordinate system $(q^\mu)$ is
\beq \Tis_{\mu \nu} := (1\!-\!2\xi)\partial_\mu \Fis\!\circ\partial_\nu \Fis\!
- \!\l({1\over 2}\!-\!2\xi\!\r)\!\eta_{\mu\nu}\!
\l(\!\partial^\lam\Fis\partial_\lam \Fis\!+\!V(\Fis)^2\r)
- 2 \xi \, \Fis\!\circ \nabla_{\!\mu \nu} \Fis \label{tiquansq} \feq
($\nabla_{\mu}$ is the covariant derivative induced by the metric
\rref{dsq}). For any scalar function $f$ there hold ($\ga^{k}_{ij}$
are the Christoffel symbols for the spatial metric $(a_{ij}(\bq))$)
\begin{equation}\begin{split}
& \hspace{0.6cm} \nabla_{\!\mu} f = \partial_\mu f ~, \qquad
\nabla_{ij} f = D_{ij}f = \partial_{ij} f - \ga^{k}_{ij} \partial_k f ~,\\
& \nabla_{0 i} f = \partial_0(\partial_{i}f) =  \partial_i(\partial_0 f) =
\nabla_{i 0}f ~, \qquad \nabla_{0 0} f = \partial_{0 0} f ~.
\label{compu}
\end{split}\end{equation}
Many results in the previous subsections are readily adapted to the
variations considered in this subsection for the space domain.
\section{The case of a massless field between parallel hyperplanes}\label{PPSubsec}
\subsection{Introducing the problem for arbitrary boundary conditions.}
As mentioned in Section 6 of Part I, the segment configuration
can be considered as the $d=1$ case of a general, $d$-dimensional
configuration with two parallel hyperplanes; this is the subject we
are now going to analyze (with no external potential). So, we assume
\beq \Om := (0,a) \times \reali^{d-1} ~, \quad
\mbox{$a>0$}~, \qquad V = 0 ~; \label{pp} \feq
these choices correspond to a massless scalar field confined between
the two parallel hyperplanes
({\footnote{\label{Foot15} There holds comments analogous to the ones
in footnote 18 on page 49 of Part I; namely, in place of the standard
Cartesian coordinates $\bx \equiv (x^i)$, we could have used the set
of rescaled coordinates (best fitting the features of the present configuration)
$$ \bx_\star \equiv (x^i_\star)_{i=1,...,d}~, \qquad \mbox{with} \quad
x^1_\star := x^1/a ~, \quad x^i_\star := x^i ~~ \mbox{for $i \in \{2,...,d\}$} ~. $$
Also in this case, we choose not to employ the above rescaled coordinate
system in order to render comparison with kwnown results easier.}})
\beq \pi_0 = \{ \bx \in \reali^d \;|\; x^1 = 0\} ~, \quad
\pi_a = \{ \bx \in \reali^d \;|\; x^1 = a\} ~. \feq
As we did in Section 6 of Part I for the segment configuration,
we are going to consider separately the cases where the field fulfills
Dirichlet, Neumann or periodic boundary conditions on the hyperplanes
$\pi_0,\pi_a$. Throughout this section we assume
\beq d~ \mbox{odd}~, ~~ d \geqs 3~; \label{assud} \feq
this hypothesis is purely technical and
will be motivated later (see the comments after Eq.
\rref{DirRPP2}); note that the case $d=1$, here excluded, has been already
discussed in Part I. \parn
In passing let us notice that, for $d = 3$ and Dirichlet boundary
conditions (see subsection \ref{DDPPSec}), the above configuration
is the one most typically considered when dealing with the (scalar)
Casimir effect \cite{Bord,Mil,Zim1}. The case with Dirichlet boundary
conditions on one plane and Neumann conditions on the other (discussed in
subsection \ref{DNPPSec}) was originally considered in the electromagnetic
case by Boyer \cite{Boy}, who derived the total energy; later, computations
of the total energy and boundary forces for a (massless or massive)
scalar field at both zero and non-zero temperature were performed
by Pinto et al. \cite{Pinto,Pin} and Santos et al. \cite{PPDN}
(see also \cite{Actor,Bord}). Finally, let us also mention the monography
by Fulling \cite{Ful} where the stress-energy VEV for the model with
periodic boundary conditions is given; see also \cite{AAP0,10AppZ}
and, again, \cite{Bord,Mil} for the derivation of the total energy
in the same configuration. \salto
For any one of the boundary conditions mentioned above, keeping in
mind the considerations of subsection \ref{HiNPBC}, we proceed in
the manner explained in the following subsection. Before moving on,
let us remark that, due to the results on slab configurations
reported in subsection \ref{slabSubsec}, we just have to study the
reduced one-dimensional problem based on
\beq \Om_1 := (0,a) \subset \reali ~, \qquad \AA_1 := -\partial_{x^1 x^1} ~, \feq
keeping into account the boundary conditions in $x^1 = 0$ and $x^1 = a$\,;
in consequence of this, we can resort to the results of Section 6
in Part I. \parn
In subsections \ref{DirTPP}-\ref{foSeg} we will present some general
results on the configuration under analysis, holding for all the types
of boundary conditions mentioned before. In subsections \ref{DDPPSec}-\ref{PPPerSubsec}
we will consider specific boundary conditions, with a special attention
for the case $d=3$.
\vspace{-0.4cm}
\subsection{The reduced Dirichlet and cylinder kernels.}\label{DirTPP}
According to Eq.s (\ref{TmnDirRid1}-\ref{TmnDirRid3}), the basic
ingredients for the analysis of the $d$-dimensional problem are
the Dirichlet kernel $\Dir^{(1)}_s$ of the reduced $1$-dimensional
problem at $s=(\s - d)/2$, and its derivatives at $s=(\s-d+2)/2$.
On the other hand, these functions can be expressed in terms of the
$1$-dimensional cylinder kernel $T^{(1)}(\t\,;x^1,y^1)$, which has
been determined in Section 6 of Part I (where it was
indicated simply with $T(\t\,;x^1,y^1)$); let us recall that this
kernel and all its derivatives, when evaluated on the diagonal
$y^1 = x^1$, are meromorphic functions of $\t$ in a neighborhood of
the positive real half-axis with a unique singularity in $\t=0$,
and they vanish exponentially for $\Re\t\to +\infty$. For the
reduced Dirichlet kernel and for the derivatives in which we are
interested, Eq. \rref{DirHankCyl} yields the following expressions:
\beq \Dir_{\s - d \over 2}^{(1)} (x^1,y^1) =
{e^{-i\pi(\s-d)}\,\Ga(d\!+\!1\!-\!\s) \over 2 \pi i} \!\int_\Hank d\t\;
\t^{\s - d - 1}\,T^{(1)}(\t\,;x^1,y^1) ~; \label{DirRPP1} \feq
\begin{equation}\begin{split}
& \partial_{z w}\Dir_{\!{\s-d+2 \over 2}}^{(1)}(x^1\!,y^1) =
{e^{-i\pi(\s-d)}\,\Ga(d\!-\!1\!-\!\s) \over 2 \pi i}\!\int_\Hank  d\t\;
\t^{\s-d+1}\,\partial_{z w}T^{(1)}(\t\,;x^1\!,y^1) \\
& \hspace{4.5cm} \mbox{for $z,w \in \{x^1,y^1\}$} ~. \label{DirRPP2}
\end{split}\end{equation}
Consider the above relations along with Eq.s (\ref{TmnDirRid1}-\ref{TmnDirRid3}),
relating the $1$-dimensional functions to the $d$-dimensional Dirichlet
kernel $\Dir_{\s-1 \over 2}(\bx,\by)$ or to the derivatives of
$\Dir_{\s+1 \over 2}(\bx,\by)$; these equations yield the sought-for
meromorphic continuations in $\s$ to the whole complex plane.
By direct inspection of the expressions thus obtained
it appears that, with the assumption \rref{assud} on $d$, these meromorphic continuations are analytic for $\s$
in a neighborhood of the origin; so, the zeta strategy for renormalization
is implemented by simply setting $\s=0$.
In conclusion, we have the following integral representations for the
renormalized Dirichlet kernel and for its renormalized derivatives:
\beq \l. \Dir_{-{1 \over 2}}(\bx,\by) \r|_{\by = \bx} =
- {\Cd \over 2 \pi i} \int_\Hank {d\t \over \t^{d+1}}\;
T^{(1)}(\t\,;x^1,y^1)\Big|_{y^1 = x^1} ~;
\label{DirP1} \feq
\beq \!\l.\partial_{z w}\Dir_{1 \over 2}\!(\bx,\by)\r|_{\by = \bx}\!\! =
-{\Cd \over (d\!-\!1)2 \pi i}\! \int_\Hank\! {d\t \over \t^{d-1}}\,
\partial_{z w}T^{(1)}(\t\,;x^1\!,y^1)\Big|_{y^1 = x^1} \;
\mbox{for $z,\!w\!\in\!\{x^1\!,y^1\}$}\,;\!\!\!\! \label{DirPP2} \feq
\begin{equation}\begin{split}
& \l.\partial_{x_2^i y_2^j}\Dir_{1\over 2}(\bx,\by)\r|_{\by = \bx}\!\! =
-\,\l.\partial_{x_2^i x_2^j}\Dir_{1\over 2}(\bx,\by)\r|_{\by = \bx}\!\! =
-\,\l.\partial_{y_2^i y_2^j}\Dir_{1\over 2}(\bx,\by)\r|_{\by = \bx}\!\! = \\
& \hspace{2.5cm} = \de_{ij}\, {\Cd \over 2 \pi i} \int_\Hank
{d\t \over \t^{d+1}}\; T^{(1)}(\t\,;x^1,y^1)\Big|_{y^1 = x^1}
\quad \mbox{for $i,j \!\in\! \{1,...,d\!-\!1\}$}\,; \label{DirPP3}
\end{split}\end{equation}
for the sake of brevity, in the above we have set
(\footnote{To justify the expression \rref{Ad}, some identities
regarding the gamma function must be used.})
\beq \Cd := (-\pi)^{-{d-1 \over 2}}\,\Ga\!\l({d\!+\!1 \over 2}\r) ~. \label{Ad} \feq
Eq.s (\ref{DirP1}-\ref{DirPP3}) are completed with Eq. \rref{TmnDirRid2a},
stating the vanishing of certain mixed derivatives. \salto
Finally, recall that Eq. \rref{Ered1} (here employed with
$d_2 = d\!-\!1$) for the regularized, reduced bulk energy
(see the subsequent subsection \ref{RedEnPP}) requires the
evaluation of the trace $\Tr\AA_1^{d - \s \over 2}$.
To this purpose, we first consider the one-dimensional cylinder
trace $T^{(1)}(\t)$, which has also been determined in Section
6 of Part I (where it was indicated simply with $T(\t)$);
we recall that, for all the boundary conditions considered in
the following applications, the map $\t \mapsto T^{(1)}(\t)$
admits a meromorphic extension to a neighborhood of $[0,+\infty)$
which only has a pole at $\t = 0$ and vanishes exponentially for
$\Re\t \to +\infty$. \parn
A discussion similar to the one carried over above for the
Dirichlet kernel allows us to derive an explicit expression
for the analytic continuation of $\Tr\AA_1^{d - \s \over 2}$
at $\s = 0$; more precisely, using for the reduced problem a
relation analogous to one in Eq. \rref{TrAHankCyl}, we conclude
\beq \Tr \AA_1^{d/2} = {(-1)^d \,\Ga(d\!+\!1) \over 2\pi i}
\int_\Hank d\t\;\t^{-(d+1)}\,T^{(1)}(\t) ~. \label{TrA1d} \feq
\vspace{-0.8cm}
\subsection{The stress-energy tensor.}
Substituting the relations \rref{TmnDirRid2a} and (\ref{DirP1}-\ref{DirPP3})
into Eq.s (\ref{Tidir00}-\ref{Tidirij}), we straightforwardly
deduce the contour integral representations for the non-vanishing
components of the renormalized stress-energy VEV; moreover, due to
the meromorphic nature of the cylinder kernel (and of its derivatives),
the resulting integrals along the Hankel contour can be explicitly
evaluated via the residue theorem. The final expressions for the
(non-zero) components of the renormalized stress-energy VEV are the
following ones:
\begin{equation}\begin{split}
& \la 0 | \Ti_{00}(\bx)|0\ra_{ren} = -\,\Cd \;\Res\l(\t^{-(d+1)}\!
\l[\l(\xi - {d\!-\!2 \over 4d}\r)\!d\;T^{(1)}(\t\,;x^1\!,y^1) ~ + \r.\r.\\
& \hspace{5.cm}\l.\l. +  {\t^2 \over d\!-\!1} \l({1 \over 4}-\xi\r)
\partial_{x^1 y^1}T^{(1)}(\t\,;x^1\!,y^1)\r]_{y^1 = x^1}\!;0\r) ; \label{T00PP}
\end{split}\end{equation}
\begin{equation}\begin{split}
& \la 0 | \Ti_{11}(\bx)|0\ra_{ren} = -\,\Cd\;\Res\l(\t^{-(d+1)}\!
\l[\l({1 \over 4} - \xi\r)\!d\;T^{(1)}(\t\,;x^1\!,y^1)~+ \r.\r. \\
& \hspace{3.5cm}\l.\l. +\, {\t^2 \over d\!-\!1}\l({1 \over 4}\;\partial_{x^1 y^1}
- \xi\,\partial_{x^1 x^1} \r)\!T^{(1)}(\t\,;x^1\!,y^1)\r]_{y^1 = x^1}\!;0\r) ; \label{T11PP}
\end{split}\end{equation}
\begin{equation}\begin{split}
& \la 0 | \Ti_{ij}(\bx)|0\ra_{ren} = \de_{ij}\;\Cd \;\Res\l(\t^{-(d+1)}\!
\l[\l(\xi - {d\!-\!2 \over 4d}\r)\!d\;T^{(1)}(\t\,;x^1\!,y^1) ~ + \r.\r.\\
& \hspace{1.5cm}\l.\l. + \,{\t^2 \over d\!-\!1} \l({1 \over 4}-\xi\r)
\partial_{x^1 y^1}T^{(1)}(\t\,;x^1\!,y^1)\r]_{y^1 = x^1}\!;0\r) \quad
\mbox{for $i,j \!\in\!\{2,...,d\}$}\,.\!\! \label{TijPP}
\end{split}\end{equation}
We repeat that, in the above, $d$ is an arbitrary odd dimension $>1$.
Starting from subsection \ref{DDPPSec}, for each one of the previously
mentioned boundary conditions we will report the explicit expressions
for the the stress-energy components arising from Eq.s (\ref{T00PP}-\ref{TijPP})
in the case of spatial dimension $d = 3$; again, we will give the final
results in the form described in subsection \ref{ConfSubsec}
(see, in particular, Eq. \rref{TRinCo}), noting that \rref{xic} gives
\beq \xi_3 = {1 \over 6} ~. \label{xi3}\feq
Some additional details related to these computations will be given, as
examples, in the case of Dirichlet and periodic boundary conditions.
\vspace{-0.4cm}
\subsection{The reduced energy.}\label{RedEnPP}
Let us recall again that this is the energy per unit volume in the
free dimensions and that it can be expressed in terms of the reduced
bulk and boundary energies (see subsection \ref{totenslab}). \parn
The general identity \rref{Ered1} allows us to represent the reduced
bulk energy in terms of the renormalized trace $\Tr \AA_1^{d/2}$,
corresponding to the reduced operator $\AA_1$; on the other hand,
Eq. \rref{TrA1d} gives an explicit expression for the latter
quantity in terms of the reduced cylinder trace $T^{(1)}(\t)$.
Evaluating the integral along the Hankel contour in the cited
equation via the residue theorem (compare with Eq. \rref{DirHankCylTr}),
we conclude
\beq E_1^{ren} = {(-1)^{d+1} \,\Ga(d\!+\!1)\,\Ga(-{d \over 2}\,)\over
2\,(4\pi)^{d/2}}\; \Res\Big(\t^{-(d+1)}\,T^{(1)}(\t)\,; 0\Big) ~.
\label{E1RenPP} \feq
Let us also mention that, for any one of the boundary conditions to
be considered in the following, the (regularized and renormalized)
reduced boundary energy always vanishes identically. \parn
Finally we point out that, at least in spatial dimension $d = 3$,
the results obtained via Eq. \rref{E1RenPP} for the renormalized,
reduced bulk energy coincide with the integral over the interval
$(0,a)$ of the conformal part of the corresponding renormalized
energy density $\la 0|\Ti_{00}|0 \ra_{ren}$; on the contrary, the
non-conformal part of the latter appears to diverge in a non-integrable
manner near the planes $\pi_0,\pi_a$. These facts closely resemble
the ones pointed out for the total energy of the segment configuration
in Section 6 of Part I; as in that case, they will be
checked by direct computation in subsections \ref{DDPPSec}-\ref{PPPerSubsec}.
\vspace{-0.4cm}
\subsection{The boundary forces.}\label{foSeg} The situation we meet in
the present situation is of the kind described in general in subsection
\ref{pressuretmunu}, and already faced for the segment configuration in
Part I: in principle, we can define the renormalized pressure on the plates
$\pi_0,\pi_a$ in two alternative ways. \parn
Let $\bn(\bx)$ denote the unit ``outer normal'' at points on the
boundary, so that $\bn(\bx) = (-1,0,...,0)$ on $\pi_0$ and $\bn(\bx)
= (1,0,...,0)$ on $\pi_a$.
Then, on the one hand we can put
\beq p^{ren}_i(\bx) := \l.\la 0|\Tis_{i j}(\bx)|0\ra \r|_{\s = 0}\, n^j(\bx)
= \de_{i1}\!\l.\la 0|\Tis_{11}(\bx)|0\ra \r|_{\s = 0} ~; \label{alt1PP} \feq
on the other hand, we have the alternative definition
\begin{equation}\begin{split}
& p^{ren}_i(\bx) := \l(\lim_{\bx'\in\Om, \bx'\to\bx}
\la 0|\Ti_{i j}(\bx') |0\ra_{ren} \r) n^j(\bx) \\
& \hspace{0.8cm} = \de_{i1}\l(\lim_{\bx'\in\Om, \bx'\to\bx}
\la 0|\Ti_{11}(\bx') |0\ra_{ren} \r) \,. \label{alt2PP}
\end{split}\end{equation}
As a matter of fact, \textsl{definitions \rref{alt1PP} and \rref{alt2PP}
will be found by explicit computations to yield the same result} for any
one of the boundary conditions to be considered in the next subsections.
\vspace{-0.4cm}
\subsection{Dirichlet boundary conditions.}\label{DDPPSec} Let us first
consider the case where the field fulfills Dirichlet boundary conditions
on both the hyperplanes $\pi_0,\pi_a$, meaning that
\beq \Fi(t,\bx) = 0 \qquad \mbox{for $t \in \reali$,
$\bx \in \pi_0$ or $\bx \in \pi_a$} ~. \feq
Recall that in this case the cylinder kernel associated to the reduced
problem is (see Eq. (6.20) in Part I)
\beq {~}\hspace{-0.4cm} T^{(1)}(\t\,;x^1,y^1) =
{1 \over 2a}\!\l[{\cos({\pi \over a}(x^1\!-\!y^1)) - e^{-{\pi \over a}\t}
\over \cosh({\pi \over a}\t)\!-\!\cos({\pi \over a}(x^1\!-\!y^1))}
- {\cos({\pi \over a}(x^1\!+\!y^1)) - e^{-{\pi \over a}\t} \over
\cosh({\pi \over a}\t)\!-\!\cos({\pi \over a}(x^1\!+\!y^1))} \r]\!.\! \label{TDD}\feq
To obtain the renormalized stress-energy VEV in any spatial dimension $d$,
it suffices to substitute Eq. \rref{TDD} into Eq.s (\ref{T00PP}-\ref{TijPP})
and to compute the residues therein. Let us explicitate the final
results for $d = 3$; in this case the residues in (\ref{T00PP}-\ref{TijPP}),
involving $T^{(1)}(\t\,;x^1,y^1)|_{y^1 = x^1}$, can be derived
from the $\t \to 0$ expansion
\begin{equation}\begin{split}
& \hspace{5cm} T^{(1)}(\t\,;x^1,y^1)\Big|_{y^1 = x^1} = \\
& {1\over \pi\t} + {\pi(3\!-\!\sin^2({\pi \over a}\,x^1))\over
12a^2 \sin^2({\pi \over a}\,x^1)}\;\t\, +
{\pi^3(15(2\!+\!\cos({2\pi \over a}\,x^1))-\sin^4({\pi \over a}\,x^1))
\over 720 a^4 \sin^4({\pi \over a}\,x^1)}\;\t^3 + O(\t^5) ~.
\end{split}\end{equation}
Proceeding in a similar manner where the spatial derivatives of
$T^{(1)}$ appear, we obtain this final result for the $d=3$
renormalized VEV of the stress-energy tensor:
$$ \la 0 | \Ti_{\mu\nu}(\bx) | 0 \ra_{ren} \Big|_{\mu,\nu = 0,1,2,3}
\hspace{-0.1cm} = A \! \l(\!\!\barray{cccc}
-1\!& \!0   &   0   &   0   \\
0\! & \!-3  &   0   &   0   \\
0\! & \!0   &   1   &   0   \\
0\! & \!0   &   0   &   1   \farray \!\!\r)\!\!
- \!\l(\!\xi\!-\!{1 \over 6}\r)\! B (x^1) \!
\l(\!\!\barray{cccc}
-1\!&   0   &   0   &   0   \\
0\! &   0   &   0   &   0   \\
0\! &   0   &   1   &   0   \\
0\! &   0   &   0   &   1   \farray\!\!\r) ,  $$
\beq A = {\pi^2 \over 1440 a^4}~, \qquad
B(x^1) = {\pi^2 \over 8 a^4}\, {3-2\sin^2 ({\pi \over a}\,x^1)
\over \sin^4 ({\pi \over a}\,x^1)} \quad \mbox{for $x^1 \!\in\! (0,a)$} ~.
\label{TmnDD} \feq
This is a classical result, whose earlier derivations used point
splitting methods rather than zeta regularization (see, e.g.,
the paper by Esposito et al. \cite{Esp} or the already cited
monographies \cite{BirDav,Ful,Mil}); in our previous work \cite{ptp}
the expression \rref{TmnDD} was obtained via zeta regularization,
but the required analytic continuations were derived by \textsl{ad hoc}
considerations, strictly related to the peculiar configuration under
analysis. On the contrary, here the analytic continuation arises
automatically from the general schemes that we have developed for
an arbitrary geometry ({\footnote{In connection with the present
approach, see also \cite{tes}.}}). \salto
Next, let us pass the reduced bulk energy. To this purpose,
recall that the cylinder trace associated to the reduced problem
under analysis is (see Eq. (6.25) in Part I)
\beq T^{(1)}(\t) = {1 \over e^{{\pi \over a}\t} - 1} ~; \label{TDDTrPP}\feq
now, the general relation \rref{E1RenPP} allows us to infer, for $d = 3$,
\beq E_1^{ren} = -\,{\pi^2 \over 1440 a^3} ~. \label{EnDDPP} \feq
Finally, using either definition \rref{alt1PP} or \rref{alt2PP}, we
obtain for the non-vanishing component of the pressure on the planes
$\pi_0$ and $\pi_a$ the following expressions, respectively:
\beq p^{ren}_1(\bx)\Big|_{\pi_0} = 3A ~, \qquad
p^{ren}_1(\bx)\Big|_{\pi_a} = - 3A \label{PDir}\feq
where $A$ as in Eq. \rref{TmnDD}. This means that in the present case,
the forces on the boundary planes produced by the field in the interior
region are attractive
({\footnote{\label{FootFoExt} Clearly, we are not taking into account
any effect related to the outer region (see the comments at the end
of subsection \ref{pressuretmunu} of Part I). As a matter of fact,
the forces produced by a massless scalar field in this region,
fulfilling either Dirichlet or Neumann boundary conditions on the
planes, vanish identically for $d = 3$; this result can be derived
using the methods that will be developed in the following Section
\ref{SecPerp}.}}).
\vspace{-0.4cm}
\subsection{Dirichlet-Neumann boundary conditions.}\label{DNPPSec} Consider now the
parallel hyperplanes configuration where the field fulfills Dirichlet
and Neumann boundary conditions, respectively, on the hyperplanes
$\pi_0$ and $\pi_a$; explicitly,
\beq \Fi(t,\bx) = 0 ~~ \mbox{for $(t,\bx) \in \reali\!\times\!\pi_0$} ~,
\qquad \partial_{x^1}\Fi(t,\bx) = 0 ~~
\mbox{for $(t,\bx) \in \reali\!\times\!\pi_a$} ~. \feq
The cylinder kernel associated to the reduced operator $\AA_1$ is\!
(see Eq.\!\! (6.30)\! in Part\! I)
\beq {~}\hspace{-0.4cm} T^{(1)}(\t\,;x^1,y^1) = {1 \over a}\!
\l[{\sinh({\pi \over 2a}\,\t)
\cos({\pi \over 2a}(x^1\!-\!y^1)) \over \cosh({\pi \over a}\,\t)\!
- \!\cos({\pi \over a}(x^1\!-\!y^1))} - {\sinh({\pi \over 2a}\,\t)
\cos({\pi \over 2a}(x^1\!+\!y^1)) \over \cosh({\pi \over a}\,\t)\!
- \!\cos({\pi \over a}(x^1\!+\!y^1))}\r]\!.\! \feq
Employing this kernel along with relations (\ref{T00PP}-\ref{TijPP}),
we can evaluate the renormalized VEV of the stress-energy tensor;
in particular, for $d = 3$ we obtain
$$ \la 0 | \Ti_{\mu\nu}(\bx) | 0 \ra_{ren} \Big|_{\mu,\nu = 0,1,2,3}
\hspace{-0.1cm} = A \! \l(\!\!\barray{cccc}
1   &   0   &   0\!     &   \!0 \\
0   &   3   &   0\!     &   \!0 \\
0   &   0   &   -1\!    &   \!0 \\
0   &   0   &   0\!     &   \!-1\farray \!\!\r)\!
-\!\l(\!\xi\!-\!{1 \over 6}\r)\! B(x^1)\!
\l(\!\!\barray{cccc}
-1\!&   0   &   0   &   0   \\
0\! &   0   &   0   &   0   \\
0\! &   0   &   1   &   0   \\
0\! &   0   &   0   &   1   \farray\!\!\r) , $$
\beq A = {7\pi^2 \over 11520 a^4}~, \quad
B (x^1) = {\pi^2 \over 64\,a^4}\;{23 \cos({\pi \over a}\,x^1)
\!+\! \cos({3\pi \over a}\,x^1) \over \sin^4({\pi\over a}\,x^1)}
\quad\! \mbox{for $x^1 \!\in\! (0,a)$} ~. \label{TND} \feq
To the best of our knowledge, the full stress-energy VEV for
the present configuration has never been given in the preceeding
literature; only the global results on energy and pressure
discussed hereafter have
previously been considered (see the citations below). \parn
Let us recall that the reduced cylinder trace in this case is
(see Eq. (6.33) in Part I)
\beq T^{(1)}(\t) = {e^{{\pi \over 2a}\t}\over e^{{\pi \over a}\t}-1}~.
\label{TDNTr} \feq
The above expression, along with prescription \rref{E1RenPP}, allows
us to determine the renormalized, reduced bulk energy; for example, for
$d = 3$ we obtain
\beq E_1^{ren} = {7\pi^2 \over 11520 a^3} ~. \label{EnDN} \feq
Concerning the non-vanishing component of the boundary pressure,
both definitions \rref{alt1PP} \rref{alt2PP} give
(with $A$ as in Eq. \rref{TND})
\beq p^{ren}_1(\bx)\Big|_{\pi_0} = -3A ~, \qquad
p^{ren}_1(\bx)\Big|_{\pi_a} = 3A ~; \label{PDN} \feq
let us stress that, similarly to the results found for the segment
configuration in Part I (see Eq.s (6.27) (6.35) therein),
the expressions in Eq. \rref{PDN} have the opposite sign with
respect to those in Eq. \rref{PDir}, holding in the case of
Dirichlet conditions on both the planes $\pi_0$ and $\pi_a$\,.
This means that in the present Dirichlet-Neumann case, the forces
on the boundary planes are repulsive
({\footnote{Recall the considerations of the previous footnote
\ref{FootFoExt}.}}). \parn
The results in Eq.s \rref{EnDN} \rref{PDN} agree with those reported, e.g.,
in \cite{Pinto,Pin} and Santos et al. \cite{PPDN} (see also \cite{Actor,Bord}).
\vspace{-0.4cm}
\subsection{Neumann boundary conditions.} Assume that
\beq \partial_{x^1}\Fi(t,\bx) = 0 \qquad \mbox{for $t \in \reali$,
and $\bx \in \pi_0$ or $\bx \in \pi_a$} ~. \feq
Recall that in this case, according to the considerations of subsection
\ref{HiNPBC}, the reduced operator $\AA_1$ must be viewed as acting
in the Hilbert space $\L2m0(0,a)$ of square integrable functions on
$(0,a)$ with mean zero (see Eq. \rref{mean0}). The cylinder kernel of
$\AA_1$ is (see Eq. (6.38) in Part I)
\beq T^{(1)}(\t\,;x^1,y^1) = {1 \over 2a} \l[{\cos({\pi \over a}(\bx\!-\!\by))
- e^{-\t}\over \cosh\t - \cos({\pi \over a}(\bx\!-\!\by))}
+ {\cos({\pi \over a}(\bx\!+\!\by)) - e^{-\t} \over \cosh\t
- \cos({\pi \over a}(\bx\!+\!\by))} \r] . \feq
Using this kernel along with relations (\ref{T00PP}-\ref{TijPP}),
we obtain the renormalized stress-energy VEV; in particular,
for $d = 3$ the residue evaluation yields
\begin{equation}\begin{split}
& \hspace{2.7cm} \la 0 | \Ti_{\mu\nu}(\bx) | 0 \ra_{ren}\Big|_{\mu,\nu = 0,1,2,3} = \\
& = A \l(\!\!\barray{cccc}
-1\!& \!0   &   0   &   0   \\
0\! & \!-3  &   0   &   0   \\
0\! & \!0   &   1   &   0   \\
0\! & \!0   &   0   &   1   \farray \!\!\r)
+ \l(\!\xi\! -\!{1 \over 6}\r) B(x^1)
\l(\!\!\barray{cccc}
-1\!&   0   &   0   &   0   \\
0\! &   0   &   0   &   0   \\
0\! &   0   &   1   &   0   \\
0\! &   0   &   0   &   1   \farray\!\!\r) ,
\end{split}\end{equation}
where $A$ and $B(x^1)$ are defined as in Eq. \rref{TmnDD}. \parn
Let us pass to the reduced bulk energy; in subsection 6.8
of Part I we noticed that in this case the cylinder trace
$T^{(1)}(\t)$ coincides with the one of Eq. \rref{TDDTrPP},
corresponding to the case of Dirichlet boundary conditions.
In consequence of this the renormalized, reduced bulk energy
for $d = 3$ is the same as in the case of Dirichlet boundary
conditions (see Eq. \rref{EnDDPP}):
$$ E_1^{ren} = -\,{\pi^2 \over 1440 a^3} ~. $$
Furthermore, concerning the pressure on the boundary, both definitions
\rref{alt1PP} and \rref{alt2PP} also give the same result we derived
in the case of Dirichlet boundary conditions (see Eq. \rref{PDir};
again, $A$ is the coefficient of Eq. \rref{TmnDD})
$$ p^{ren}_1(\bx)\Big|_{\pi_0} = 3A ~, \qquad
p^{ren}_1(\bx)\Big|_{\pi_a} = - 3A ~. $$
\vspace{-0.8cm}
\subsection{Periodic boundary conditions.}\label{PPPerSubsec} The last
case we consider is the one where
$$ \Fi(t,0,x^2,...,x^d) = \Fi(t,a,x^2,...,x^d) ~, \qquad
\partial_{x^1}\Fi(t,0,x^2,...,x^d) = \partial_{x^1}\Fi(t,a,x^2,...,x^d) $$
\beq \mbox{for $t,x^2,...,x^d \in \reali$} ~. \feq
Similarly to what was said for the segment with periodic boundary
conditions, note that, as explained in subsection \ref{curvSubsec},
this configuration would be more properly formulated in terms of a free
scalar field on the flat manifold $\Om := \Toro^1_a \times \reali^{d-1}$\,,
where the first factor is the torus $\Toro^1_a := \reali/(a \interi)$\,. \parn
As in the case of Neumann boundary conditions, the basic Hilbert space
for the reduced problem is $\L2m0(\Toro^1_a)$ (see subsection \ref{HiNPBC},
Eq. \rref{mean0}). We know that the cylinder kernel associated to $\AA_1$
in this case is (see Eq. (6.44) in Part I)
\beq T^{(1)}(\t\,;x^1,y^1) = {\cos({2\pi\over a}(x^1\!-\!y^1)) - e^{-{2\pi\over a}\t}
\over a \l[\cosh({2\pi\over a}\t) - \cos({2\pi\over a}(x^1\!-\!y^1))\r]} ~. \feq
Again, we can employ Eq.s (\ref{T00PP}-\ref{TijPP}) to evaluate the
renormalized stress-energy VEV. Differently from the previous subcases,
this time the expressions appearing in intermediate steps of the required
calculations are simple enough to be reported; as an example, let us focus
on the evaluation of the component $\la 0| \Ti_{00}(\bx)|0\ra_{ren}$\,.
First of all, note that
\beq T^{(1)}(\t\,;x^1,x^1) = {1 \over a} \,
\Big[\coth\Big({\pi \over a}\,\t\Big) - 1\Big] ~; \feq
\beq \partial_{x^1 y^1} T^{(1)}(\t\,;x^1,x^1) = {2\pi^2 \over a^3} \,
\coth\Big({\pi \over a}\,\t\Big)\,
\mbox{csch}^2\Big({\pi \over a}\,\t\Big) ~. \feq
So, after some simple algebraic manipulations, Eq. \rref{T00PP} yields
\begin{equation}\begin{split}
& \la 0 | \Ti_{00}(\bx)|0\ra_{ren} = -{\Cd \over 4a}\,\Res\!
\l({2\!-\!(1\!-\!4\xi)d \over \t^{d+1}}\,
\Big[\coth\Big({\pi \over a}\,\t\Big) - 1\Big] ~+ \r. \\
& \hspace{4.5cm} +\! \l. {2(1\!-\!4\xi) \over (d\!-\!1)\,\t^{d+1}}
\l({\pi \over a}\,\t\r)^{\!\!2} \coth\!\Big({\pi \over a}\,\t\Big)
\mbox{csch}^2\Big({\pi \over a}\,\t\Big) ; 0\r) . \label{T00PPRes}
\end{split}\end{equation}
The function in the above expression, whose residue in $\t = 0$ is
required, is easily seen to be meromorphic with a pole of order $d+2$
in $\t = 0$;\! more precisely, its Laurent expansion is
\begin{equation}\begin{split}
& \hspace{1.7cm} -{d(d\!-\!3)\!+\!4(d^2\!-\!d\!-\!2)\xi \over 4\pi(d\!-\!1)}\,
{1\over\t^{d+2}} + {(2\!-\!d)\!+\!4d\xi \over 4a}\,{1 \over \t^{d+1}}\; + \\
& + {\pi((2\!-\!d)\!+\!4d\xi) \over 12 a^2}\,{1 \over \t^d} -
{\pi^3(-(d^2\!-\!3d\!-\!4)\!+\!4(d^2\!-\!d\!-\!6)\xi) \over
180(d\!-\!1)\,a^4}\,{1 \over \t^{d-2}} + O(\t^{4-d}) ~.
\end{split}\end{equation}
For example, for $d=3$ Eq. \rref{T00PPRes} yields ($A_3 = -1/\pi$,
see Eq. \rref{Ad})
\beq \la 0 | \Ti_{00}(\bx)|0\ra_{ren} = -\,{\pi^2 \over 90\,a^4} ~;
\label{T00PP3} \feq
proceeding similarly for the other components of the renormalized
stress-energy VEV (for $d = 3$), we obtain
\beq \la 0 | \Ti_{\mu\nu}(\bx) | 0 \ra_{ren} \Big|_{\mu,\nu = 0,1,2,3}
= {\pi^2 \over 90 a^4} \! \l(\!\!\barray{cccc}
-1\!& \!0   &   0   &  \,0   \\
0\! & \!-3  &   0   &  \,0   \\
0\! & \!0   &   1   &  \,0   \\
0\! & \!0   &   0   &  \,1   \farray \!\!\r) \,. \label{TP} \feq
Notice that the renormalized stress-energy tensor \rref{TP} does not depend
explicitly on the periodic coordinate $x^1$. This comes as no surprise;
indeed, it reflects the invariance of the theory under translations
$x^1 \mapsto x^1\!+\!\al$ (for arbitrary $\al \in \reali$). \parn
To conclude, let us consider the total bulk energy. First, recall
that the reduced cylinder trace in this case is (see Eq. (6.47)
in Part I)
\beq T(\t) = {2 \over e^{{2\pi \over a}\t} - 1} ~; \label{TPTr}\feq
so, using once more the prescription \rref{E1RenPP} we obtain
\beq E^{ren} = -\,{\pi^2 \over 90 a^3} ~. \label{EnP} \feq
In conclusion, let us stress that the above expression for the
total energy agrees with the results in \cite{Bord,AAP0,10AppZ,Mil}.
\section{The case of massive field constrained by perpendicular hyperplanes}\label{SecPerp}
\subsection{Introducing the problem.} We consider a scalar field of
nonzero mass $m$ fulfilling either Dirichlet or Neumann boundary
conditions on $d_1$ orthogonal hyperplanes in $d = d_1\!+\!d_2$
spatial dimension. More precisely, we are interested in the case
\beq \Om = (0,+\infty)^{d_1} \times \reali^{d_2} ~, \qquad
V(\bx) = m^2 \quad (m > 0) ~. \label{ddim} \feq
The domain $\Omega$ is bounded by the hyperplanes $\{ x^1 = 0 \}$,...,
$\{ x^d = 0 \}$; its boundary is the union of the faces
\beq \pi_n := \{\bx \in \partial\Om ~|~ x^n \!=\! 0\} ~,
\qquad \mbox{($n \in \{1,...,d_1\}$)} \feq
and, for each one of them, either Dirichlet or Neumann boundary conditions
are prescribed.
As anticipated in the Introduction, the same configuration with
at most three faces was considered by Actor et al. in \cite{ActBox2,Actor};
when possible we will establish connections with these works,
making direct comparison. \parn
Before proceeding, let us stress that also in this case we can use
the results of subsection \ref{slabSubsec} on slab configurations;
because of this, we will consider the reduced problem based on
\beq \Om_1 = (0,+\infty)^{d_1} ~, \qquad \AA_1 := -\Delta_1 + m^2 ~, \label{reduc} \feq
with the appropriate boundary conditions.
%% \vspace{-0.4cm}
\subsection{The reduced heat kernel.} In our approach, a basic step
for the analysis of the reduced problem \rref{reduc} is the computation
of the heat kernel $K^{(1)}(\t\,;\bx_1,\by_1)$ associated to $\AA_1$.
The result is
({\footnote{Here is one way to derive Eq. \rref{K2PP}. First, notice
that a complete orthonormal system of (improper) eigenfunctions of
$\AA_1 = -\Delta_1 + m^2$ on $\Om_1 = (0,+\infty)^{d_1}$ fulfilling
the prescribed (either Dirichlet or Neumann) boundary conditions on
$\partial\Om_1$ is given by
$$ F_{\bk}(\bx_1) := {1 \over (2\pi)^{d_1/2}}\prod_{n = 1}^{d_1}
\Big(e^{i k_n x_1^n} + \al_n\,e^{-i k_n x_1^n} \Big)\,, \quad
\om_{\bk} := \sqrt{|\bk|^2\!+\!m^2} \quad \mbox{for $\bk
\in \KK \equiv (0,+\infty)^{d_1}$} ~; $$
here, for any $n \in \{1,...,d_1\}$, $\al_n$ is defined according to
Eq. \rref{an}. Then, using the eigenfunction expansion \rref{eqheat}
of the heat kernel, we obtain
\begin{equation*}\begin{split}
K^{(1)}(\t\,; \bx_1, \by_1) & = {e^{-m^2 \t} \over (2\pi)^{d_1}} \prod_{n = 1}^{d_1}
\int_0^{+\infty}\!\!dk \; e^{-\t\,k^2} \Big(e^{i k x_1^n} +
\al_n\,e^{-i k x_1^n} \Big) \Big(e^{-i k y_1^n} + \al_n\,e^{i k y_1^n} \Big) = \\
& = {e^{-m^2 \t} \over (2\pi)^{d_1}} \prod_{n = 1}^{d_1}\int_{-\infty}^{+\infty}
\!\!dk\;e^{-\t\,k^2}\Big(e^{i k (x_1^n-y_1^n)} + \al_n\,e^{i k (x_1^n+y_1^n)}\Big)~;
\end{split}\end{equation*}
Eq. \rref{K2PP} follows by explicitly evaluating every single Gaussian
integral in the product.}})
\beq K^{(1)}(\t\,;\bx_1,\by_1) = {e^{-m^2 \t} \over (4\pi \t)^{d_1/2}}
\prod_{n = 1}^{d_1}\!\l(e^{-{(x_1^n-y_1^n)^2 \over 4\t}} +
\al_n\, e^{-{(x_1^n + y_1^n)^2\over 4\t}}\r) \,, \label{K2PP} \feq
where, for any $n \in \{1,...,d_1\}$, $\al_n \in \reali$ is a parameter
distinguishing between Dirichlet and Neumann boundary conditions on the
face $\pi_n$; more precisely,
\beq \al_n := \l\{\!\!\barray{ll} -1 & \mbox{for Dirichlet B.C. on $\pi_n$} \\
+1 & \mbox{for Neumann B.C. on $\pi_n$} \farray\r. \qquad
(n \in \{1,...,d_1\}) ~. \label{an} \feq
Eq. \rref{K2PP} can be re-written as
({\footnote{\label{footProdSum} This result depends on the identity
$$ \prod_{n = 1}^d (a_n + b_n) = \sum_{n = 0}^d {1 \over (d\!-\!n)!n!}
\sum_{\si \in S_d}\! \l(\prod_{i=1}^n\, a_{\si(i)}\r)\!
\l(\prod_{j= n+1}^d b_{\si(j)}\r) $$
holding for any $d \in \{1,2,3,...\}$, $a_n,b_n \in \reali$ ($n\in\{1,2,...,d\}$),
where by convention we intend $\prod_{i=1}^0 a_{\si(i)} :=
\prod_{j=d+1}^d b_{\si(j)} := 1$\,.}})
\begin{equation}\begin{split}
& \hspace{5.3cm} K^{(1)}(\t\,;\bx_1,\by_1) = \\
& {e^{-m^2 \t} \over (4\pi \t)^{d_1/ 2}}
\sum_{n = 0}^{d_1} {1 \over (d_1\!-\!n)!n!} \sum_{\si \in S_{d_1}}\!\al_{n,\si}\,
e^{-{1 \over 4\t}\l(\sum_{i=1}^n\l(x_1^{\si(i)}\!-y_1^{\si(i)}\r)^{\!2}
+ \sum_{j= n+1}^{d_1}\l(x_1^{\si(j)}\!+y_1^{\si(j)}\r)^{\!2}\r)} ~. \label{KPP}
\end{split}\end{equation}
Here and in the following $S_{d_1}$ denotes the symmetric group with
$d_1$ elements and, by convention, the sums over $i$ and $j$ in \rref{KPP}
are zero for $n = 0$ and $n = d_1$, respectively;
moreover, for any $\si \in S_{d_1}$, we put
\beq \al_{n,\si} := \prod_{l = n+1}^{d_1} \! \al_{\si(l)} \quad
\mbox{for $n \in \{0,...,d_1\!-\!1\}$}~, \qquad \al_{d_1,\si} := 1 ~.  \feq
\vspace{-0.8cm}
\subsection{The reduced Dirichlet kernel.} In the following we will use
Eq. \rref{DirHeat} to express the Dirichlet kernel $\Dir_s^{(1)}(\bx_1,\by_1)$
associated to $\AA_1$, along with its analytic continuation, in terms of
$K^{(1)}$. Substituting the expression \rref{KPP} for the heat kernel
into Eq. \rref{DirHeat}, we obtain
\beq \Dir_s^{(1)}(\bx_1,\by_1) = {1 \over (4\pi)^{d_1/2}\,\Ga(s)}
\sum_{n = 0}^{d_1} {1 \over (d_1\!-\!n)!n!} \sum_{\si \in S_{d_1}}
\al_{n,\si}\;\II_{s,n,\si}(\bx_1,\by_1) ~, \label{Dir2PP} \feq
where, for each $n\!\in\!\{0,...,d_1\}$ and each $\si\!\in\!S_{d_1}$,
\beq \II_{s,n,\si}(\bx_1,\by_1) := \int_0^{+\infty}\!\!d\t\;
\t^{s-{d_1 \over 2}-1}\, e^{-m^2 \t -{1 \over 4\t}
\Ns^2_{n,\si}(\bx_1,\by_1)} ~, \label{IntII} \feq
\beq \Ns_{n,\si}(\bx_1,\by_1) := \l(\sum_{i=1}^n
(x_1^{\si(i)}\!-y_1^{\si(i)})^2 + \sum_{j= n+1}^{d_1}
\!(x_1^{\si(j)}\!+ y_1^{\si(j)})^2\r)^{\!\!1/2} ~. \feq
Clearly $\Ns_{n,\si}(\bx_1,\by_1) \geqs 0$\,, so that convergence
conditions for the integral in Eq. \rref{IntII} can be readily infered;
more precisely, if $\Ns_{n,\si}(\bx_1,\by_1) > 0$ the integral converges
for any $s \in \complessi$ while if $\Ns_{n,\si}(\bx_1,\by_1) = 0$
(which, in the interior of $\Om$, happens if and only if $n\!=\!d_1$
and $\by_1\!=\!\bx_1$) it only converges for
\beq \Re s > {d_1 \over 2} ~. \label{Resd2}\feq
Under these assumptions for convergence, Eq. \rref{IntII} is strictly
related to a known integral representation of the modified Bessel
function of the second kind $K_{\nu}$ (see, e.g., \cite{NIST}, pag.253,
Eq.10.32.10); using this representation we obtain, for any $\bx_1,\by_1
\in (0,+\infty)^{d_1}$,
\beq \II_{s,n,\si}(\bx_1,\by_1) = 2^{{d_1 \over 2}+1-s}\,m^{d_1 - 2s}\;
\GD_{s-{d_1\over 2}}(m^2\,\Ns^2_{n,\si}(\bx_1,\by_1))~, \label{IIks} \feq
where, for the sake of brevity, we have put
\beq \GD_\nu:[0,+\infty) \to \complessi ~, \qquad
u \mapsto \GD_\nu(z) := z^{\nu/2} K_\nu(\sqrt{z})~. \label{eqz} \feq
In conclusion
\beq \Dir_s^{(1)}(\bx_1,\by_1)\! =\! {2^{1-s}m^{d_1-2s}
\over (2\pi)^{d_1/2}\,\Ga(s)} \sum_{n=0}^{d_1}{1 \over (d_1\!-\!n)!n!}
\!\!\sum_{\si \in S_{d_1}}\!\!\al_{n,\si}\,\GD_{\!s-{d_1\over 2}}
(m^2\Ns^2_{n,\si}(\bx_1,\by_1)) \,. \label{Dir2PPGD}\feq
The derivatives of $\Dir_s^{(1)}$ of any order can be computed using
the identity
({\footnote{We already considered the map $\GD_\nu$ of Eq. \rref{eqz}
in Appendix D of Part I; therein we also showed how to derive
the relations \rref{idender} \rref{GDzero}.}})
\beq {d^n \GD_\nu \over dz^n}\,(z) = \l(\!-{1 \over 2}\r)^{\!\!n}
\GD_{\nu-n}(z) \qquad \mbox{for $n \in \{1,2,3,...\}$} ~; \label{idender} \feq
moreover, in the cases where $\Ns_{n,\si}(\bx_1,\by_1) = 0$, we can
resort to the relation
\beq \GD_\nu(0) = 2^{\nu-1} \Ga(\nu) \label{GDzero}\qquad
\mbox{for $\nu \in \complessi$, $\Re \nu > 0$} ~. \feq
\subsection{The $\boma{d}$-dimensional Dirichlet kernel.} Using the
previous results and Eq.s (\ref{TmnDirRid1}-\ref{TmnDirRid3}), we
obtain the following expressions for the Dirichlet kernel $\Dir_s(\bx,\by)$
of the $d$-dimensional problem \rref{ddim} and for its derivatives,
along the diagonal $\by = \bx$\,:
\begin{equation}\begin{split}
\Dir_{\s \pm 1 \over 2}(\bx,\by) \Big|_{\by = \bx} & =
{2^{2 \mp 1 - \s \over 2} m^{d \mp 1 - \s} \over (2\pi)^{d/2}\,
\Ga({\s \pm 1 \over 2})}\sum_{n = 0}^{d_1} {1\over(d_1\!-\!n)!n!} \;\cdot \\
& \hspace{2cm} \cdot \sum_{\si \in S_{d_1}}\! \al_{n,\si}\,
\GD_{\!{\s-d \pm 1 \over 2}}(m^2\Ns^2_{n,\si}(\bx_1,\by_1))
\Big|_{\by_1 = \bx_1} ~; \label{DirPM1}
\end{split}\end{equation}
\begin{equation}\begin{split}
& \partial_{z_1^i w_1^j} \Dir_{\s+1\over 2}(\bx,\by)\Big|_{\by = \bx} =
{2^{{1 - \s \over 2}}\,m^{d-1-\s} \over (2\pi)^{d/2}\,\Ga({\s+1\over 2})}
\sum_{n = 0}^{d_1}{1 \over (d_1\!-\!n)!n!} \;\cdot \\
& \hspace{4.5cm}\cdot \sum_{\si \in S_{d_1}}\!\al_{n,\si}\,\partial_{z_1^i w_1^j}
\GD_{\!{\s-d + 1 \over 2}}(m^2\Ns^2_{n,\si}(\bx_1,\by_1))\Big|_{\by_1 = \bx_1} \\
& \hspace{3.3cm} \mbox{for $z,w \in \{x,y\}$ and $i,j \in \{1,...,d_1\}$} ~ ;
\label{DirPM2}
\end{split}\end{equation}
\begin{equation}\begin{split}
& \partial_{x_2^i y_2^j}\Dir_{\s+1\over 2}(\bx,\by)\Big|_{\by = \bx}\!\! =
- \partial_{x_2^i x_2^j}\Dir_{\s+1\over 2}(\bx,\by)\Big|_{\by = \bx}\!\! =
- \partial_{y_2^i y_2^j}\Dir_{\s+1\over 2}(\bx,\by)\Big|_{\by = \bx}\!\! = \\
& \hspace{0.1cm} = \de_{ij}{2^{{1 - \s \over 2}} m^{d + 1 - \s} \over
(2\pi)^{d/2}\,\Ga({\s+1 \over 2})}\!\sum_{n = 0}^{d_1}\!{1\over(d_1\!-\!n)!n!}\!
\sum_{\si \in S_{d_1}}\!\!\al_{n,\si}\,\GD_{\!{\s-d - 1 \over 2}}
(m^2\Ns^2_{n,\si}(\bx_1,\by_1))\Big|_{\by_1 = \bx_1} \! \\
& \hspace{5.4cm} \mbox{for $i,j \in \{1,...,d_2\}$} ~. \label{DirPM3}
\end{split}\end{equation}
Note that, in each of the sums over $n$ appearing in the above expressions,
the terms corresponding to $n \in \{0,1,...,d_1 - 1\}$ are analytic functions
of $\s$ on the whole complex plane since $\Ns^2_{n,\si}(\bx_1,\bx_1)>0$
for any $\bx_1 \in (0,+\infty)^{d_1}$. On the contrary, the terms
corresponding to $n = d_1$ deserve particular attention; indeed,
$\Ns^2_{d_1,\si}(\bx_1,\bx_1) = 0$ for any $\bx_1 \in (0,+\infty)^{d_1}$
so that, in order to evaluate these contributions, we have to resort to
Eq. \rref{GDzero} (also recalling Eq. \rref{idender}). In this way we obtain
\beq \GD_{\!{\s-d \pm 1 \over 2}}(m^2\Ns^2_{d_1,\si}(\bx_1,\by_1))
\Big|_{\by_1 = \bx_1} \!= 2^{{\s - d\pm 1\over 2}-1}\,
\Ga\!\l({\s\!-\!d\!\pm\! 1 \over 2}\r) \,; \label{GG0}\feq
\begin{equation}\begin{split}
& \partial_{z_1^i w_1^j} \GD_{\!{\s-d + 1 \over 2}}
(m^2\Ns^2_{d_1,\si}(\bx_1,\by_1))\Big|_{\by_1 = \bx_1}\!\! =
-\,\de_{ij}\,(2\de_{zw}\!-\!1)m^2\,2^{{\s-d-3\over 2}}\,
\Ga\!\l(\!{\s\!-\!d\!-\!1 \over 2}\!\r) \\
& \hspace{3cm} \mbox{for $z,w \in \{x,y\}$ and
$i,j \in \{1,...,d_1\}$} ~. \label{DzwGG0}
\end{split}\end{equation}
In principle, the above equations hold with the limitations on $\s$
arising from Eq. \rref{GDzero}; more precisely, Eq. \rref{GG0} holds
for $\Re\s > d\!\mp\!1$ and Eq. \rref{DzwGG0} for $\Re \s > d\!+\!1$.
However, the right-hand sides of these equations are well defined
and analytic on the whole complex plane with the exception of simple
poles placed at
\beq \s = d + 1 - 2 \ell \qquad \mbox{for $\ell \in \{0,1,2,...\}$} ~;
\label{pole} \feq
this remark gives the meromorphic continuation in $\s$ of the
functions in Eq.s \rref{GG0} \rref{DzwGG0} and, consequently,
of the terms with $n = d_1$ in Eq.s (\ref{DirPM1}-\ref{DirPM3}).
\vspace{0.1cm}
\subsection{The stress-energy tensor.} \label{quellachee}
Using Eq.s (\ref{Tidir00}-\ref{Tidirij}) along with
Eq.s (\ref{DirPM1}-\ref{DirPM3}) (and Eq.s (\ref{GG0}-\ref{DzwGG0})
for the terms with $n = d_1$), we obtain the analytic continuation
to a meromorphic function of each component of the regularized
stress-energy VEV, required in order to implement the local zeta
approach. In particular, we have the following expressions for
the non-vanishing components:
\begin{equation}\begin{split}
& \la 0 | \Tis_{0 0}(\bx) | 0 \ra =
{2^{1 - \s \over 2}\, m^{d + 1} \over (2\pi)^{d/2}\,\Ga({\s + 1 \over 2})}
\l({m \over \mm}\r)^{\!\!-\s} \sum_{n = 0}^{d_1} {1\over(d_1\!-\!n)!n!}
\sum_{\si \in S_{d_1}}\! \al_{n,\si}\, \cdot \\
& \hspace{2cm} \cdot \l[\!\l(\!{d\!-\!d_1\!-\!1\!+\!\s \over 4}
- (d\!-\!d_1\!+\!1\!-\!\s)\xi\!\r)\!
\GD_{\!{\s-d - 1 \over 2}}(m^2\Ns^2_{n,\si}(\bx_1,\by_1)) \; + \r. \\
& \hspace{3.2cm} \l. + \l(\!{1 \over 4}\!-\!\xi\!\r)\!
(1 + m^{-2}\partial^{x_1^\ell} \partial_{y_1^\ell})\,\GD_{\!{\s-d+1 \over 2}}
(m^2\Ns^2_{n,\si}(\bx_1,\by_1))\r]_{\by_1 = \bx_1}\! ; \label{T00PPM}
\end{split}\end{equation}
\vbox{
\begin{equation*}\begin{split}
& \la 0 | \Tis_{i j}(\bx) | 0 \ra =
\la 0 | \Tis_{j i}(\bx) | 0 \ra = {2^{1 - \s \over 2}\, m^{d + 1} \over
(2\pi)^{d/2}\,\Ga({\s + 1 \over 2})} \l({m \over \mm}\r)^{\!\!-\s}
\sum_{n = 0}^{d_1} {1\over(d_1\!-\!n)!n!}\sum_{\si \in S_{d_1}}\! \al_{n,\si}\, \cdot \\
& \cdot \!\l[\!-\!\l(\!{1\over 4}\!-\!\xi\!\r)\!\de_{i j}\!
\l((d\!-\!d_1\!+\!1\!-\!\s)\,\GD_{\!{\s-d-1 \over 2}}(m^2\Ns^2_{n,\si}(\bx_1,\by_1))
+ \GD_{\!{\s-d + 1 \over 2}}(m^2\Ns^2_{n,\si}(\bx_1,\by_1)) \r) + \r. \\
& \;\l. +\, m^{-2}\!\l(\!-\!\l(\!{1\over 4}\!-\!\xi\!\r)\!\de_{i j}
\partial^{x_1^{\ell}}\partial_{y_1^{\ell}}\!
+ \!\l(\!{1\over 2}\!-\!\xi\!\r)\! \partial_{x_1^i y_1^j} -
\xi\,\partial_{x_1^i x_1^j}\!\r)\! \GD_{\!{\s-d + 1 \over 2}}
(m^2\Ns^2_{n,\si}(\bx_1,\by_1))\r]_{\!\by_1 = \bx_1}
\end{split}\end{equation*}
\beq \mbox{for $i,j \in \{1,...,d_1\}$} ~ ; \label{TijPPM11} \feq
}
\begin{equation*}\begin{split}
& \la 0 | \Tis_{i j}(\bx) | 0 \ra =
\la 0 | \Tis_{j i}(\bx) | 0 \ra = -\de_{ij}\, {2^{1 - \s \over 2}\, m^{d + 1}
\over (2\pi)^{d/2}\,\Ga({\s + 1 \over 2})}\!\l({m \over \mm}\r)^{\!\!-\s}
\sum_{n = 0}^{d_1} {1\over(d_1\!-\!n)!n!}\!\sum_{\si \in S_{d_1}}\!\! \al_{n,\si}\, \cdot \\
& \hspace{2cm} \cdot \!\l[\!\l(\!{d\!-\!d_1\!-\!1\!-\!\s \over 4} - (d\!-\!d_1\!+\!1\!-\!\s)\xi\!\r)\!
\GD_{\!{\s-d - 1 \over 2}}(m^2\Ns^2_{n,\si}(\bx_1,\by_1))\, + \r. \\
& \hspace{3.2cm} \l. + \l(\!{1\over 4}\!-\!\xi\!\r)\!
(1 + m^{-2}\partial^{x_1^{\ell}}\partial_{y_1^{\ell}})
\GD_{\!{\s-d + 1 \over 2}}(m^2\Ns^2_{n,\si}(\bx_1,\by_1))\r]_{\by_1 = \bx_1}
\end{split}\end{equation*}
\beq \mbox{for $i,j \in \{d_1\!+\!1,...,d_1\!+\!d_2 \equiv d\}$} ~ . \label{TijPPM22} \feq
The renormalized VEV of the stress-energy tensor is obtained sending
$\s$ to zero in the above expressions; the only singularities appear in
the terms corresponding to $n = d_1$, which must be treated resorting
to Eq.s (\ref{GG0}-\ref{pole}). \salto
The conclusions are the following: \parn
i) For \textsl{$d$ even} each component of the regularized stress-energy
VEV is an analytic function of $\s$ near $\s = 0$; thus its renormalized
version is obtained via the restricted zeta approach, i.e., by simply
evaluating Eq.s (\ref{T00PPM}-\ref{TijPPM22}) at $\s = 0$. \parn
ii) For \textsl{$d$ odd} the regularized stress-energy VEV has a
simple pole in $\s = 0$, so that we have to resort to the extended
zeta approach and consider the regular part at $\s = 0$. \salto
The manipulations indicated in i) are trivial; the ones indicated
in ii) could be performed in principle for an arbitrary odd dimension,
but the final expressions are too lengthy to be reported here. For this
reason, we prefer to exemplify ii) in two special cases with $d = 3$
(see subsections \ref{PlaneMass} and \ref{2planes}).
\vspace{-0.4cm}
\subsection{The boundary forces.}\label{pressPPer} As in the previous
section, following the general framework of subsection \ref{pressuretmunu},
we can give two alternative definitions for the pressure acting on the
boundary of the spatial domain $\Om$\,. \parn
Let us consider a point $\bx \in \partial\Om$; if $d_1 > 1$ we exclude
$\bx$ to be on a corner, where the outer normal is ill-defined. To
fix our ideas, we assume that $\bx$ is an inner point of the face
\beq \pi_1 := \{x^1 = 0\} \cap \partial\Om ~; \feq
let $\bn(\bx)$ denote the unit outer normal at $\bx$, so that
$\bn(\bx) = (-1,0,...,0)$. On the one hand, we can define
\beq p^{ren}_i(\bx) := RP\Big|_{\s = 0}\la 0|\Tis_{i j}(\bx)|0\ra\,n^j(\bx)
= - RP\Big|_{\s = 0}\la 0|\Tis_{i1}(\bx)|0\ra ~. \label{alt1Perp} \feq
On the other hand, we can consider the alternative definition
\beq p^{ren}_i(\bx) := \!\l(\lim_{\bx'\in\Om, \bx'\to\bx}
\la 0|\Ti_{i j}(\bx') |0\ra_{ren} \r)\!n^j(\bx) = -\l(\lim_{\bx'\in\Om, \bx'\to\bx}
\la 0|\Ti_{i1}(\bx') |0\ra_{ren} \r) .\! \label{alt2Perp} \feq
As a matter of fact, \textsl{the alternatives \rref{alt1Perp} \rref{alt2Perp}
give the same result for the renormalized pressure}; the rest of
the present subsection is mainly devoted to the justification
of this statement, which requires a nontrivial analysis. \parn
Consider the expression \rref{TijPPM11} of $\la 0|\Tis_{i1}|0\ra$.
Due to the considerations in the previous subsections, it is
apparent that the terms in \rref{TijPPM11} deserving special
attention when comparing the definitions \rref{alt1Perp}
\rref{alt2Perp} for the pressure at $\bx = (\bx_1,\bx_2) \in \pi_1$
are those with $n,\si$ such that
\beq \mbox{$\Ns^2_{n,\si}(\bx_1,\bx_1) = 0$} \quad \mbox{and}
\quad \mbox{$\Ns^2_{n,\si}(\bx'_1,\bx'_1) \neq 0$ for
$\bx'\equiv (\bx'_1, \bx'_2) \in \Om$} ~; \label{NsiZer}\feq
these are easily seen to correspond to the choices
\beq \mbox{$n = d_1\!-\!1$ and $\si \in S_{d_1}$ such that
$\si(d_1) = 1$} ~. \label{nsi} \feq
Indeed, all the terms corresponding to values of $n,\si$ different
from the above ones are straightforwardly seen to yield the same
results according to both the prescriptions \rref{alt1Perp} and
\rref{alt2Perp}. Now, let us focus on the potentially troublesome
terms described in Eq. \rref{TijPPM11}, corresponding to a choice
of the form \rref{nsi}; in the sequel we will show that these terms
\textsl{do not contribute to $p^{ren}_i(\bx)$} for both the alternatives
\rref{alt1Perp} \rref{alt2Perp}. In order to prove this, let us denote
with $\bx' = (\bx'_1,\bx'_2)$ either the previously mentioned boundary
point $\bx \in \pi_1$ or a point of $\Om$. In the expression \rref{TijPPM11}
for $\la 0|\Tis_{i1}(\bx')|0\ra$ we pick up any problematic term with
$n,\si$ as in Eq. \rref{nsi}; this reads
\beq  \l[-\!\l(\!{1\over 4}\!-\!\xi\!\r)\!\de_{i 1}\!
\l((d\!-\!d_1\!+\!1\!-\!\s)\,\GD_{\!{\s-d-1 \over 2}}(m^2\Ns^2_{d_1-1,\si})
+ \GD_{\!{\s-d + 1 \over 2}}(m^2\Ns^2_{d_1-1,\si}) \r) + \r. \label{d1si} \feq
$$ \l. +\, m^{-2}\!\l(\!-\!\l(\!{1\over 4}\!-\!\xi\!\r)\!\de_{i 1}
\partial^{{x'}_1^{\ell}}\partial_{{y'}_1^{\ell}}\!
+ \!\l(\!{1\over 2}\!-\!\xi\!\r)\! \partial_{{x'}_1^i {y'}_1^1}\! -
\xi\,\partial_{{x'}_1^i {x'}_1^1}\!\r)\! \GD_{\!{\s-d + 1 \over 2}}
(m^2\Ns^2_{d_1-1,\si})\r]_{\!{\by'}_1 = {\bx'}_1} $$
(some of the arguments have been suppressed for the sake of brevity).
With some effort, noting that $\Ns^2_{d_1-1,\si}({\bx'}_1,{\by'}_1) =
({x'}_1^1 + {y'}_1^1)^2 + \sum_{i=2}^{d_2} ({x'}_1^i-{y'}_1^i)^2$
(due to $\si(d_1) = 1$) and using Eq. \rref{idender}, we can re-write
expression \rref{d1si} as
\begin{equation}\begin{split}
& -\de_{i1} \l({1\over 4} - \xi \r)\!\l[(d\!+\!1\!-\!\s)
\GD_{\s-d-1 \over 2}(z^2) + \GD_{\s-d+1 \over 2}(z^2)
- z^2 \GD_{\s-d-3 \over 2}(z^2)\r]_{z = 2m {x'}_1^1} \\
& \hspace{6cm} \equiv f(\s,{x'}^1_1) ~. \label{expd}
\end{split}\end{equation}
We now claim that
\beq f(\s,0)\Big|_{\s = 0} = 0 \label{fs0X} \feq
and
\beq \lim_{{x'}^1_1 \to 0}\bigg(f(\s,{x'}^1_1)\Big|_{\s = 0}\,\bigg) = 0~;
\label{fs0Y} \feq
let us remark that in both the above equations the prescription of taking
the regular part is superfluous, and $|_{\s = 0}$ indicates the analytic
continuation at $\s = 0$. Eq.s \rref{fs0X} \rref{fs0Y} state that the
problematic terms do not contribute to $p_i^{ren}(\bx)$ as defined by
Eq.s \rref{alt1Perp} and \rref{alt2Perp}, respectively. \parn
In order to prove Eq. \rref{fs0X} we note that
\beq f(\s,0) = -\de_{i1}\!\l(\!{1\over 4}\!-\!\xi\!\r)\!2^{\s-d-1 \over 2}\!
\l[{d\!+\!1\!-\!\s \over 2}\; \Ga\Big({\s\!-\!d\!-\!1\over 2}\Big)
+ \Ga\Big({\s\!-\!d\!+\!1 \over 2}\Big)\r] = 0 \label{expd0} \feq
where in the first passage we used Eq. \rref{GDzero}, while
equality to zero follows from the well-known relation
$\Ga(z\!+\!1) = z\,\Ga(z)$\,. \parn
To prove Eq. \rref{fs0Y}, recalling the definition \rref{eqz} we infer
(for all ${x'}_1^1 > 0$)
\beq f(\s,{x'}^1_1)\Big|_{\s = 0} = \label{expd1} \feq
$$ = -\de_{i1}\!\l(\!{1\over 4}\!-\!\xi\!\r)\!
\l[z^{-{d+1 \over 2}}\Big((d\!+\!1)K_{d+1 \over 2}(z)+ z\,K_{d-1 \over 2}(z)
- z\, K_{d + 3 \over 2}(z)\Big)\r]_{z = 2m {x'}_1^1} = 0 ~; $$
in this case equality to zero follows from the identity below,
holding for Bessel functions $K_\nu$ of any order (see \cite{NIST},
p.251, Eq.10.29.1):
\beq z\,K_{\nu + 1}(z) - z\,K_{\nu - 1}(z) = 2\nu\;K_\nu(z) ~. \label{BesselK}\feq
In the above we assumed $\bx$ to belong to the face with $x^1 = 0$
but, of course, similar considerations also hold for all the other
boundary points not on the corners. \salto
Now, let us spend a few words about points on the corners of $\partial\Om$,
which appear if $d_1 > 1$; we already noticed that the outer normal is
ill-defined at these points, so that the notion of pressure is itself
problematic. The natural strategies that could be guessed to overcome
the problem make apparent some pathologies that we prefer to describe
in an example, rather than in general: see subsection \ref{2planes}.
\salto
In passing, let us anticipate that an analysis similar to the one of
this subsection will also be given in Part IV (see subsection 3.6 therein),
for the case of a massless field confined within a $d$-dimensional
box and fulfilling Dirichlet boundary conditions.
As in the present setting, the alternative definitions \rref{preren}
\rref{alt} will be found to agree at all boundary points except those
on the corners, where pathologies appear.
\vspace{-0.4cm}
\subsection{Introducing two examples.} The framework developed in the
previous subsections will be illustrated heafter, for $d=3$, in these
cases: $\Om := (0,+\infty) \times \reali^{2}$, representing a half-space,
and $\Om := (0,+\infty)^2 \times \reali$, representing a wedge bounded
by orthogonal half-planes. In both cases, we consider Dirichlet and/or
Neumann boundary conditions.
\vspace{-0.4cm}
\subsection{A half-space in spatial dimension $\boma{d = 3}$.} \label{PlaneMass}
Let
\beq \Om := (0,+\infty) \times \reali^2 ~; \label{Om1Pla} \feq
this is the subcase of the general setting \rref{ddim} corresponding to $d=3$ and
\beq d_1 = 1 ~, \qquad d_2 = 2 ~. \feq
With the above choices, the symmetric group appearing in the general framework
of subsection \ref{quellachee} consists of the sole idendity ($S_{d_1} = S_{1}
= \{id\}$). We have $\bx_1 = (x_1^1) \equiv x^1$ (and the analogous relations
for $\by_1$)\,; besides, $\Ns_{0,id}(\bx_1,\by_1) = |x^1\!+\!y^1|$ and
$\Ns_{1,id}(\bx_1,\by_1) = |x^1\!-y^1|$\,. Using the relations
(\ref{DirPM1}-\ref{DirPM3}), (\ref{Tidir00}-\ref{Tidirij}) and Eq.s
\rref{idender} \rref{GDzero} and \rref{BesselK}, with some simple algebraic
manipulations we obtain the following expressions for the non-vanishing
components of the regularized stress-energy VEV (where $\bx = (x^1,x^2,x^3)$):
\begin{equation}\begin{split}
& \la 0 | \Tis_{0 0}(\bx) | 0 \ra = -\,{m^4 \over 32 \pi^{3/2}\,\Ga({\s+1 \over 2})}
\l({m \over \mm}\r)^{\!\!-\s} \l[(1\!-\!\s)\,\Ga\Big({\s\!-\!4 \over 2}\Big)\; + \r. \\
& \hspace{1.5cm} \l. -\,2^{5-{\s \over 2}}\,\al_1 \l(\!\Big({1 \over 4} - \xi\Big)
\GD_{\s-2 \over 2}(z^2) \!+\! \Big({1 \over 2} - (3\!-\!\s)\xi\Big)
\GD_{\s-4 \over 2}(z^2)\!\r)_{\!\!z = 2 m x^1}\r];
\end{split}\end{equation}
\beq \la 0 | \Tis_{1 1}(\bx) | 0 \ra = {m^4 \over 32 \pi^{3/2}\,\Ga({\s + 1 \over 2})}
\l({m \over \mm}\r)^{\!\!-\s} \Ga\Big({\s\!-\!4 \over 2}\Big) ~; \feq
\begin{equation}\begin{split}
& \la 0 | \Tis_{2 2}(\bx) | 0 \ra = \la 0 | \Tis_{3 3}(\bx) | 0 \ra =
{m^4 \over 32 \pi^{3/2}\,\Ga({\s + 1 \over 2})} \l({m \over \mm}\r)^{\!\!-\s}
\l[\Ga\Big({\s\!-\!4 \over 2}\Big) \,+ \r. \\
& \hspace{1.cm}\l. -\,2^{5-{\s \over 2}} \al_1 \l(\!\Big({1 \over 4}-\xi\Big)
\GD_{\s - 2 \over 2}(z^2)\! +\! \Big({2\!-\!\s \over 4}-(3\!-\!\s)\xi\Big)
\GD_{\s - 4 \over 2}(z^2)\!\r)_{\!\!z = 2 m x^1}\r] .
\end{split}\end{equation}
The above expressions are easily seen to give the meromorphic continuation
of the regularized stress-energy VEV to the whole complex plane, with poles
determined by terms with the gamma function. In particular, all the above
components have a simple pole in $\s = 0$; thus, we follow the extended
version of the zeta approach and define the renormalized quantities to be
the regular parts in $\s= 0$. Recalling again that Eq. \rref{xic} gives
$$ \xi_3 = {1 \over 6} ~, $$
we write the final results in the form \rref{TRinCo}, obtaining
\begin{equation}\begin{split}
& \la 0 | \Ti_{0 0}(\bx) | 0 \ra_{ren} = {m^4 \over 384 \pi^2}
\l(\!3\Big(4 \ln\!\Big({m \over 2\mm}\Big)\!+\!1\Big)
+ 32\,\al_1\,{K_1(2 m x^1) \over 2 m x^1} \r) + \\
& \hspace{3.5cm} - \al_1 \l(\!\xi\!-\!{1 \over 6}\r) {m^4 \over \pi^2}\!
\l({(2m x^1)K_1(2 m x^1) + 3 K_2(2 m x^1) \over (2 m x^1)^2}\r) ; \label{T001pM}
\end{split}\end{equation}
\begin{equation}\begin{split}
& \la 0 | \Ti_{1 1}(\bx) | 0 \ra_{ren} =
- {m^4 \over 128 \pi^2} \l(4 \ln\!\Big({m \over 2\mm}\Big) - 3\r) ;
\end{split}\end{equation}
$$ \la 0 | \Ti_{2 2}(\bx) | 0 \ra_{ren} \!=\! \la 0 | \Ti_{3 3}(\bx)|0\ra_{ren}
\!= - {m^4 \over 384 \pi^2}\!\l(\!3\Big(4 \ln\!\Big({m \over 2\mm}\Big)\!-\!3\Big)\!
+ 32\,\al_1\,{K_1(2 m x^1) \over 2 m x^1} \r)\! + $$
\beq - \,\al_1 \l(\!\xi\!-\!{1 \over 6}\r) {m^4 \over \pi^2}\!
\l({(2m x^1)K_1(2 m x^1) + 3 K_2(2 m x^1) \over (2 m x^1)^2}\r) . \label{T221pM} \feq
Let us comment briefly on the above results. Firtly, note that the
renormalized VEV of the stress-energy tensor does not depend explicitly
on the spatial coordinates $x^2,x^3$\,; this comes as no surprise due
to the homogeneity with respect to these variables of the spatial
configuration considered. Besides, in agreement with the general
results of subsection \ref{pressPPer}, the components
$\la 0|\Tis_{11}|0\ra$, $\la 0|\Ti_{11}|0\ra_{ren}$ are constant
and the two alternative definitions for the pressure on the plane
$\pi_1 = \{x_1 = 0\}$ (see Eq.s \rref{alt1Perp} \rref{alt2Perp})
give the same result; more esplicitly, we obtain
\beq p^{ren}_i = - \de_{i 1} \la 0|\Ti_{11}|0\ra_{ren} \qquad
(i \in \{1,2,3\}) ~. \label{press1P}\feq
In conclusion, let us make a comparison with the results derived in
\cite{Actor} for the configuration with a single plane in arbitrary
spatial dimension (to be considered here with $d = 3$); therein the
attention is restricted to the ``minimal'' ($\xi = 0$) and
``conformal'' ($\xi = 1/6$) settings. In both cases the results
derived here are found to agree with those reported in \cite{Actor}
(let us remark that the mass scale $\mm$ employed here does not
coincide with the one considered therein; the latter is proportional,
via a numerical coefficient, to $m^2/\mm$).
\vspace{-0.4cm}
\subsection{The rectangular wedge.}\label{2planes}
Let us pass to the case of a wedge in $\reali^3$, bounded by two
perpendicular half-planes; this is represented as
\beq \Om := (0,+\infty)^2 \times \reali~, \label{Om2Pla}\feq
corresponding to the general framework \rref{ddim} with $d=3$ and
\beq d_1=2~, \qquad d_2 = 1 ~. \feq
In passing, let us mention that the rectangular wedge model is also
considered by Actor and Bender \cite{Actor}, for arbitrary spatial
dimension; yet, these author restrict the attention to a massless
field, a case we discuss in the next subsection. \parn
In the present setting, the symmetric group ($S_{d_1} = S_2$) of
subsection \ref{quellachee} consists of two elements, i.e., the
identity $id$ and the exchange $\sip$:
\beq S_{d_1} \equiv S_{2} = \{id,\sip\}~, \qquad id(1) = 1\,, ~~
id(2) = 2~, \quad \sip(1) = 2\,,~~ \sip(2) = 1 ~. \feq
Moreover, with $\bx_1 = (x_1^1,x_1^2) \equiv (x^1,x^2)$
and the analogous relations for $\by_1$, we have
\beq \barray{c} \Ns_{0,\si}(\bx_1,\by_1) =
\Big((x^1\!\!+\!y^1)^2\!+(x^2\!\!+\!y^2)^2\Big)^{\!\!1/2}
\\ \Ns_{2,\si}(\bx_1,\by_1) = \Big((x^1\!\!-\!y^1)^2\!
+(x^2\!\!-\!y^2)^2\Big)^{\!\!1/2} \farray
\qquad \mbox{for $\si \in S_2 = \{id,\sip\}$} ~; \feq
$\phantom{.}$ \vspace{-0.6cm}
$$ \Ns_{1,id}(\bx_1,\by_1) = \Big((x^1\!\!-\!y^1)^2\!+
(x^2\!\!+\!y^2)^2\Big)^{\!\!1/2}\,,
\;\; \Ns_{1,\sip}(\bx_1,\by_1) = \Big( (x^1\!\!+\!y^1)^2\!
+(x^2\!\!-\!y^2)^2\Big)^{\!\!1/2}\,.$$
Also in this case, we can use the relations (\ref{Tidir00}-\ref{Tidirij}),
(\ref{DirPM1}-\ref{DirPM3}) and the identities \rref{idender} \rref{GDzero}
and \rref{BesselK} to deduce expressions for the non-vanishing components
of the regularized stress-energy VEV. More precisely, we obtain the following
(with $\bx = (x^1,x^2,x^3)$) \vspace{0.4cm}\\
\vbox{
\beq \la 0 | \Tis_{0 0}(\bx) | 0 \ra = - \,{m^4 \over 32 \pi^{3/2}\,
\Ga({\s + 1 \over 2})} \l({m \over \mm}\r)^{\!\!-\s} \l[ (1\!-\!\s)\,
\Ga\Big({\s\!-\!4 \over 2}\Big)\, + \r. \label{Tis00P2}\feq
\begin{equation*}\begin{split}
& \hspace{0.7cm} - 2^{5-{\s\over 2}} \sum_{i=1,2} \al_i
\l(\!\Big({1 \over 2}-(3\!-\!\s)\xi\Big) \GD_{\s - 4 \over 2}(z^2)
+ \!\Big({1 \over 4}-\xi\Big) \GD_{\s - 2 \over 2}(z^2)\!\r)_{\!\!z = 2m x^i}\!+ \\
& \l. -\, 2^{5-{\s\over 2}}\,\al_1 \al_2 \l(\!\Big({1 \over 4}-(2\!-\!\s)\xi\Big)
\GD_{\s - 4 \over 2}(z^2) +\! \Big({1 \over 4}-\xi\Big)
\GD_{\s - 2 \over 2}(z^2)\!\r)_{\!\!z = 2m \sqrt{(x^1)^2\!+ (x^2)^2}}\,\r] ;
\end{split}\end{equation*}
} \\
\vbox{
\beq \la 0 | \Tis_{i j}(\bx) | 0 \ra = \la 0 | \Tis_{j i}(\bx) | 0 \ra =
{m^4 \over 32 \pi^{3/2}\, \Ga({\s + 1 \over 2})} \l({m \over \mm}\r)^{\!\!-\s}
\l[\de_{i j} \; \Ga\Big({\s\!-\!4 \over 2}\Big)\, + \r. \label{TisijP2} \feq
}
\begin{equation*}\begin{split}
& \hspace{0.5cm}- 2^{5-{\s \over 2}} \,\al_{\bj}\; \de_{i j}
\l(\!\Big({2\!-\!\s \over 4}-(3\!-\!\s)\xi\Big)\GD_{\s- 4 \over 2}(z^2)
+ \Big({1 \over 4}-\xi\Big)\GD_{\s - 2 \over 2}(z^2)\!\r)_{\!\!z = 2 m x^{\bj}}\!+ \\
& - 2^{5-{\s \over 2}} \,\al_1 \al_2 \l(\!{1 \over 4}\!-\!\xi\!\r)\!\l(\!\de_{i j}\,
\Big((3\!-\!\s) \GD_{\s-4 \over 2}(z^2) + \GD_{\s-2 \over 2}(z^2)\Big)+ \r. \\
& \hspace{4.5cm} \l.\l. - \,4\,m^2 x^i x^j\, \GD_{\s-6 \over 2}(z^2)
\r)_{\!\!z = 2 m \sqrt{(x^1)^2+(x^2)^2}}\,\r] \hspace{0.5cm} \mbox{for $i,j \in \{1,2\}$}\,;
\end{split}\end{equation*}
\beq \la 0 | \Tis_{3 3}(\bx) | 0 \ra = {m^4 \over 32 \pi^{3/2}\,
\Ga({\s + 1 \over 2})} \l({m \over \mm}\r)^{\!\!-\s}
\l[\Ga\Big({\s\!-\!4 \over 2}\Big)\, + \r. \label{Tis33P2}\feq
\begin{equation*}\begin{split}
& \hspace{0.6cm} - 2^{5-{\s \over 2}} \sum_{i=1,2}\al_i\!
\l(\!\Big({2\!-\!\s \over 4}-(3\!-\!\s)\xi\Big)\GD_{\s - 4 \over 2}(z^2)
+ \Big({1 \over 4}-\xi\Big) \GD_{\s - 2 \over 2}(z^2)\!\r)_{\!\!z = 2 m x^i}\! + \\
& \l. -\,2^{5-{\s \over 2}}\,\al_1\al_2
\l(\!\Big({1\!-\!\s \over 4}-(2\!-\!\s)\xi\Big) \GD_{\s - 4 \over 2}(z^2)
+ \Big({1 \over 4}-\xi\Big) \GD_{\s - 2 \over 2}(z^2)\!
\r)_{\!\!z = 2 m \sqrt{(x^1)^2+(x^2)^2}}\r].
\end{split}\end{equation*}
The above expressions give the meromorphic continuation of the regularized
stress-energy VEV to the whole complex plane, with poles determined by
terms with the gamma function. In particular all components
(\ref{Tis00P2}-\ref{Tis33P2}) have a simple pole in $\s = 0$, and we must
resort to the extended zeta approach taking again the regular parts at $\s = 0$.
Recalling once more that Eq. \rref{xic} gives
$$ \xi_3 = {1 \over 6} ~, $$
hereafter we report separately the conformal and non-conformal parts of
each component of the renormalized stress-energy VEV; these are
\begin{equation}\begin{split}
& \la 0|\Ti^{\Co}_{0 0}(\bx)|0\ra_{ren} = {m^4 \over 384 \pi^2}
\l[3\Big(4\ln\!\Big({m \over 2\mm}\Big)\!+1\Big)\!
+ 32 \sum_{i = 1,2}\! \al_i\l({K_1(z) \over z}\r)_{\!\!z = 2m x^i}\! + \r. \\
& \hspace{4.5cm}\l. + \,32\,\al_1 \al_2 \l({z\,K_1(z)\!-\!K_2(z) \over z^2}
\r)_{\!\!z = 2 m \sqrt{(x^1)^2\!+ (x^2)^2}} \r] , \label{T00Rin2Pla}
\end{split}\end{equation}
\begin{equation}\begin{split}
& \la 0|\Ti^{\NCo}_{0 0}(\bx)|0\ra_{ren} = -\,{m^4 \over \pi^2}
\l[\,\sum_{i =1,2}\al_i \l({z\,K_1(z)+3 K_2(z) \over z^2}\r)_{\!\!z = 2m x^i} + \r. \\
& \hspace{5cm} \l. + \,\al_1\al_2 \l({z\, K_1(z)\!+\!2 K_2(z) \over z^2}
\r)_{\!\!z = 2m \sqrt{(x^1)^2\!+ (x^2)^2}} \r] ;
\end{split}\end{equation}
\begin{equation*}\begin{split}
& \la 0 | \Ti^{\Co}_{i j}(\bx) | 0 \ra_{ren} =
-\,{m^4 \over 384\pi^2}\l[3\,\de_{i j}\! \l(4\ln\Big({m \over 2\mm}\Big)\!-\!3\r)\!
+ 32\,\al_{\bj}\, \de_{i j}\!\l({K_1(z) \over z}\!\r)_{\!\!z = 2 m x^{\bj}}\! + \r. \\
& \hspace{1.cm}\l. +\, 32\,\al_1 \al_2 \l(\!\de_{i j}\,{z\,K_1(z)\!+\!3 K_2(z) \over z^2}
- 4 m^2 x^i x^j\,{K_3(z) \over z^3} \r)_{\!\!z = 2 m \sqrt{(x^1)^2+(x^2)^2}}\r]
\end{split}\end{equation*}
\beq \mbox{for $i,j \in \{1,2\}$} ~; \label{TijCoRin2Pla} \feq
\begin{equation}\begin{split}
& \la 0 | \Ti^{\NCo}_{i j}(\bx) | 0 \ra_{ren} = {m^4 \over \pi^2} \l[\al_{\bj}\,\de_{i j}
\l({z\,K_1(z)\!+\!3 K_2(z) \over z^2} \r)_{\!\!z = 2 m x^{\bj}} \!+ \r.\\
& \hspace{1.cm} \l. +\, \al_1 \al_2 \l(\de_{i j}\,{z\,K_1(z)\!+\!3 K_2(z) \over z^2}
- \,4 m^2 x^i x^j\; {K_3(z) \over z^3}\r)_{\!\!z = 2 m \sqrt{(x^1)^2+(x^2)^2}} \r] \\
& \hspace{5.5cm} \mbox{for $i,j \in \{1,2\}$} ~; \label{TijNCoRin2Pla}
\end{split}\end{equation}
\begin{equation}\begin{split}
& \la 0 | \Ti^{\Co}_{3 3}(\bx) | 0 \ra_{ren} = - {m^4 \over 384 \pi^2}
\l[3 \l(4\ln\!\Big({m \over 2\mm}\Big)\!-\!3\r)\! +
32 \sum_{i=1,2}\al_i\l({K_1(z) \over z} \r)_{\!\!z = 2 m x^i}\! + \r. \\
& \hspace{4.5cm} \l. + \,32\, \al_1 \al_2 \l({z\,K_1(z)\!-\!K_2(z) \over z^2}
\r)_{\!\!z = 2 m \sqrt{(x^1)^2+(x^2)^2}}\r] ;
\end{split}\end{equation}
\begin{equation}\begin{split}
& \la 0 | \Ti^{\NCo}_{3 3}(\bx) | 0 \ra_{ren} = {m^4 \over \pi^2} \l[\,
\sum_{i=1,2}\al_i \l({z\,K_1(z)\!+\!3 K_2(z) \over z^2}\r)_{\!\!z = 2 m x^i}\! + \r. \\
& \hspace{5cm} \l. +\, \al_1 \al_2 \l({z\,K_1(z)\!+\! 2 K_2(z) \over z^2}
\r)_{\!\!z = 2 m \sqrt{(x^1)^2+(x^2)^2}}\r]. \label{T33Rin2Pla}
\end{split}\end{equation}
As was to be expected, since the configuration \rref{Om2Pla} is invariant
under translation along the $x^3$ direction, none of the expressions
(\ref{T00Rin2Pla}-\ref{T33Rin2Pla}) depends on the spatial coordinate
$x^3$. \salto
Let us now discuss the pressure at points in the half-plane $\pi_1 =
\{x^1 = 0,\, x^2 > 0\}$; note that, we are excluding points on the axis
$\zeta := \{x^1\!=\!x^2\!= 0\}$. The two alternative definitions \rref{alt1Perp}
\rref{alt2Perp} are easily seen to give the same result for the renormalized
version of this quantity, in agreement with the general results of subsection
\ref{pressPPer}: indeed, we can equivalently put $x^1 = 0$ in Eq. \rref{TisijP2}
for $\la 0|\Tis_{i1}(\bx)|0\ra$ ($i \in \{1,2,3\}$) and then analytically continue
up to $\s = 0$, or directly evaluate the renormalized expressions \rref{TijCoRin2Pla}
\rref{TijNCoRin2Pla} for $\la 0|\Ti_{i1}(\bx) |0\ra_{ren}$ in $x^1 = 0$\,.
In both ways, we obtain
$$ p^{ren}_i(\bx)\Big|_{\pi_1} = \de_{i 1}\l[{m^4 \over 384\pi^2}
\l(3\!\l(4\ln\Big({m \over 2\mm}\Big)\!-\!3\r)\!
+ 32\,\al_2\,{(1\!+\!\al_1) z\,K_1(z)\!+\!3 \al_1 K_2(z) \over z^2}\r)\!+ \r. $$
\beq \l.- \l(\!\xi\!-\!{1 \over 6}\!\r){m^4 \over \pi^2}\,(1\!+\!\al_1)\al_2 \;
{z\,K_1(z)\!+\!3 K_2(z) \over z^2}\,\r]_{\!z = 2 m x^2} \quad
(i \in \{1,2,3\}) ~.\hspace{-1cm} \label{press2P} \feq
Let us stress that, differently from all the configurations considered so
far within this paper, in this case the renormalized pressure on the boundary
depends in general on the parameter $\xi$; more precisely, this happens
whenever Neumann boundary conditions are imposed on the half-plane $\pi_1$
(so that $\al_1 = +1$)\,. \vspace{0.1cm}\\
Now, let us consider the axis $\zeta = \{x^1\!=\!x^2\!= 0\}$; at any point
of this axis the outer normal is ill defined, so that there is a basic
obstruction to speaking of the pressure. However, we can discuss what
happens if a point $\bx = (0,x^2, x^3) \in \pi_1$ moves towards the axis
$\zeta$, i.e., if we consider the limit $x^2\to 0^{+}$. In this limit
$p^{ren}_i(\bx)$ is found to diverge; more precisely, Eq. \rref{press2P}
and the known asymptotic behaviour of the Bessel function $K_\nu$ near
zero (see \cite{NIST}, p.252, Eq.10.30.2) give the following:
\beq p^{ren}_1(\bx) = O\left({1 \over (x^2)^4}\right) \qquad
\mbox{for $\bx \in \pi_1$, $x^2 \to 0^+$} ~. \feq
\vspace{-0.9cm}
\subsection{The previous examples in the zero mass limit.}\label{zermassPP}
Let us first consider the half-space configuration \rref{Om1Pla} bounded by
the plane $\pi_1$, analysed in subsection \ref{PlaneMass}. The expressions
(\ref{T001pM}-\ref{T221pM}) for the components of $\la 0|\Ti_{\mu\nu}|0\ra_{ren}$
give, in the limit $m \to 0^{+}$, \vspace{-0.4cm}\\
\beq \la 0 | \Ti_{\mu\nu}(\bx) | 0 \ra_{ren} \Big|_{\mu,\nu = 0,1,2,3}
\hspace{-0.1cm} = \l(\!\xi\!-\!{1 \over 6}\r){3\,\al_1 \over 8\pi^2 (x^1)^4}
\l(\!\!\barray{cccc}
-1\!&   0   &   0   &   0   \\
0\! &   0   &   0   &   0   \\
0\! &   0   &   1   &   0   \\
0\! &   0   &   0   &   1   \farray\!\r) \, ; \label{TP3} \feq
as for the pressure on a point $\bx \in \pi_1$, starting with Eq. \rref{press1P}
and taking the limit $m \to 0^+$, it is trivial to infer \vspace{-0.2cm}\\
\beq p_i^{ren}(\bx) = 0 \qquad (i \in \{1,2,3\}) ~. \label{press1Pzer}\feq
Let us pass to the case of the rectangular wedge (see subsection \ref{2planes}).
Eq.s (\ref{T00Rin2Pla}-\ref{T33Rin2Pla}) for the renormalized stress-energy VEV
give, in the limit $m \to 0^{+}$, \vspace{-0.2cm}\\
\beq \la 0 | \Ti_{\mu\nu}(\bx) | 0 \ra_{ren} \Big|_{\mu,\nu = 0,1,2,3}
\hspace{-0.1cm} = {\al_1 \al_2 \over 96 \pi^2 \rho^4}
\! \l(\!\!\barray{cccc}
-1\!&   \!0\!           &  \!0          &   \!0   \\
0\! &   \!A_1(\bx)\!    &  \!B(\bx)     &   \!0   \\
0\! &   \!B(\bx)\!      &  \!A_2(\bx)   &   \!0   \\
0\! &   \!0\!           &  \!0          &   \!1   \farray\!\r) + \label{T2P3} \feq
$$ - \l(\!\xi\!-\!{1\over 6}\!\r)\!\l[
{\al_1 \al_2 \over 8 \pi^2 \rho^4} \! \l(\!\!\barray{cccc}
-1\!&   \!0\!           &  \!0          &   \!0   \\
0\! &   \!A_1(\bx)\!    &  \!B(\bx)     &   \!0   \\
0\! &   \!B(\bx)\!      &  \!A_2(\bx)   &   \!0   \\
0\! &   \!0\!           &  \!0          &   \!1   \farray\!\r)\!
- {3 \over 8 \pi^2} \! \l(\!\!\barray{cccc}
-C_0(\bx)\!\!   &   \!\!0\!\!       &  \!\!0\!\!        &   \!\!0   \\
0\!\!           &   \!\!C_1(\bx)\!\!&  \!\!0\!\!        &   \!\!0   \\
0\!\!           &   \!\!0\!\!       &  \!\!C_2(\bx)\!\! &   \!\!0   \\
0\!\!           &   \!\!0\!\!       &  \!\!0\!\!        &   \!\!C_0(\bx) \farray\!\!\r)\r] ; $$
$$ A_i(\bx) := 1-{4 (x^{\sip(i)})^2 \over \rho^2} ~, \quad
C_i(\bx) := {\al_{\sip(i)} \over (x^{\sip(i)})^4} ~, \qquad \mbox{for $i=1,2$} ~; $$
$$ B(\bx) := {4 x^1 x^2 \over \rho^2} ~, \qquad C_0(\bx) := {\al_1 \over (x^1)^4}
+ {\al_2 \over (x^2)^4} + {\al_1\al_2 \over \rho^4} ~, \qquad
\rho := \sqrt{(x^1)^2\!+\!(x^2)^2} $$
(recall that $\sip(1)=2$, $\sip(2)=1$). Next, consider the expression
\rref{press2P} for the pressure acting on a point $\bx$ in the half-plane
$\pi_1 = \{x^1 = 0,\,x^2 > 0\}$, for $m > 0$; using the asymptotic
behaviour of the Bessel function $K_\nu$ near zero (see \cite{NIST},
p.252, Eq.10.30.2) we infer, in the limit $m \to 0^+$,
\beq p_i^{ren}(\bx) = \de_{i1}\l[{\al_1 \al_2 \over 32 \pi^2 (x^2)^4}
- \l(\!\xi\!-\!{1 \over 6}\!\r) {3(1\!+\!\al_1)\al_2 \over 8 \pi^2 (x^2)^4}\r]
\qquad (i \in \{1,2,3\}) ~. \label{press2Pzer} \feq
Notice that, as for the massive analogue \rref{press2P}, the above
expression for the renormalized pressure depends explicitly on $\xi$
if we assume Neumann boundary conditions on $\pi_1$
(so that $\al_1 = + 1$)\,. \parn
In passing, let us remark that both results \rref{press1Pzer} and
\rref{press2Pzer} for the pressure on $\pi_1$ could be determined
equivalently via the prescription \rref{alt}; according to the
latter, we should have first considered the renormalized
stress-energy VEV inside the corresponding spatial domain in the
limit of zero mass (see Eq.s \rref{TP3} \rref{T2P3}) and then move
to the boundary (half-)plane $\pi_1$, i.e., take the limit $x^1 \to 0^+$
({\footnote{Notice that not all of the components in Eq.s \rref{TP3}
\rref{T2P3} are finite for $x^1 \to 0^+$, but only those involved
in the computation of the pressure on $\pi_1$\,; for example, in
both cases we have
$$ \lim_{x^1 \to 0^+}\la 0 | \Ti_{2 2}(\bx) | 0 \ra_{ren} = \infty ~. $$}}). \salto
Let us comment briefly on the construction described above, namely, that
of taking the zero mass limit ($m \to 0^{+}$) of the renormalized results
(\ref{T001pM}-\ref{T221pM}) \rref{press1P} and (\ref{T00Rin2Pla}-\ref{T33Rin2Pla})
\rref{press2P} for the theory with a massive field. As a matter of fact,
this procedure corresponds to studying the case of a massless scalar field
(in the same spatial configurations) with the technique of subsection
\ref{AAeps}. Indeed, in the massless case ($m = 0$) the spectrum of the
fundamental operator $\AA = - \Delta$ is $[0,+\infty)$, for both the settings
\rref{Om1Pla} and \rref{Om2Pla}; according to the framework of the cited
subsection, we could treat these cases using the deformed operator
$\AA_\eps := \AA + \eps^2$ (see Eq. (5.14)), and eventually taking
the limit $\eps \to 0^+$. On the other hand, if $\eps$ is identified with
$m$, we recover the present constructions
({\footnote{The same comments could be made in general for any spatial
dimension $d$ and for any number $d_1$ of faces.}}). \parn
Summing up: Eq.s (\ref{TP3}-\ref{press1Pzer}) and (\ref{T2P3}-\ref{press2Pzer})
yield the renormalized VEVs of the stress-energy tensor and pressure for
a $d=3$ massless field, respectively confined within a half-space and a
rectangular wedge, fulfilling either Dirichlet ($\al_i = -1$) or Neumann
($\al_i = +1$) boundary conditions. \parn
We point out that, in the Dirichlet case, the above results are found to
agree with those derived by Actor and Bender \cite{Actor} for $\xi = 0$
and $\xi = 1/6$, via a different version of the zeta approach
(also involving, essentially, a subtraction of divergent contributions;
see subsections 3.1.1 and 4.1 of the cited paper, setting $d = 3$ therein).
Let us also mention that the massless half-space configuration in the
case of Dirichlet boundary conditions was considered as well in our
previous work \cite{ptp}; therein the same results were obtained starting
from the renormalized stress-energy VEV of a massless field between
parallel planes, and taking the limit of infinite distance between the
planes (see Eq. (5.7) in \cite{ptp}; after an exchange $1 \leftrightarrow 3$
in the coordinate labels this becomes the present Eq. \rref{TP3},
with $\al_1 = -1$). \salto
In the next section we show that the renormalized expressions
(\ref{TP3}-\ref{press2Pzer}), here deduced as the zero mass limit of
a massive theory, can be obtained equivalently as particular cases
of the theory of a massless scalar field confined within two half-planes
forming an angle $\tez$ of arbitrary width; more precisely, the present
configurations with one single plane and two orthogonal planes correspond,
respectively, to the limits $\tez \to \pi$ and $\tez \to \pi/2$\,.
\section{The case of massless field in a $\boma{3}$-dimensional wedge} \label{secwedge}
\subsection{Introducing the problem for arbitrary boundary conditions.}\label{IntroWed}
In this section we consider the case of a scalar field (with no
external forces) confined within a $3$-dimensional wedge, meaning
that the spatial domain $\Om$ is the portion of the space $\reali^3$
enclosed by two half-planes $\pi_0, \pi_\tez$ forming an angle
$\tez \in (0, 2 \pi]$\,; we assume $\pi_0 = \{x^2\!=\!0,\,x^1\!\geqs\!0\}$.
Suitable boundary conditions will be specified in the following. \parn
We choose to confine our attention to the massless case ($V=0$) since,
in this case, we are able to perform the explicit computations with a
moderate effort for arbitrary values of the angle $\al \in (0, 2 \pi]$;
in the massive case ($V = m^2$) we could give an exhaustive analysis
for rational values of $\al/\pi$\, but this would require a big
computational effort and produce cumbersome expressions for the final
results. \parn
We already pointed out in the Introduction that the present framework
has been previously considered by Dowker \cite{DowKen,Dow}, Deutsch
and Candelas \cite{Deu}, Saharian et al. \cite{SahWed00,SahWed} and by
Fulling et al. \cite{FulWed}, amongst other; a more detailed comparison
between our results and these works will be performed in subsections
\ref{DirWed}-\ref{string}. \parn
In passing let us notice that, in case of either Dirichlet or Neumann
boundary conditions, the spatial domain under analysis corresponds for
$\tez = \pi$ and $\tez = \pi/2$, respectively, to the configurations
with a boundary made of a single plane and of two orthogonal half-planes.
Besides, let us also stress that for $\tez = 2\pi$ the two half-planes
$\pi_0, \pi_\tez$ overlap; because of this, for Dirichlet or Neumann
boundary conditions, the boundary consists of a single half-plane,
while, in the case of periodic boundary conditions, the spatial domain
$\Om$ can be identified with $\reali^3$. So, in the latter case one is
actually considering a massless scalar field on the whole Minkowski
spacetime. \parn
We will comment further on each specific case in the next subsections.
In particular, we will show that in the cases with $\al = \pi$ and
$\al = \pi/2$ one recovers the results of subsection \ref{zermassPP}. \salto
In order to deal with the present configuration, it is advisable to
pass to the system of cylindrical coordinates
\beq \bx \mapsto \bq(\bx) \equiv (\rho(\bx),\te(\bx), z(\bx)) \in
(0,+\infty)\!\times\!(0,2\pi)\!\times\!\reali~, \feq
whose inverse will be written $\bq \mapsto \bx(\bq)$.
These coordinates are chosen so that $z(\bx) = x^3$ and the boundary
$\partial\Om$ corresponds to the limit values $\te = 0$ and $\te = \tez$;
the spatial line element reads
\beq d\ell^2 = d\rho^2 + \rho^2 d\te^2 + dz^2 ~. \label{dellWed}\feq
In order to avoid clumsy notations, given any function $\Om \to Y$, $\bx
\mapsto f(\bx)$ (with $Y$ any set), the composition $\bq \in \bq(\Om)
\mapsto f(\bx(\bq))$ will be written as $\bq \mapsto f(\bq)$. \parn
Since curvilinear coordinates are being employed, one should
refer to the framework of subsection \ref{curvSubsec}. Writing $\rho,\te,z$
for the coordinate labels, we see that the only non-vanishing Christoffel
symbols associated to the line element \rref{dellWed} are $\ga^{\rho}_{\te\te}
= -\rho$ and $\ga^{\te}_{\rho \te} = \ga^{\te}_{\te \rho} = {1 \over \rho}$;
so, the second order covariant derivatives of any scalar function $f$ are given by
\begin{equation}\begin{split}
& \!\nabla_{\!\rho\rho}f = \partial_{\rho\rho}f ~, \qquad
\nabla_{\!\rho\te}f = \partial_{\rho\te}f - {1 \over \rho}\,\partial_{\te}f ~,
\qquad \nabla_{\!\rho z}f = \partial_{\rho z}f ~, \\
& \nabla_{\!\te\te}f = \partial_{\te\te}f + \rho \partial_{\rho}f ~,
\qquad \nabla_{\!\te z}f = \partial_{\te z}f ~, \qquad
\nabla_{\!zz}f = \partial_{z z} f ~. \vspace{0.5cm}
\end{split}\end{equation}
In conclusion, let us stress that the configuration under analysis could
be dealt with as a slab configuration where $\Om = \Om_1 \times\reali$,
and $\Om_1 \subset \reali^2$ corresponds to $(0,+\infty)\times(0,\al)$ in
terms of the coordinates $(\rho,\te)$; yet, this approach is not convenient.
In fact, if one works directly on the $3$-dimensional spatial domain $\Om$
(for any of the boundary conditions to be considered in the following),
the modified cylinder kernel $\Tm$ defined in Eq. \rref{SKer} associated
to the fundamental operator $\AA = -\Delta$ can be expressed in terms of
elementary functions (moreover, it is meromorphic in the variable $\t$ on
the whole complex plane). On the contrary, to treat the problem as a slab
configuration we should use the analogous kernel $\Tm$ for the reduced
operator $\AA_1$ on $\Om_1$, or any other integral kernel related to the
latter; these kernels do not possess simple expressions (only integral
representations are available), so that the whole analysis would become
a lot more involved.
\vspace{-0.4cm}
\subsection{The Dirichlet kernel.} For any of the boundary conditions to
be considered in the following, it turns out that the spectrum of $\AA$
is $[0,+\infty)$. Since $\{0\}$ is a non-isolated point of the spectrum,
we must resort to the methods discussed in subsection \ref{AAeps} to
determine the renormalized Dirichlet kernel (and its derivatives); we
regularize the theory using the deformed operator $\AA_\eps = (\sqrt{\AA}+\eps)^2$
(see Eq. (5.15)), whose choice is found, \textsl{a posteriori}, to
be more effective from the computational viewpoint. In the sequel (see
subsections \ref{DirWed}-\ref{string}) we derive the explicit expression
of the modified cylinder kernel $\Tm$ of $\AA$, for several types of
boundary conditions; in any case $\Tm$ is found to be a meromorphic
function of $\t$, decreasing faster than $\t^{-1}$ for $\Re\t \to +\infty$;
this allows us to  proceed as explained in subsection \ref{AAeps}. In
this way we obtain, for the Dirichlet kernels and its derivatives, these
renormalized expressions at $s = \pm 1/2$, respectively (see Eq. \rref{Res1}):
\beq \Dir_{-{1\over 2}}(\bq,\bp)\Big|_{\bp = \bq} =
\Res \Big(2\,\t^{-3}\,\Tm(\t\,;\bq,\bp)\Big|_{\bp = \bq}; 0 \Big) ~;
\label{Res1b} \feq
\begin{equation}\begin{split}
& \nabla_{\!z w}\Dir_{\!+{1\over 2}}(\bq,\bp) \Big|_{\bp = \bq} \!\! =
\Res\Big(\t^{-1}\nabla_{\!v w}\Tm(\t\,;\bq,\bp) \Big|_{\bp = \bq}\!; 0 \Big)~, \\
& \hspace{0.8cm} \mbox{for $v,w$ any two cylindrical coordinates} ~. \label{Res2b}
\end{split}\end{equation}
Here and in the sequel, we are using the following notations:
\begin{equation}\begin{split}
& \hspace{2.15cm} \bq \equiv (\rho,\te,z) ~, \qquad \bp \equiv (\rho',\te',z') ~; \\
& f(\bx(\bq), \bx(\bp)) \equiv f(\bq, \bp) \quad
\mbox{for any $f: \Om \times \Om \to Y$ ($Y$ a set)} ~.
\end{split}\end{equation}
\vspace{-0.8cm}
\subsection{The stress-energy tensor.} Relations \rref{Res1b} \rref{Res2b},
along with Eq.s (\ref{Tidir00}-\ref{Tidirij}), can be used to obtain
the following expressions for the renormalized VEV of the stress-energy tensor:
\begin{equation}\begin{split}
& \la 0 | \Ti_{0 0}(\bq) | 0 \ra_{ren} = \Res\!\l(\t^{-3} \l[\l(\!
\frac{1}{2}\!+\!2\xi\!\r) \Tm(\t\,;\bq,\bp)\; + \r.\r. \\
& \hspace{5.5cm}\l.\l. + \l(\!\frac{1}{4}\!-\!\xi\!\r) \t^2\;
\partial^{q^\ell}\!\partial_{p^\ell}\Tm(\t\,;\bq,\bp)\r]_{\bp = \bq}; 0\r)\,;
\label{T00W}
\end{split}\end{equation}
\begin{equation}\begin{split}
& \la 0|\Ti_{ij}(\bq)|0\ra_{ren} = \la 0|\Ti_{ji}(\bq)|0\ra_{ren} =
\Res\l(\t^{-3}\l[\Big({1\over 2} - 2\xi\Big)\,\de_{i j}\;\Tm(\t\,;\bq,\bp)\;
+ \r.\r. \\
& \hspace{0.4cm} \l.\l.\!+\,\t^2\!\l(\!\Big({1\over 4}\!-\!{\xi\over 2}\Big)
\partial_{q^i p^j}\!-\!{\xi\over 2}\,\nabla_{\!q^i q^j}\!
- \Big({1\over 4}\!-\!\xi\Big)\de_{i j}\partial^{\,q^\ell}\!\partial_{p^\ell}\!\r)\!
\Tm(\t\,;\bq,\bp)\r]_{\bp = \bq};0 \r) \\
& \hspace{5cm} \mbox{for $i,j \!\in\!\{\rho,\te,z\}$} ~. \label{TijW}
\end{split}\end{equation}
In the following subsections we will first compute $\Tm$ and then
explicitly evaluate the residues appearing in Eq.s \rref{T00W} \rref{TijW}
for the cases where either Dirichlet, Neumann or periodic boundary
conditions are prescribed; we will present the final results in the
form \rref{TRinCo}, recalling once again that Eq. \rref{xic} implies
\beq \xi_3 = {1 \over 6} ~. \feq
\vspace{-1cm}
\subsection{The boundary forces.} In order to discuss this topic, since
the spectrum of $\AA$ contains $\{0\}$ as a non-isolated point, we must
resort once more to the framework of subsection \ref{AAeps}; again, we
choose to use the deformed operator $\AA_\eps := (\sqrt{\AA}+\eps)^2$.
We can consider the two alternative definitions \rref{altm} \rref{altt}
for the renormalized pressure on the boundary. Contrary to the
configurations analized in the previous sections, in this case the
two mentioned alternatives \textsl{do not yield the same result}. \parn
As an example, let us focus on the pressure acting on the half-plane
$\pi_\tez$. We indicate with $\bn(\bq)$ the unit outer normal at a
point of this half-plane with coordinates $\bq = (\rho,\tez,z)$;
this has components $(n^\rho, n^{\te}, n^z)(\bq) = (0,1/\rho,0)$\,.
On the one hand, prescription \rref{altm} corresponds to put,
for $i \in \{\rho,\te,z\}$\,,
\beq p^{ren}_i(\bq) := \!\lim_{\eps \to 0^+} \Big(RP\Big|_{\s = 0}\!
\la 0|\Tiseps_{i j}(\bq)|0\ra\, n^j(\bq)\Big) = {1 \over \rho}\;
\lim_{\eps \to 0^+}\! RP\Big|_{\s = 0}\!\la 0|\Tiseps_{i \te}(\bq)|0\ra
\label{alt1W0} \feq
(first consider the regularized stress-energy VEV on $\pi_\tez$, then
analytically continue at $\s=0$ and finally take the limit $\eps\to 0^+$).
Similarly to Eq. \rref{TijW}, the prescription \rref{alt1W0} yields
\begin{equation}\begin{split}
& \hspace{2cm} p^{ren}_i(\bq) = {1 \over \rho}\;
\Res\Bigg(\t^{-3}\l[\Big({1\over 2} - 2\xi\Big)\,\de_{i \te}\;\Tm(\t\,;\bq,\bp)\;
+ \r. \\
& \l.\!+\,\t^2\!\l(\!\Big({1\over 4}\!-\!{\xi\over 2}\Big)
\partial_{q^i \te'}\!-\!{\xi\over 2}\,\nabla_{\!q^i \te}\!
- \Big({1\over 4}\!-\!\xi\Big)\de_{i \te}\partial^{\,q^\ell}\!\partial_{p^\ell}\!\r)\!
\Tm(\t\,;\bq,\bp)\r]_{\bp = \bq \in \pi_\tez}\! ;0 \Bigg) ~; \label{alt1W}
\end{split}\end{equation}
let us stress that in the above expression, we first perform the
evaluation on the boundary and then compute the residue. \parn
On the other hand, the alternative prescription \rref{altt} yields,
for $i \in \{\rho,\te,z\}$\,,
\begin{equation}\begin{split}
& p^{ren}_i(\bq) := \l(\lim_{\bq' \to \bq, \bx(\bq') \in \Om}
\la 0|\Ti_{i j}(\bq') |0\ra_{ren} \r) n^j(\bq) = \\
& \hspace{1.2cm} = {1 \over \rho} \l(\lim_{\bq'\to\bq, \bx(\bq') \in \Om}
\la 0|\Ti_{i\te}(\bq') |0\ra_{ren} \r)\,. \label{alt2W}
\end{split}\end{equation}
In the cases to be considered in subsections \ref{DirWed}-\ref{NeuWed},
it will be apparent that the explicit expressions obtained for the
renormalized stress-energy VEV inside the wedge diverge when approaching
the boundary, in such a way to make divergent the renormalized pressure
defined by \rref{alt2W}. Because of this, we will always refer to
Eq. \rref{alt1W} to deal with the pressure on the boundary.
\vspace{-0.4cm}
\subsection{Some remarks.}\label{genremW} For the computation of $\Tm$
we will often refer to \cite{FulWed}, where this kernel was already
determined for the present configuration (but used in a different
renormalization scheme, based on point splitting as an alternative to
zeta regularization; we note that the kernel $\overline{T}$ in \cite{FulWed}
is the opposite of the kernel $\Tm$ considered here). \parn
At the end of each subsection dealing with Dirichlet and/or Neumann
boundary conditions, we will comment briefly on the results obtained
for the renormalized stress-energy VEV (\ref{T00W}-\ref{TijW}) and
pressure \rref{alt2W} in the cases $\tez = \pi$ and $\tez = \pi/2$,
respectively, describing a massless scalar field on a half-space and
inside a wedge with orthogonal half-planes. Recall that these very
same configurations were analysed as the zero mass limit of a
corresponding massive theory in subsection \ref{zermassPP}; indeed,
we will find that the same results obtained therein can be re-obtained
from Eq.s (\ref{T00W}-\ref{TijW}) and \rref{alt2W}, returning to
the Cartesian coordinates $x^1 = \rho \sin\te$, $x^2 = \rho \cos\te$,
$x^3 = z$ and considering the appropriate transformation laws
({\footnote{Clearly, $\la 0|\Ti_{\mu\nu}(\bq)|0\ra_{ren}$ and $p^{ren}_i(\bq)$
transform, respectively, as a rank-two tensor and a vector.}}).
\vspace{-0.4cm}
\subsection{Dirichlet boundary conditions.}\label{DirWed} Let us
first consider the case where the field fulfills Dirichlet boundary
conditions on the half-planes $\pi_0$, $\pi_\tez$.
In this case a complete orthonormal system of (improper) eigenfunctions
$(\Fk)_{k \in \KK}$ of $\AA = - \Delta$, with eigenvalues $(\om_k^2)_{k \in \KK}$,
is given by
\begin{equation}\begin{split}
& \hspace{0.5cm}\Fk(\bq) := {\sqrt{\om \over \pi \tez}} \,
J_{\lam_n}(\om\rho)\sin(\lam_n \te)\, e^{i h z}\,, \qquad
\lam_n := {n \pi \over \tez} ~, \\
& \om_k^2 := \om^2 + h^2  \qquad \mbox{for $k \equiv (n,\om,h)
\in \KK \equiv \naturali\!\times\!(0,+\infty)\!\times\!\reali$}
\end{split}\end{equation}
(here and elsewhere we are considering the set of positive integers with
$\naturali := \{1,2,3,...\}$; $\KK$ is endowed with the counting measure
on $\naturali$ times the standard Lebesgue measure on $(0,+\infty) \times\reali$,
meaning that $\int_\KK dk \equiv \sum_{n = 1}^{+\infty}$ $\int_0^{+\infty}d\om$
$\int_{-\infty}^{+\infty}dh$). The modified cylinder kernel $\Tm$ can be
evaluated starting from its eigenfunction expansion \rref{SKer}, which
in the present setting reads
\beq \Tm(\t\,;\bq,\bp) = \int_\KK dk\;{e^{-t\om_k} \over \om_k}
\;\Fk(\bq)\Fkc(\bp) = \feq
$$ {1 \over \pi\tez}\sum_{n=1}^{+\infty}\sin(\lam_n\te)\sin(\lam_n\te')
\int_0^{+\infty}\hspace{-0.2cm}d\om\;\om\,J_{\lam_n}(\om\rho)\,
J_{\lam_n}(\om\rho') \int_{-\infty}^{+\infty}\hspace{-0.2cm}dh\;
{e^{-t\sqrt{\om^2+h^2}} \over \sqrt{\om^2\!+\!h^2}}\;e^{i h (z-z')} ~. $$
With some effort, the integrals over $h$ and $\om$ in the above expression
can be explicitly evaluated to yield
\begin{equation}\begin{split}
& \hspace{0.45cm} \Tm(\t\,;\bq,\bp) = {1\over\pi\tez\,\rho\rho'\sinh \uu}\,
\sum_{n = 1}^{+\infty}\sin(\lam_n\te)\,\sin(\lam_n\te')\;
e^{-\lam_n \uu} ~, \\ \label{TWSer}
& \uu := - \ln\!\l({r_{+}-r_{-} \over r_{+}+r_{-}}\r), \qquad
r_{\pm} := \sqrt{(\rho\pm \rho')^2 + (z\!-\!z')^2 + t^2}~.
\end{split}\end{equation}
We refer to \cite{FulWed} for the detailed computations giving the above
result; the notations $\uu, r_{\pm}$ are mutuated from this reference
(also see \cite{LukWed}).
The series in Eq. \rref{TWSer} can be re-expressed via four geometric
series writing the trigonometric functions in terms of complex exponentials;
in this way we obtain
\begin{equation}\begin{split}
& \hspace{5.8cm} \Tm(\t\,;\bq,\bp) =  \label{TWDir} \\
& {1 \over 4\pi\tez\,\rho \rho' \sinh \uu}
\l({\cos({\pi \over \tez}(\te\!-\!\te')) - e^{-{\pi \over \tez} \uu} \over
\cosh({\pi \over \tez} \uu) -\cos({\pi \over \tez}(\te\!-\!\te'))}
- {\cos({\pi \over \tez}(\te\!+\!\te')) - e^{-{\pi \over \tez} \uu} \over
\cosh({\pi \over \tez} \uu) -\cos({\pi \over \tez}(\te\!+\!\te'))} \r) .
\end{split}\end{equation}
Now, we resort to Eq.s (\ref{T00W}-\ref{TijW}) for the renormalized
stress-energy VEV; evaluating the residues therein we obtain
$$ \la 0 | \Ti_{\mu\nu}(\bq) | 0 \ra_{ren}
\Big|_{\mu,\nu = 0,\rho,\te,z} = $$
$$ A(\bq) \! \l(\!\!\barray{cccc}
-1& 0   &   \!\!0\!         &   \!0 \\
0 & 1   &   \!\!0\!         &   \!0 \\
0 & 0   &   \!\!-3\rho^2\!  &   \!0 \\
0 & 0   &   \!\!0\!         &   \!1 \farray \!\!\r)\!
- \l(\!\xi\! -\!{1 \over 6}\r)\!\!\l(\!\!\barray{cccc}
-(B(\bq)\!+\!C(\bq))\!\!\!\!& \!\!\!0\!             & 0\!\!             & \!\!\!0 \\
0\!\!\!\!                   & \!\!\!B(\bq)\!        & -\rho\,E(\bq)\!\! & \!\!\!0 \\
0\!\!\!\!                   & \!\!\!-\rho\,E(\bq)\! & \rho^2\,C(\bq)\!\!& \!\!\!0 \\
0\!\!\!\!                   & \!\!\!0\!             & 0\!\!             & \!\!\!B(\bq)\!+\!C(\bq) \farray\!\!\r) , $$
$$ A(\bq) := {\pi^4\!-\!\tez^4 \over 1440\pi^2\tez^4\,\rho^4} \,, \quad
B(\bq) := {9\pi^4\!-\!3\pi^2(2\pi^2\!+\!\tez^2)\sin^2({\pi\te \over \tez})\!
+\!\tez^2 (\pi^2\!-\!\tez^2)\sin^4({\pi\te \over \tez}) \over
24 \pi^2\tez^4 \sin^4({\pi\te \over \tez})\,\rho^4} ~, $$
\beq C(\bq) := {3\pi^2\!-\!(\pi^2\!\!-\!\tez^2)\sin^4({\pi\te \over \tez})
\over 8 \pi^2 \tez^2 \sin^2({\pi\te\over\tez})\,\rho^4} ~,
\qquad  E(\bq) := {3\pi\cos({\pi\te \over \tez})
\over 8\tez^3\sin^3({\pi\te \over \tez})\,\rho^4} ~. \label{TDW} \feq
It can be easily checked that the above result agrees with the
one derived by Saharian and Tarloyan by means of point-splitting
regularization in \cite{SahWed} (see, in particular, Section 3 therein);
see also the former papers by Dowker et al. \cite{DowKen,Dow} and
by Deutsch and Candelas \cite{Deu}, where point-splitting regularization
is used for the computation of the conformal stress-energy VEV alone. \parn
Let us briefly comment on the explicit expression \rref{TDW} obtained
for the renormalized VEV $\la 0|\Ti_{\mu\nu}(\bq)|0 \ra_{ren}$.
First of all, notice that the function $A(\bq)$ multipling the conformal
part of the renormalized VEV is positive for $\tez < \pi$, negative for
$\tez > \pi$ and vanishes for $\tez = \pi$.
Next, let us stress that both the conformal and non-conformal parts diverge
quartically in $\rho$ in the proximity of the axis $\{\rho = 0\}$. \parn
The non-conformal part also diverges near the half-planes
$\pi_0,\pi_\tez$, that is for $\te \to 0,\tez$; because of this,
the pressure on these half-planes evaluated according to Eq. \rref{alt2W}
is infinite. On the other hand, the alternative definition
(\ref{alt1W0}-\ref{alt1W}) (first move to the boundary, and then take
the analytic continuation) gives a finite pressure on $\pi_\tez$
with components
\beq p^{ren}_i(\bq) = -\,\de_{i\te}\;{\pi^4\!-\!\tez^4 \over 480 \pi^2 \tez^4 \rho^3}
~. \label{pressWDir} \feq
To conclude, we consider the special cases $\al = \pi$ and $\al = \pi/2$
(a space domain bounded by a plane or by two perpendicular half-planes),
and compare the present results with the ones of subsection \ref{zermassPP}.
This can be done with the procedure outlined in subsection \ref{genremW}
(i.e., returning to Cartesian coordinates via the appropriate transformation
rules for tensor coefficients). Indeed, the expressions \rref{TDW} \rref{pressWDir}
(giving the renormalized stress-energy VEV and pressure) are easily seen to
yield for $\tez = \pi$ and $\tez = \pi/2$, respectively, Eq.s \rref{TP3}
\rref{press1Pzer} with $\al_1 = -1$, and Eq.s \rref{T2P3} \rref{press2Pzer}
with $\al_1 = \al_2 = -1$.
\vspace{-0.5cm}
\subsection{Dirichlet-Neumann boundary conditions.} Let us now pass
to the analysis of the wedge configuration where Dirichlet and Neumann
boundary conditions are prescribed, respectively, on the half-planes
$\pi_0$ and $\pi_\tez$;
let us point out that, to the best of our knowledge, this case has
never been considered before in the literature. \parn
A complete orthonormal system of (improper) eigenfunctions
$(\Fk)_{k \in \KK}$ of the fundamental operator $\AA$, with
eigenvalues $(\om_k^2)_{k \in \KK}$, is given by
\begin{equation}\begin{split}
& \Fk(\bq) := {\sqrt{\om \over \pi \tez}} \,
J_{\lam_n}(\om\rho)\sin(\lam_n \te)\, e^{i h z}\,, \qquad
\lam_n := \l(\!n\!+\!{1\over 2}\r) {\pi \over \tez} ~, \\
& \,\om_k^2 := \om^2 + h^2  \qquad \mbox{for $k \equiv (n,\om,h)
\in \KK \equiv \naturali_0\!\times\!(0,+\infty)\!\times\!\reali$}
\end{split}\end{equation}
($\naturali_0 := \{0,1,2,...\}$ is the set of non-negative integers;
the measure on the label space $\KK$ is such that $\int_\KK dk \equiv
\sum_{n = 0}^{+\infty}$ $\int_0^{+\infty}d\om$ $\int_{-\infty}^{+\infty}dh$).
Resorting again to Eq. \rref{SKer} and proceeding similarly to the case
with Dirichlet boundary conditions, we can express the modified cylinder
kernel as \vspace{-0.2cm}\\
\begin{equation}\begin{split}
& \hspace{5.8cm} \Tm(\t\,;\bq,\bp) = \label{TWDirNeu} \\
& {1 \over 2\pi\tez\,\rho \rho' \sinh \uu}
\l({\sinh({\pi \over 2\tez} \uu)\cos({\pi \over 2\tez}(\te\!-\!\te'))
\over \cosh({\pi \over \tez} \uu) -\cos({\pi \over \tez}(\te\!-\!\te'))}
- {\sinh({\pi \over 2\tez} \uu)\cos({\pi \over 2\tez}(\te\!+\!\te'))
\over \cosh({\pi \over \tez} \uu) -\cos({\pi \over \tez}(\te\!+\!\te'))}\r) ,
\end{split}\end{equation}
where $\uu$ si defined as in Eq. \rref{TWSer}. Using Eq.s
(\ref{T00W}-\ref{TijW}) we obtain the renormalized VEV of the
stress-energy tensor:
$$ \la 0 | \Ti_{\mu\nu}(\bq) | 0 \ra_{ren}
\Big|_{\mu,\nu = 0,\rho,\te,z} = $$
$$ A(\bq) \! \l(\!\!\barray{cccc}
1\! & 0   &   \!\!0\!        &   \!\!0 \\
0\! & -1  &   \!\!0\!        &   \!\!0 \\
0\! & 0   &   \!\!3\rho^2\!  &   \!\!0 \\
0\! & 0   &   \!\!0\!        &   \!\!-1 \farray \!\!\r)\!
- \l(\!\xi\! -\!{1 \over 6}\r)\!\!\l(\!\!\barray{cccc}
-(B(\bq)\!+\!C(\bq))\!\!\!\!& \!\!\!0\!             & 0\!\!             & \!\!\!0 \\
0\!\!\!\!                   & \!\!\!B(\bq)\!        & -\rho\,E(\bq)\!\! & \!\!\!0 \\
0\!\!\!\!                   & \!\!\!-\rho\,E(\bq)\! & \rho^2\,C(\bq)\!\!& \!\!\!0 \\
0\!\!\!\!                   & \!\!\!0\!             & 0\!\!             & \!\!\!B(\bq)\!+\!C(\bq) \farray\!\!\r) , $$
$$ A(\bq) := {7\pi^4\!+\!8\tez^4 \over 11520\pi^2\tez^4\rho^4} ~, $$
$$ B(\bq) := {-3 \pi^2 \cos({\pi\te \over \tez})
(11 \pi^2\!-\!2\tez^2\!+\!(\pi^2\!+\!2\tez^2)\cos({2\pi\te \over \tez}))
+2\tez^2 (\pi^2\!+\!2\tez^2)\sin^4({\pi\te \over \tez}) \over
96 \pi^2\tez^4 \sin^4({\pi\te \over \tez}) \,\rho^4} ~, $$
\beq C(\bq) := {6 \pi^2\!\cos({\pi\te \over \tez})\!+\!(\pi^2\!+\!2\tez^2)
\sin^2({\pi\te \over \tez}) \over 16 \pi^2 \tez^2 \sin^2({\pi\te \over \tez})\,\rho^4}~,
\quad E(\bq) := {3 \pi (3\!+\!\cos({2\pi\te \over \tez}))
\over 32\tez^3 \sin^3({\pi \te \over \tez})\,\rho^4}~. \label{TDNW} \feq
Let us compare the above results with the ones of Eq. \rref{TDW},
holding for the case of Dirichlet boundary conditions on both the
half-planes $\pi_0,\pi_\tez$\,. As in Eq. \rref{TDW}, both the
conformal and non-conformal parts of the renormalized stress-energy
VEV diverge for $\rho \to 0$\,; the latter also diverges for
$\te \to 0,\tez$, so that Eq. \rref{alt2W} yields again a divergent
pressure on the boundary\,. On the other hand, the present results
differ from the ones derived in the previous subsection because of
some crucial features; in particular, the conformal part has an
overall minus sign and the function $A(\bq)$ in  Eq. \rref{TDNW} is
always strictly positive (whereas the one in Eq. \rref{TDW} changes
sign for $\al < \pi$ and $\al > \pi$). \salto
As for the boundary forces on $\pi_\tez$, resorting to Eq. \rref{alt1W},
in this case we obtain
\beq p^{ren}_i = {\de_{i\te} \over 8\pi^2\rho^3}\l[{7\pi^4\!+\!8\tez^4
\over 480 \tez^4} - \l(\!\xi\!-\!{1 \over 6}\r)
{\pi^2\!+\!2\tez^2 \over \tez^2}\r] . \label{pressWDN} \feq
We notice that also in this case the parameter $\xi$ appears in the
final expression for the renormalized pressure; because of this the
resulting boundary forces can be either attractive or repulsive,
depending on the value of $\xi$\,. \salto
Again, we conclude comparing the results obtained for the renormalized
stress-energy VEV and pressure for $\tez = \pi/2$ with the analogous
ones deduced in subsection \ref{zermassPP}. Also this time, Eq.s \rref{TDNW}
\rref{pressWDN} (with $\tez = \pi/2$) are found to give, respectively,
Eq.s \rref{T2P3} \rref{press2Pzer} with $\al_1 = 1$, $\al_2 = -1$.
\vspace{-0.4cm}
\subsection{Neumann boundary conditions.}\label{NeuWed} We now analyze
the case where the field fulfills Neumann boundary conditions on both
the half-planes $\pi_0$, $\pi_\tez$\,.
A complete orthonormal system of (improper) eigenfunctions $(\Fk)_{k \in \KK}$
in $L^2(\Om)$ of the fundamental operator $\AA$, with related eigenvalues
$(\om_k^2)_{k \in \KK}$, is
\begin{equation}\begin{split}
& \hspace{0.6cm}\Fk(\bq) = {\sqrt{\om \over \pi \tez}} \,J_{\lam_n}(\om\rho)
\cos(\lam_n \te)\, e^{i h z}\,, \qquad \lam_n := {n \pi \over \tez} ~, \\
& \hspace{0cm} \om_k^2 = \om^2 + h^2 \qquad
\mbox{for $k \equiv (n,\om,h) \in \KK \equiv \naturali_0\!\times\!
(0,+\infty)\!\times\!\reali$}
\end{split}\end{equation}
(recall that $\naturali_0 := \{0,1,2,...\}$; again, we assume the measure on
the label space $\KK$ is such that $\int_\KK dk \equiv \sum_{n = 0}^{+\infty}$
$\int_0^{+\infty}d\om$ $\int_{-\infty}^{+\infty}dh$). Also in this case, the
modified cylinder kernel $\Tm$ can be evaluated according to Eq. \rref{SKer}.
More precisely, proceeding as we did in subsection \ref{DirWed} for the
case of Dirichlet boundary conditions (see, in particular, the derivation of
Eq. \rref{TWDir}), we obtain
\beq \Tm(\t\,;\bq,\bp) = \label{TWNeu} \feq
$$ {1 \over 4\pi\tez\,\rho \rho' \sinh \uu}
\l({e^{{\pi \over \tez} \uu} - \cos({\pi \over \tez}(\te\!-\!\te')) \over
\cosh({\pi \over \tez} \uu) -\cos({\pi \over \tez}(\te\!-\!\te'))}
+ {e^{{\pi \over \tez} \uu} - \cos({\pi \over \tez}(\te\!+\!\te')) \over
\cosh({\pi \over \tez} \uu) -\cos({\pi \over \tez}(\te\!+\!\te'))} \r) ; $$
again, $\uu$ si defined as in Eq. \rref{TWSer}. Now, we can resort
once more to Eq.s (\ref{T00W}-\ref{TijW}) to evaluate the renormalized
VEV of the stress-energy tensor; the result is
\beq \la 0 | \Ti_{\mu\nu}(\bq) | 0 \ra_{ren}
\Big|_{\mu,\nu = 0,\rho,\te,z} = A(\bq) \! \l(\!\!\barray{cccc}
-1\,    &   0   &   \!\!0\!         &   \!0 \\
0\,     &   1   &   \!\!0\!         &   \!0 \\
0\,     &   0   &   \!\!-3\rho^2\!  &   \!0 \\
0\,     &   0   &   \!\!0\!         &   \!1 \farray \!\!\r) + \label{TDWNeu} \feq
$$ + \!\l(\!\xi\! -\!{1 \over 6}\!\r)\!\!\l[\!\l(\!\!\!\barray{cccc}
-(B(\bq)\!+\!C(\bq))\!\!\!\!& \!\!\!\!0\!               & \!0\!\!\!             & \!\!\!0 \\
0\!\!\!\!                   & \!\!\!\!B(\bq)\!          & \!-\rho\,E(\bq)\!\!\! & \!\!\!0 \\
0\!\!\!\!                   & \!\!\!\!-\rho\,E(\bq)\!   & \!\rho^2\,C(\bq)\!\!\!& \!\!\!0 \\
0\!\!\!\!                   & \!\!\!\!0\!               & \!0\!\!\!             & \!\!\!B(\bq)\!+\!C(\bq) \farray\!\!\!\r)
\!\!
+ G(\bq)\!\!\l(\!\!\barray{cccc}
-2\!\!&   \!0\! &   \!\!0\!         &   \!0 \\
0\!\! &   \!-1\!&   \!\!0\!         &   \!0 \\
0\!\! &   \!0\! &   \!\!3\rho^2\!   &   \!0 \\
0\!\! &   \!0\! &   \!\!0\!         &   \!2 \farray \!\!\r)\!\r]\! , $$
where the functions $A,B,C,E$ are defined as in Eq. \rref{TDW} and we set
\beq G(\bq) := {\pi^2\!-\!\tez^2 \over 12\pi^2 \tez^2\,\rho^4} ~. \feq
We notice that, in accordance with the existing literature (see, e.g.,
\cite{Deu}), the conformal part of the renormalized stress-energy VEV
coincides with the analogous contribution derived for Dirichlet boundary
conditions in subsection \ref{DirWed}; besides, comments analogous to
those made at the end of the cited subsection also hold in this case.
Let us only remark that the VEV $\la 0 | \Ti_{\mu\nu}(\bq) | 0 \ra_{ren}$
has an additional term proportional to the function $G(\bq)$; this function
changes sign for either $\tez < \pi$ or $\tez > \pi$ and diverges for
$\rho \to 0$\,. \salto
Concerning the pressure on the boundary, also in this case Eq. \rref{alt2W}
clearly yields a divergent result; on the other hand, using Eq. \rref{alt1W},
we obtain
\beq p^{ren}_i = -{\de_{i\te} \over 4\pi^2\rho^3}\l[{\pi^4\!-\!\tez^4 \over 120\tez^4}
- \l(\!\xi\!-\!{1 \over 6}\r){\pi^2\!-\!\tez^2 \over \tez^2}\r] .
\label{pressWN} \feq
As in the previous subsection, we find that the renormalized pressure
depends of the parameter $\xi$\,. \salto
Proceeding as explained in subsection \ref{genremW}, the renormalized
stress-energy VEV \rref{TDWNeu} and pressure \rref{pressWN} are easily
seen to give for $\tez = \pi$ and $\tez = \pi/2$, respectively,
Eq.s \rref{TP3} \rref{press1Pzer} with $\al_1 = 1$, and Eq.s \rref{T2P3}
\rref{press2Pzer} with $\al_1 = \al_2 = 1$.\\
$\phantom{a}$\vspace{-0.85cm}
\subsection{Periodic boundary conditions (the cosmic string).}\label{string}
Finally, let us consider the case where the field fulfills periodic
boundary conditions on the half-planes $\pi_0,\pi_\tez$, meaning that
\begin{equation}\begin{split}
& \Fi(t,\rho,0,z) = \Fi(t,\rho,\tez,z) ~, \qquad
\partial_\te\Fi(t,\rho,0,z) = \partial_\te\Fi(t,\rho,\tez,z) \\
& \hspace{3cm} \mbox{for $t,z \in \reali$, $\rho \in (0,+\infty)$} ~.
\end{split}\end{equation}
In passing, let us mention that the same framework was also analysed
by Dowker \cite{Dow} and by Fulling et al. \cite{FulWed}, both
employing a point-splitting approach; more precisely, in \cite{Dow}
the conformal part of the energy density alone is computed, while
in \cite{FulWed} the authors only report the graphs of the energy
density and pressure for $\xi = 1/4$. \parn
Similarly to the cases of the segment and parallel hyperplanes
configurations with periodic boundary conditions (considered,
respectively, in subsection 6.9 of Part I and in subsection
\ref{PPPerSubsec} of the present paper), the spatial domain $\Om$
for the present setting is more properly addressed as a flat
Riemannian manifold. The manifold $\Om$ has a global coordinate
system $\bq = (\rho, \te, z): \Om \to (0,+\infty) \times \Toro^1_\al \times \reali$,
$\bx \mapsto \bq(\bx)$ where the second factor is the one-dimensional
torus $\Toro^1_\al := \reali/(\al \interi)$; the line element in these
coordinates has the form \rref{dellWed}
(\footnote{In other words, $\Om$ is a quotient space of the
Dowker manifold $\Om_{\infty}$; this is an infinite-sheeted
Riemannian surface that can be described in terms of a global
coordinate system
$$ \bq : \Om_\infty \to (0,+\infty)\times \reali \times \reali ~,
\qquad \bx \mapsto \bq(\bx) = (\rho(\bx),\te(\bx),z(\bx))  $$
(and of the line element \rref{dellWed}).
}).
The corresponding spacetime $\reali\times\Om$ (with the line element
$ds^2 = -dt^2+d\ell^2$) is usually described in terms of a ``cosmic
string'' due to the presence of a $1$-dimensional topological defect
coinciding with the axis $\{\rho=0\}$. \parn
A complete orthonormal system of (improper) eigenfunctions
$(\Fk)_{k \in \KK}$ of $\AA$ in $L^2(\Om)$, with the related
eigenvalues $(\om_k^2)_{k \in \KK}$, is given by
\begin{equation}\begin{split}
& \hspace{0.5cm}\Fk(\bq) := \sqrt{\om \over 2\pi\tez} \,
J_{|\lam_n|}(\om\rho)\;e^{i \lam_n\,\te}\, e^{i h z} \,,
\qquad \lam_n := {2 n \pi \over \tez} ~, \\
& \hspace{0cm} \om_k^2 := \om^2 + h^2 \qquad \mbox{for $k \equiv (n,\om,h)
\in \KK \equiv \interi\!\times\! (0,+\infty) \!\times\!\reali$}
\end{split}\end{equation}
(similarly to the previous subsections, we are assuming $\KK$
to be a measure space such that $\int_\KK dk\equiv\sum_{n = -\infty}^{+\infty}$
$\int_0^{+\infty}d\om$ $\int_{-\infty}^{+\infty}dh$). The modified
cylinder kernel $\Tm$ can then be evaluated starting from its
eigenfunction expansion \rref{SKer}:
\begin{equation}\begin{split}
& \hspace{3.cm}\Tm(\t\,;\bq,\bp) = \int_\KK dk\;{e^{-t\om_k} \over \om_k}
\;\Fk(\bx)\Fkc(\by) = \\
& {1 \over 2\pi\tez}\sum_{n = -\infty}^{+\infty}\!e^{i\lam_n(\te-\te')}
\!\int_0^{+\infty}\hspace{-0.5cm} d\om\;\om\,J_{\lam_n}(\om\rho)\,J_{\lam_n}(\om\rho')
\!\int_{-\infty}^{+\infty} \hspace{-0.5cm}dh\,{e^{-t\sqrt{\om^2+h^2}}\over
\sqrt{\om^2\!+\!h^2}}\;e^{ih(z-z')}~.
\end{split}\end{equation}
Evaluating the integrals in $h$ and $\om$ as in \cite{FulWed} and
considering separately the terms with positive and negative values
of $n$, we obtain the expression
\beq \Tm(\t\,;\bq,\bp) = {1\over 2\pi\tez\rho\rho'\sinh \uu}\l(1 +
\sum_{n = 1}^{+\infty}e^{-\lam_n \uu}\,e^{i \lam_n(\te-\te')} +
\sum_{n = 1}^{+\infty}e^{-\lam_n \uu}\, e^{-i \lam_n(\te-\te')} \r)\!,
\label{TWSerP} \feq
which in turn, summing the geometric series, yields
\beq \Tm(\t\,;\bq,\bp) = {1 \over 2\pi\tez\,\rho \rho' \sinh \uu}\;
{\sinh({2\pi \over \tez} \uu) \over \cosh({2\pi \over \tez} \uu) -
\cos({2\pi \over \tez}(\te\!-\!\te'))} ~
\label{TWP} \feq
(again, $\uu$ is defined as in Eq. \rref{TWSer}). Using Eq.s
(\ref{T00W}-\ref{TijW}) once more, we find the following
expression for the renormalized stress-energy VEV:
\begin{equation*}\begin{split}
& \hspace{3.5cm} \la 0|\Ti_{\mu\nu}(\bq)|0\ra_{ren}\Big|_{\mu,\nu = 0,\rho,\te,z}= \\
& A(\bq) \! \l(\!\!\barray{cccc}
-1\,& 0   &   \!\!0\!         &   \!0 \\
0\, & 1   &   \!\!0\!         &   \!0 \\
0\, & 0   &   \!\!-3\rho^2\!  &   \!0 \\
0\, & 0   &   \!\!0\!         &   \!1 \farray \!\!\r)\!
+ \l(\!\xi\! -\!{1 \over 6}\r)G(\bq)\!\l(\!\!\barray{cccc}
-2\!\!  &   0\! &   \!0\!       &   0 \\
0\!\!   &   -1\!&   \!0\!       &   0 \\
0\!\!   &   0\! &   \!3\rho^2\! &   0 \\
0\!\!   &   0\! &   \!0\!       &   2 \farray \!\r) ,
\end{split}\end{equation*}
\beq A(\bq) = {(2\pi)^4-\al^4 \over 1440\pi^2 \al^4 \rho^4} ~, \label{TPW} \qquad
G(\bq) = {(2\pi)^2-\al^2 \over 24 \pi^2 \al^2 \rho^4} ~. \feq
Let us observe that the above result does not depend explicitly
on the angular variable $\te$; this was to be expected due to the
homogeneity of the considered configuration with respect to this
coordinate. Besides, as for the cases of Dirichlet and Neumann
boundaries, both the conformal and non-conformal part of the
renormalized VEV of the stress-energy tensor diverge near the
axis $\{\rho = 0\}$, that is in the proximity of the cosmic string. \parn
In conclusion, we notice that for $\tez = 2\pi$, in which case
the considered configuration is equivalent to that of a scalar
field on the whole Minkowski spacetime, Eq. \rref{TPW} gives
$A(\bq) = G(\bq) = 0$. So, we have this result with its own interest:
when zeta regularization is applied to a massless scalar field
on the whole $(3+1)$-dimensional Minwkowski spacetime, the
renormalized stress-energy VEV vanishes identically.
\vskip 0.5cm \noindent
\textbf{Acknowledgments.}
This work was partly supported by INdAM, INFN and by MIUR, PRIN 2010
Research Project  ``Geometric and analytic theory of Hamiltonian systems in finite and infinite dimensions''.
\vfill \eject \noindent
%%%Appendici Parte II
\vfill \eject \noindent
%%Bibliografia Parte II

%%%Fine Parte II%%%%%%%%%%%%%%%%%%%%%%%%%%%

\end{document}